\documentclass[rmp,makeidx,showidx,tabularx,multicol,aps,epsf]{revtex4}

\usepackage{perpage}
\MakePerPage{footnote}

\usepackage{graphicx}
\usepackage{setspace}
\newenvironment{itemize*}%
  {\begin{itemize}%
    \setlength{\itemsep}{0pt}%
    \setlength{\parskip}{0pt}}%
  {\end{itemize}}

\newenvironment{enumerate*}%
  {\begin{enumerate}%
    \setlength{\itemsep}{0pt}%
    \setlength{\parskip}{0pt}}%
  {\end{enumerate}}

\usepackage{amssymb}

\usepackage[makeroom]{cancel}
\usepackage{bm}  
\usepackage{graphicx}
\usepackage{amssymb}
\usepackage{color}
\usepackage{amscd}
\usepackage{epsfig}
\usepackage{epstopdf}
\usepackage{enumerate}
\usepackage{hyperref}
\usepackage{amsmath,eufrak,amsthm,amscd}
\usepackage{eucal}
\usepackage{eufrak}
\usepackage{chemarr}

\numberwithin{equation}{section} 
\numberwithin{figure}{section}

\newcommand{\DD}[0]{\bm{D_\mathrm{c}}}
\newcommand{\RR}[0]{\bm{R}}
\newcommand{\pp}[0]{\bm{p}}
\newcommand{\kB}[0]{k_{\mathrm{B}}}
\newcommand{\ex}[1]{\mathrm{e}^{#1}}
\newcommand{\XX}[0]{\bm{X}}
\newcommand{\FF}[0]{\bm{F}}
\newcommand{\xx}[0]{\bm{x}}
\newcommand{\uu}[0]{\bm{u}}
\newcommand{\nn}[0]{\bm{n}}
\newcommand{\dd}[0]{\mathrm{d}}
\newcommand{\cc}[0]{\mathrm{c}}
\newcommand{\ee}[0]{\mathrm{e}}
\newcommand{\sss}[0]{\mathrm{s}}
\newcommand{\tot}[0]{\mathrm{tot}}
\newcommand{\A}[0]{\mathrm{A}}
\newcommand{\B}[0]{\mathrm{B}}
\newcommand{\bmu}[0]{\boldsymbol{\mu}}
\newcommand{\rotop}[0]{{\bm {\mathcal R}}}

\newcommand{\rr}[0]{\boldsymbol{r}}
\newcommand{\qq}[0]{\boldsymbol{q}}

\newcommand{\Dc}[0]{D_\mathrm{c}}

\newcommand{\beq}{\begin{equation}}
\newcommand{\eeq}{\end{equation}}
\newcommand{\beqa}{\begin{eqnarray}}
\newcommand{\eeqa}{\end{eqnarray}}
\newcommand{\bem}{\begin{math}}
\newcommand{\eem}{\end{math}}

\newcommand{\dt}[1]{{{\rm d} #1 \over {\rm d} t }}

\newcommand{\kT}{{k_{\rm B} T}}
\newcommand{\cP}{{\cal P}}

\newcommand{\bfr}{{\bm r}}

\newcommand{\bfu}{{\bm u}}
\newcommand{\bfn}{{\bm n}}

\newcommand{\bfp}{{\bm p}}

\newcommand{\bfx}{{\bm x}}
\newcommand{\bff}{{\bm f}}
\newcommand{\bfv}{{\bm v}}
\newcommand{\bfe}{{\bm e}}

\newcommand{\bfQ}{{\bm Q}}
\newcommand{\bfI}{{\bm I}}
\newcommand{\bfR}{{\bm R}}

\newcommand{\bfV}{{\bm V}}

\newcommand{\bfF}{{\bm F}}
\newcommand{\bfJ}{{\bm J}}

\newcommand{\aver}[1]{\left\langle {#1}\right\rangle}

\begin{document}
\title{{\sc Phoretic Active Matter}}

\author{Ramin Golestanian}
\affiliation{Max Planck Institute for Dynamics and Self-Organization, G\"ottingen, Germany \\ Rudolf Peierls Centre for Theoretical Physics, University of Oxford, UK}

\date{\today}

\begin{abstract}
\texttt{These notes are an account of a series of lectures I gave at the Les Houches Summer School ``Active Matter and Non-equilibrium Statistical Physics'' during August and September 2018.}

\end{abstract}

\maketitle

\tableofcontents

\newpage

\section{Introduction}

Driven motion of colloidal particles under the effect of external fields---generally termed as phoretic transport---have been studied for more than a century. The external field that leads to the driven transport could come from a gradient in electrostatic potential (electrophoresis), solute concentration (diffusiophoresis), or temperature (thermophoresis), and the motion of the particles is caused by the interaction of the ambient fluid with the modified interfacial structure near their surfaces \cite{derj47,derj87,ande82,prie82,ande84,Anderson1989}. Although ``driven'' by these external fields, the colloidal particles experience zero net force. For example, in electrophoresis the electric field applies equal and opposite forces on the charged colloid and the comoving neutralizing cloud of counterions, in diffusiophoresis the forces exerted by the asymmetric distribution of solute particles around the colloid is reacted back to the comoving cloud, and in thermophoresis similar mutual interactions are involved depending on the constituents of the system (binary mixture, charged colloid, etc). The force-free nature of the phoretic transport mechanisms suggests that they could be used in designing self-propelled particles, provided we equip them with a mechanism that could create the appropriate gradient that could lead to directed motion, e.g. by using Janus particles with built-in sources \cite{Golestanian:2007}. This design strategy has led to the development of a rich variety of microscopic self-propelled colloids that have been used to realize active matter in experiments and study their properties.

A key aspect of the nonequilibrium phoretic mechanisms that are used to design self-propulsion is that they lead to the formation of fields that mediate long-range interactions by the very nature of their nonequilibrium activity. The existence of such long-range fields implies that theoretical descriptions of self-propelled particles with short-range equilibrium-type interactions might be unrealistic when it comes to systems that rely on phoretic mechanisms for self-propulsion. Therefore, it is imperative that theoretical descriptions of the collective behaviour of such active colloids take into account phoretic interactions. As we will show here, such a description can be used to study nonequilibrium collective properties of systems that cover a wide range of length scale, from chemically active molecules such as enzymes to active colloids and chemotactic cells.

The lecture notes are organized as follows. In Sec. \ref{sec:diff}, I will discuss the nature of diffusiophoresis as a key nonequilibrium transport mechanism, followed by a microscopic account of the phenomenon in Sec. \ref{sec:MicDiff}. Section \ref{sec:self-diff} is devoted to the theoretical development of the notion of self-diffusiophoresis, and Sec. \ref{sec:stoch} follows with a description of the stochastic properties of self-phoretically active colloids. The relevant experimental developments have been reviewed in Sec. \ref{sec:exp}. In Sec. \ref{sec:apolar-steer}, the collective behaviour of apolar active colloids that are driven by a light source is discussed, where comet-like swarming appears as an emergent property. Mixtures of apolar particles are considered in Secs. \ref{sec:molecules} and \ref{sec:apolar-susp}, followed by a detailed description of the moment expansion techniques in the context of polar active colloids in Sec. \ref{sec:polar-mom} and scattering of such polar particles in Sec. \ref{sec:polar-scat}. Section \ref{sec:enz} is devoted to the nonequilibrium dynamics of catalytically active enzymes, while Sec. \ref{sec:trail} gives an account of collective chemotaxis in the limit of slow chemical diffusion. Finally, Sec. \ref{sec:growth} addresses the competition between cell division and chemotaxis and Sec. \ref{sec:concl} closes the notes with some concluding remarks.

\section{What is Diffusiophoresis} \label{sec:diff}

\begin{figure}[b]
\includegraphics[width=0.5\linewidth]{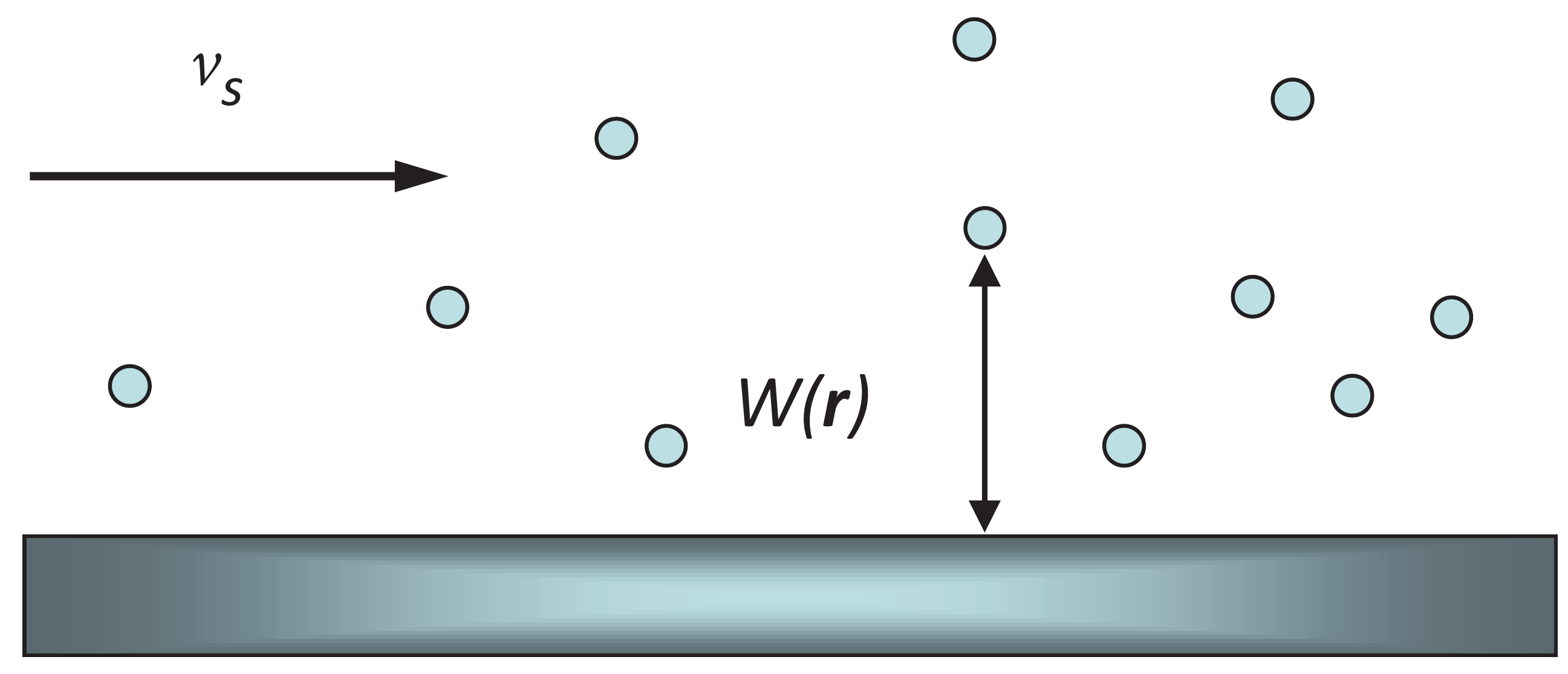}
\caption{Fluid moves past a solid surface in the presence of a solute concentration gradient due to the interaction $W(\bfr)$ between the solute particles and the solid surface. The so-called fluid slip velocity is as shown when the interaction is predominantly repulsive.\label{phoresis-1}}
\end{figure}

To understand diffusiophoresis we need to take a proper account of the effect of the boundaries on the solvent and the solute. Let us zoom in near a boundary and assume that the boundary is interacting with the solute particles with an interaction potential $W({\bfr})$, leading to the force of ${\bfF}=-\nabla W({\bfr})$ acting on individual particles see Fig.~\ref{phoresis-1}. The continuity equation for the solute concentration $\rho({\bfr},t)$  can be written as 
\begin{equation}
\partial_t \rho+\nabla \cdot {\bfJ}=0, \label{cont-eq}
\end{equation}
where the current density ${\bfJ}({\bfr},t)$ is defined as
\begin{equation}
{\bfJ}=-D \nabla \rho+\beta D \rho (-\nabla W)+\rho {\bfv},\label{j-def}
\end{equation}
where $\beta=1/(\kT)$ and Einstein relation is assumed between the mobility and diffusion coefficient. The corresponding governing equation equation for the solvent is given by Stokes equation
\begin{equation}
-\eta \nabla^2 {\bfv}=-\nabla p+{\bff},\label{stokes-1}
\end{equation}
that is complemented by the incompressibility constraint
\begin{equation}
\nabla \cdot {\bfv}=0.\label{incomp-eq}
\end{equation}
In Eq. (\ref{stokes-1}), $\eta$ is the viscosity, $p$ is the pressure, and ${\bff}$ represents the body force density acting on the solvent. Noting that the force acting on the solute particles will be transmitted to the solvent by way of force balance for each solute particle, we can write
\begin{equation}
{\bff}=\rho {\bfF}=\rho (-\nabla W).\label{f-def}
\end{equation}
This relation closes the system of equations for the solvent and the solute that should be simultaneously solved for the concentration and the velocity profiles.

In the stationary situation, the equation for the concentration reads
\begin{equation}
-D \nabla^2 \rho+{\bfv} \cdot \nabla \rho+\beta D  \nabla \cdot {\bff}=0.\label{c-eq-1}
\end{equation}
Using the incompressibility constraint, we can also find an equation for the pressure, which reads
\begin{equation}
-\nabla^2 p+\nabla \cdot {\bff}=0.\label{p-eq-1}
\end{equation}
Combining Eqs. (\ref{c-eq-1}) and (\ref{p-eq-1}) yields
\begin{equation}
\nabla^2 \left[p-\kT \rho \right]+\frac{1}{\beta D} \, {\bfv} \cdot \nabla \rho=0,\label{p-ktc-eq}
\end{equation}
which does not involve the body force that incorporates the molecular interactions.

The potential $W$ is expected to have a very short range, say $\sigma$, through which it starts from infinity---representing the impenetrability or the excluded volume effect of the surface---and decays to zero. If the wall is impenetrable to the solute particles, the normal current should be negligibly small in the vicinity of the wall, $J_\perp \simeq 0$. If the wall is impenetrable to the fluid, the normal fluid velocity should also be negligibly small near the wall, $v_\perp \simeq 0$. Then, Eq. (\ref{j-def}) requires that the singular contribution in that neighborhood due to $\nabla W$ is balanced by the gradient of concentration, namely
\begin{equation}
-k_{\rm B} T \partial_\perp \rho+\rho \,(-\partial_\perp W) \simeq 0, \label{balance-slip}
\end{equation}
within a distance $\sigma$ from the wall. This can be solved to give
\begin{equation}
\rho_{\rm s}({\bfr})=\rho_{\rm out} \;{\rm e}^{-{W({\bfr})}/{k_{\rm B} T}},\label{c-nearWall}
\end{equation}
in the ``slip'' region, where $\rho_{\rm out}$ is the concentration of the solute immediately after the wall potential has died off. Note that Eq. (\ref{c-nearWall}) implies a strong depletion of the solute particles near the wall ($\rho\left|_{\rm wall}=0\right.$).

The presence of the body force in the above equations implies that both the concentration and the pressure have singular behaviors in the vicinity of the wall. However, Eq. (\ref{p-ktc-eq}) suggests that the combination $p-k_{\rm B} T \rho$ is a smooth function through the domain of action of the wall potential, and does not entail any singular terms despite both $p$ and $\rho$ having singular behaviour near the wall. In fact, to be consistent with the above approximation scheme, the velocity term that represents advection should be neglected in this equation in the vicinity of the wall, or the so-called slip region. 

Using this smoothness property, we can relate the pressure and concentration profiles in the slip region to the those in the outer region, as follows
\begin{equation}
p_{\rm s}=p_{\rm out}+k_{\rm B} T (\rho_{\rm s}-\rho_{\rm out})=p_{\rm out}+k_{\rm B} T \rho_{\rm out} \left[{\rm e}^{-{W({\bfr})}/{k_{\rm B} T}} -1\right],\label{press-slip}
\end{equation}
where $p_{\rm out}$ is the pressure just outside the slip domain, and Eq. (\ref{c-nearWall}) is used to arrive at the second form.

\begin{figure}[t]
\includegraphics[width=0.7\linewidth]{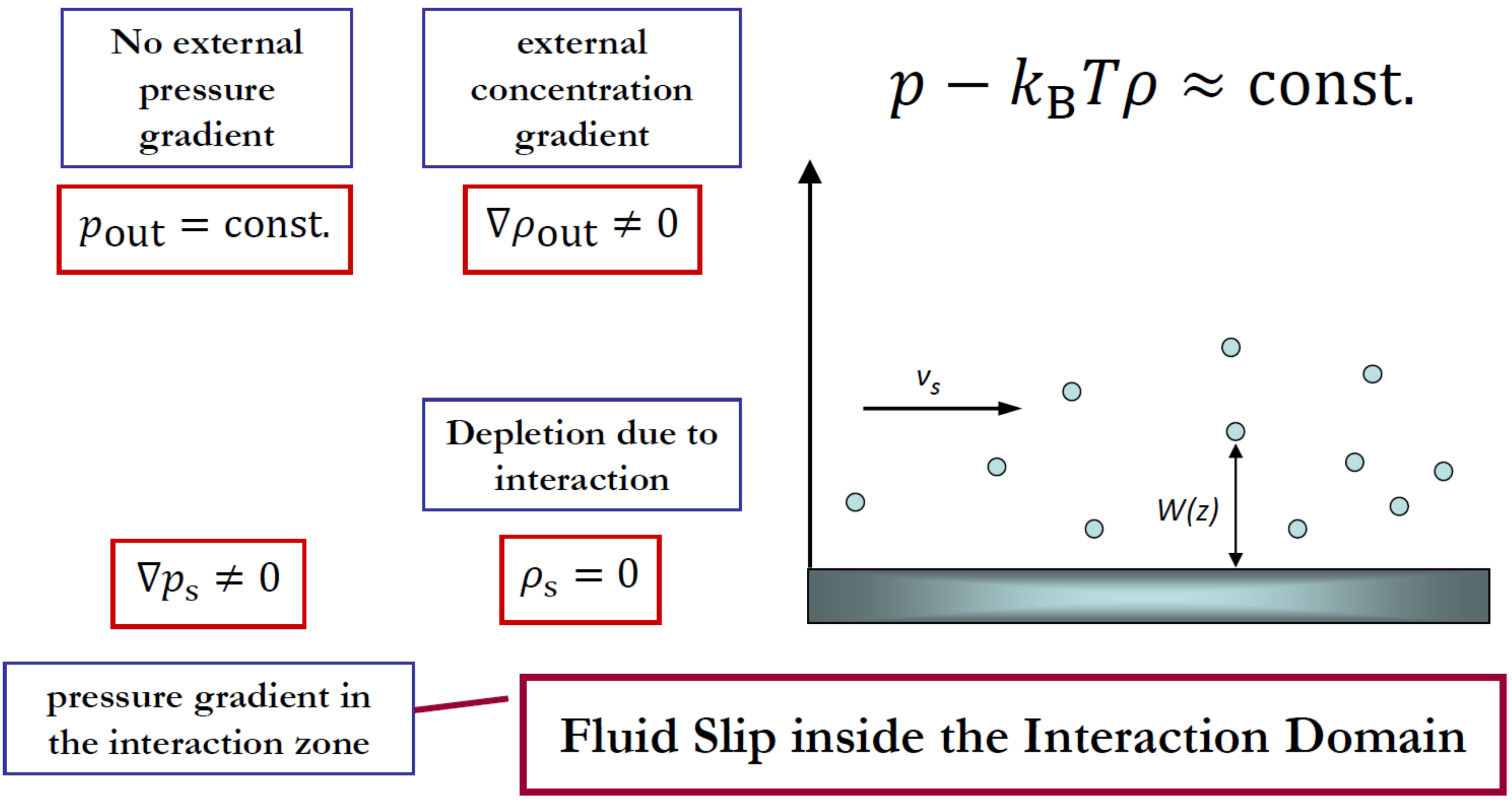}
\caption{We can divide space into the slip region near the surface, where the interaction $W(\bfr)$ is strong, and the outer region far from the surface. Whereas the pressure $p$ and the solute concentration $\rho$  behave very differently in the two regions, the combination $p-k_{\rm B} T \rho$ behaves smoothly and allows us to relate the values in the two regions, see Eq.~(\ref{press-slip}). The pressure gradient in the slip region then generates a fluid slip velocity, which has the direction shown when the interaction is predominantly repulsive. \label{phoresis-2}}
\end{figure}

The above framework suggests that we can separate the slip region (where the interaction is at work) from the outer region, see Fig.~\ref{phoresis-2}, work out the velocity of the solvent at the boundary between the two regions, and use it as a boundary condition for the outer problem.

\subsection{Slip Velocity near Surfaces} \label{sec:slipvelocity}

The flow field inside the slip region can be determined within our approximation using the Stokes equation that ensures force balance for the component of the velocity that is parallel to the surface
\begin{equation}
-\eta \nabla^2 v_\parallel=-\partial_\parallel p,\label{stokes-slip}
\end{equation}
where the body force is neglected because it is assumed to be in the perpendicular direction ($f_\parallel=0$). Since we are interested in propulsion that is driven by
concentration gradient and not external pressure gradient, we assume $\partial_\parallel p_{\rm out} = 0$. Using Eq. (\ref{press-slip}), Eq. (\ref{stokes-slip}) gives
\begin{equation}
-\nabla^2 v_\parallel=\frac{k_{\rm B} T}{\eta} (\partial_\parallel \rho_{\rm out}) \left[1-{\rm e}^{-{W(\bfr)}/{k_{\rm B} T}}\right].\label{stokes-slip-2}
\end{equation}
We note that variations in the (normal) $z$-direction are considerably stronger than variations in the parallel direction, we can implement a lubrication-like approximation $\nabla^2 \approx \partial_z^2$.  The relevant boundary conditions are $\left.\eta \partial_z v_\parallel\right|_{z \to \infty}=0$ (zero externally applied shear rate) and $\left.v_\parallel\right|_{z=0}=0$ (no-slip condition at the wall). Integrating the first moment of Eq. (\ref{stokes-slip-2}) with respect to the normal coordinate $z$ subject to the above boundary conditions, we find $-\int_0^{\infty} \dd z \, z \partial_z^2 v_\parallel=\left.v_\parallel\right|_{z \to \infty} \equiv v_{\rm s}$ on the left hand side, where the slip velocity $v_{\rm s}$ is defined as the asymptotic value of the parallel velocity. We thus obtain
\begin{equation}
v_{\rm s}=\mu \,\partial_\parallel \rho_{\rm out},\label{v-slip}
\end{equation}
where
\begin{equation}
\mu=\frac{k_{\rm B} T}{\eta} \int_{0}^\infty \dd z \, z \left[1-{\rm e}^{-{W(z)}/{k_{\rm B} T}}\right].\label{v-slip2}
\end{equation}
is the {\em phoretic mobility} of the system. The surface slip velocity outside of the slip layer can be used together with the Stokes equation to solve for the flow field. Defining the Derjaguin length via
\begin{equation}
\lambda_{\rm D}^2=\int_{0}^\infty \dd z \, z \left[1-{\rm e}^{-{W(z)}/{k_{\rm B} T}}\right].\label{Derjaguin-length}
\end{equation}
we can write the phoretic mobility as 
\begin{equation}
\mu=\frac{k_{\rm B} T}{\eta} \lambda_{\rm D}^2. \label{eq:mu-def}
\end{equation}
Note that $\lambda_{\rm D}^2 >0$ corresponds to cases where $W$ is predominantly repulsive, whereas  $\lambda_{\rm D}^2 <0$ corresponds to cases where $W$ is predominantly attractive. When $\lambda_{\rm D} \ll R$, where $R$ is the characteristic radius of curvature of the surface, the slip boundary condition on the fluid velocity can be used.

\subsection{Phoretic Drift Velocity of Colloidal Particles} \label{sec:phoretic-drift-velocity}

We can now make use of the reciprocal theorem of Lorentz \cite{Lorentz1896,Happel1981,Stone} to calculate the propulsion velocity directly from the surface slip. To do this, we start from the definition of stress tensor for an incompressible flow
\begin{equation}
\sigma_{ij}=- p \delta_{ij}+\eta (\partial_i v_j+\partial_j v_i),
\end{equation}
using which, we can write the governing equations of the Stokes flow as
\beqa
\nabla \cdot {\bfv} &=&0, \\
\nabla \cdot {\bm \sigma}&=&0.
\eeqa
If we have two solutions of the above equations in the same domain ${\cal D}(t)$, namely, $({\bfv}^{(1)},{\bm \sigma}^{(1)})$ and $({\bfv}^{(2)},{\bm \sigma}^{(2)})$, then we know from Green's theorem that the following relation holds between them
\beq
\int_{{\cal S}} \dd S \; \bfn \cdot {\bm \sigma}^{(2)} \cdot {\bfv}^{(1)}=\int_{{\cal S}} \dd S \; \bfn \cdot {\bm \sigma}^{(1)} \cdot {\bfv}^{(2)},\label{eq:Green}
\eeq
where $\bfn$ is the normal unit vector perpendicular to the surface ${\cal S}$ that defines the boundary of ${\cal D}$. Let us now choose $({\bfv}^{(1)},{\bm \sigma}^{(1)})$ to be the force-free and torque-free motion of an object with a surface slip velocity boundary condition, and $({\bfv}^{(2)},{\bm \sigma}^{(2)})$ to describe the motion of the same object when dragged through the viscous fluid by an external force $\bfF^{(2)}$ with velocity $\bfV^{(2)}$. Since ${\bfv}^{(2)}|_{\cal S}=\bfV^{(2)}$ and solution (1) is force-free, then the right hand side of Eq. (\ref{eq:Green}) vanishes. We can split the velocity of solution (1) as ${\bfv}^{(1)}|_{\cal S}=\bfV^{(1)}+\bfv_{\rm s}$, where $\bfV^{(1)}$ is a net drift velocity for the particle and the relative velocity component is given by the surface slip velocity $\bfv_{\rm s}$. With this composition, Eq. (\ref{eq:Green}) gives $\bfF^{(2)} \cdot \bfV^{(1)}=-\int_{{\cal S}(t)} \dd S \; \bfn \cdot {\bm \sigma}^{(2)} \cdot {\bfv}_{\rm s}$. Considering that for a sphere of radius $a$ we have $\bfn \cdot {\bm \sigma}^{(2)}=\frac{1}{4 \pi R^2}\bfF^{(2)}$, we find the drift velocity of the force-free and torque-free sphere as
\begin{equation}
{\bfV}=-\frac{1}{4 \pi R^2}\int_{{\cal S}} \dd S \;{\bfv}_{\rm s},\label{SS-1}
\end{equation}
where we have dropped the superscript (1). 

For a diffusiophoretic sphere, we find
\beq
{\bfV}=-\frac{1}{4 \pi R^2}\int_{{\cal S}} \dd S \;\mu \nabla_{\parallel}\rho_{\rm out}=-\mu \nabla_{\parallel}\rho_{\rm out}^{\infty},\label{SS-2}
\eeq
where a solution for diffusion equation with vanishing normal flux boundary condition around a sphere has been used to perform the integration.

In a similar manner, we can show that the angular velocity of a spherical particle is given as
\beq
{\bm \Omega}=-\frac{3}{8 \pi R^3}\int_{{\cal S}} \dd S \; \hat{\bm e}_r \times {\bfv}_{\rm s}.\label{ASS-1}
\eeq
Consider a patterned particle with axial symmetry about a given axis $\bfn$, with a mobility pattern that can be expanded in the basis of Legendre polynomials as $\mu(\theta)=\sum_\ell \mu_\ell P_\ell(\cos\theta)$. Then, Eq. (\ref{ASS-1}) yields
\beq
{\bm \Omega}=-\frac{3}{4} \frac{\mu_1}{R} \Big(\bfn \times \nabla_{\parallel}\rho_{\rm out}^{\infty}\Big),\label{ASS-2}
\eeq
which shows that the mobility pattern of the particle should have a non-vanishing first harmonic in order for diffusiophoresis to lead to an angular velocity in a concentration gradient.



\section{Microscopic Theory of Diffusiophoresis}	\label{sec:MicDiff}

The subtle interactions between solute molecules, colloids, and the incompressible solvent that lead to diffusiophoretic drift can be understood more easily from a microscopic perspective. Let us start from a reduced two-body description of the problem where we consider a colloidal particle A at position $\RR$ and a solute molecule B at position $\XX$, which interact via a the potential $W^{\A\B}(\RR-\XX)$ within the framework of the Fokker-Planck equation. The relevant governing equation for the two-body distribution reads \cite{agud18}
\begin{eqnarray}
\partial_t \rho_{\A \B}(\RR,\XX,t) & = & \nabla_{\RR}\cdot\bmu^{\A \A}\cdot \left[\kT \nabla_{\RR}\rho_{\A \B}+\left(\nabla_{\RR} W^{\A\B}\right) \rho_{\A \B}\right] \nonumber\\
& + & \nabla_{\RR}\cdot\bmu^{\A \B}\cdot \left[\kT \nabla_{\XX}\rho_{\A \B}+\left(\nabla_{\XX} W^{\A\B}\right) \rho_{\A \B}\right] \nonumber\\
& + & \nabla_{\XX}\cdot\bmu^{\B \A}\cdot \left[\kT \nabla_{\RR}\rho_{\A \B}+\left(\nabla_{\RR} W^{\A\B}\right) \rho_{\A \B}\right] \nonumber\\
& + & \nabla_{\XX}\cdot\bmu^{\B \B}\cdot \left[\kT \nabla_{\XX}\rho_{\A \B}+\left(\nabla_{\XX} W^{\A\B}\right) \rho_{\A \B}\right], 
 \label{eq:twocoup2}
\end{eqnarray}
where the $\bmu$'s are the relevant mobility coefficients that account for the hydrodynamic interactions. Integrating over $\XX$ yields
\begin{eqnarray}
\partial_t \rho_{\A}(\RR,t) & = & \nabla_{\RR}\cdot \int_{\XX} \bmu^{\A \A}(\RR,\XX) \cdot \left[\kT \nabla_{\RR}\rho_{\A \B}+\left(\nabla_{\RR} W^{\A\B}\right) \rho_{\A \B}\right] \nonumber\\
& + & \nabla_{\RR}\cdot  \int_{\XX} \bmu^{\A \B}(\RR,\XX) \cdot \left[\kT \nabla_{\XX}\rho_{\A \B}+\left(\nabla_{\XX} W^{\A\B}\right) \rho_{\A \B}\right],
 \label{eq:twocoup3}
\end{eqnarray}
which is not a closed equation because it involves both single-body and two-body distributions. Assuming the solution is dilute and B particles are point-like, $\bmu^{\A\A}$ will not depend on $\XX$; in fact, it will be a constant provided there are no boundaries in the system. Then we find
\beq
\partial_t \rho_{\A}(\RR,t) =  \kT \nabla_{\RR}\cdot \bmu^{\A \A} \cdot \nabla_{\RR}\rho_{\A}+\nabla_{\RR}\cdot  \int_{\XX} \left(\bmu^{\A \B}-\bmu^{\A \A}\right) \cdot \left[\kT \nabla_{\XX}\rho_{\A \B}+\left(\nabla_{\XX} W^{\A\B}\right) \rho_{\A \B}\right],
 \label{eq:twocoup4}
\eeq
where we have made use of $\nabla_{\RR} W^{\A\B}=-\nabla_{\XX} W^{\A\B}$ and ignored a boundary term. To close the hierarchy, we use a product approximation
\beq
\rho_{\A\B}(\RR,\XX;t)  = \rho_{\A}(\RR,t) \rho_{\B}(\XX,t) \;\ex{- W^{\A \B}(\RR-\XX)/\kT}, 
\eeq
which gives us the following simplified result:
\beq
\partial_t \rho_{\A}(\RR,t) =  \kT \nabla_{\RR}\cdot \bmu^{\A \A} \cdot \nabla_{\RR}\rho_{\A}+\kT \nabla_{\RR}\cdot \left[\int_{\XX} \left(\bmu^{\A \B}-\bmu^{\A \A}\right) \cdot \left(\nabla_{\XX}\rho_{\B}\right) \; \ex{- W^{\A \B}(\RR-\XX)/\kT} \rho_{\A}(\RR,t)\right],
 \label{eq:twocoup5}
\eeq
Noting that the mobilities are divergence-free due to the incompressibility constraint and using $\bmu^{\A\B}=\left(\bmu^{\B \A}\right)^{T}$, we find
\begin{eqnarray}
\partial_t \rho_{\A}(\RR,t) & = & \kT \nabla_{\RR}\cdot \bmu^{\A \A} \cdot \nabla_{\RR}\rho_{\A}\nonumber \\
& + & \kT \nabla_{\RR}\cdot \left\{\left[ \int_{\XX}\left(\ex{- W^{\A \B}(\RR-\XX)/\kT}-1\right)\left(\bmu^{\A \B}-\bmu^{\A \A}\right) \cdot \nabla_{\XX}\rho_{\B}\right]\rho_{\A}\right\},\label{eq:twocoup6}
\end{eqnarray}
which is in the form of a drift-diffusion equation
\beq
\partial_t \rho_{\A}(\RR,t)= \nabla_{\RR}\cdot \DD \cdot \nabla_{\RR}\rho_{\A}-\nabla_{\RR}\cdot \left[\bfV(\RR) \rho_{\A}\right],
\eeq
with the diffusivity tensor
\beq
\DD=\kT \bmu^{\A \A},
\eeq
and the phoretic drift velocity
\beq
\bfV(\RR)=- \kT\int_{\XX}\left[\ex{- W^{\A \B}(\RR-\XX)/\kT}-1\right]\left[\bmu^{\A \B}(\RR,\XX)-\bmu^{\A \A}\right] \cdot \nabla_{\XX}\rho_{\B}.\label{eq:V-phor}
\eeq

For a spherical colloid of radius $R$, we have the hydrodynamic mobility tensors as
\begin{equation}
\boldsymbol{\mu}^{\A \A} = \frac{1}{6 \pi \eta R} \boldsymbol{I},
\end{equation}
and
\begin{equation}
\boldsymbol{\mu}^{\A \B} = \frac{1}{6 \pi \eta R} \left[ \frac{1}{4} \left( 3 \, \frac{R}{r} + \frac{R^3}{r^3} \right) \left(\boldsymbol{I}-\hat{\bm e}_r \hat{\bm e}_r\right) + \frac{1}{2} \left( 3 \,\frac{R}{r} - \frac{R^3}{r^3} \right) \hat{\bm e}_r \hat{\bm e}_r\right],
\end{equation}
where $r=|\RR-\XX|$ is the distance between A and B, and $\hat{\bm e}_r$ is the radial unit vector pointing from A to B. Combining both, we obtain
\begin{equation}
\bmu^{\A \B}-\bmu^{\A \A}= \frac{1}{6 \pi \eta R} \left[  \left( -1 + \frac{3}{4} \,\frac{R}{r} + \frac{1}{4} \,\frac{R^3}{r^3} \right) \left(\boldsymbol{I}- \hat{\bm e}_r \hat{\bm e}_r\right) + \left(-1 + \frac{3}{2}\, \frac{R}{r} - \frac{1}{2}\,\frac{R^3}{r^3} \right)  \hat{\bm e}_r \hat{\bm e}_r \right],\label{eq:mu-mu}
\eeq
which is to be inserted in Eq. (\ref{eq:V-phor}). As the expression of the integrand in Eq. (\ref{eq:V-phor}) involves $\ex{- W^{\A \B}(r)/\kT}-1$, the expression in Eq. (\ref{eq:mu-mu}) can be expanded near $r=R$ when we are dealing with relatively short-range interactions. Setting $r=R+\delta$, we find 
\begin{equation}
\bmu^{\A \B}-\bmu^{\A \A}= -\frac{\delta}{4 \pi \eta R^2}\left(\boldsymbol{I}- \hat{\bm e}_r \hat{\bm e}_r\right) +O(\delta^2),\label{eq:mu-mu-small}
\eeq
which yields
\beq
\bfV(\RR)=- \frac{\kT}{\eta} \times \overbrace{\int_{0}^\infty \dd \delta \, \delta \left[1-{\rm e}^{-{W^{\A \B}(\delta)}/{\kT}}\right]}^{ \lambda_{\rm D}^2}  \times \overbrace{\int \frac{\dd \Omega}{4 \pi} \; \left(\boldsymbol{I}- \hat{\bm e}_r \hat{\bm e}_r\right) \cdot \nabla_{\XX}\rho_{\B}(R,\Omega)}^{\nabla \rho_{\B}^{\infty}}=-\mu \nabla \rho_{\B}^{\infty},\label{eq:V-phor-2}
\eeq
where $\Omega$ represents the solid angle and the phoretic mobility $\mu$ is given as defined in Eq. (\ref{eq:mu-def}) above. When there is no separation of length scales between the range of the interaction and the radius of the sphere, the full form of the expression in Eqs. (\ref{eq:V-phor}) and (\ref{eq:mu-mu}) should be used.

\section{Self-diffusiophoresis}		\label{sec:self-diff}

Since diffusiophoresis is force-free---as are all other interfacial phoretic transport mechanisms---it can be used to make self-propelled particles or microswimmers, if the system generates the gradient internally \cite{Golestanian:2005}. If we consider the case with a small Peclet number, namely, ${\rm Pe}=\frac{V R}{D} \ll 1$, we can decouple the reaction-diffusion equation that governs the dynamics of the solute molecules from the Stokes equation that governs the dynamics of the (viscous) solvent. The case with finite Peclet number poses additional technical complexities \cite{Michelin2014}.

Since the time scale for solute diffusion around the colloid is considerably shorter than the typical time scale for colloid movement, we can consider the concentration of the solute, $C(\bfr)$, to be a quasi-stationary solution of the following reaction-diffusion equation
\beq
-D \nabla^2 C=0,
\eeq
subject to the boundary condition on the surface of the colloid at position $\bfr_s$ 
\beq
\left.-D \bfn \cdot \nabla C\right|_{\bfr=\bfr_s}=\alpha(\bfr_s),
\eeq
where the {\bf activity} $\alpha$ gives the normal flux of solute particles on the surface, and is a measure of the {\it nonequilibrium activity} of the system. The resulting solution for the concentration can be used together with the slip boundary condition
\begin{equation}
{\bfv}_s=\mu(\bfr_s) \left(\boldsymbol{I}-\bfn \bfn \right) \cdot \nabla C(\bfr_s),\label{vs-mu}
\end{equation}
to obtain the propulsion (or swimming) velocity of spherical colloids as 
\begin{equation}
{\bfV}=-\frac{1}{4 \pi R^2}\int_{{\cal S}} \dd S \;\mu(\bfr_s) \left(\boldsymbol{I}-\bfn \bfn \right) \cdot \nabla C(\bfr_s),\label{V-mu-gradC}
\end{equation}
using Eq. (\ref{SS-1}). Here, we have allowed for position dependent {\bf mobility} $\mu(\bfr_s)$, which is a local measure of the fluid {\it response} to the concentration gradient. Note that $\alpha$ and $\mu$ can each be positive or negative.

For an axially symmetric distribution of activity, which can be achieved via coatings of catalytic patches with specific patterns, we can describe the activity as an expansion in appropriate harmonic modes (Legendre polynomials), namely, $\alpha(\theta)=\sum_\ell \alpha_\ell P_\ell(\cos\theta)$. The solution for the concentration profile will read $C(r,\theta)=C^{\infty}+\frac{R}{D}\sum_\ell \frac{\alpha_\ell}{\ell+1}\left(\frac{R}{r}\right)^{\ell+1} P_\ell(\cos\theta)$. Using the mobility profile $\mu(\theta)=\sum_\ell \mu_\ell P_\ell(\cos\theta)$, we can find the following expression for the propulsion velocity
\beq
{\bfV}=-\frac{\hat{\bm e}_z}{D}\sum_\ell \left(\frac{\ell+1}{2 \ell+3}\right) \alpha_{\ell+1} \left(\frac{\mu_\ell}{2 \ell+1}-\frac{\mu_{\ell+2}}{2 \ell+5}\right).
\eeq
This expression demonstrates the level of symmetry breaking in activity and mobility that is necessary to achieve self-propulsion, as a manifestation of the celebrated Curie principle \cite{Curie1894}. Calculations performed for other shapes have revealed that geometry can also play a key role in providing the necessary symmetry breaking in combination with activity and mobility \cite{Golestanian:2007,Ibrahim2018}. The symmetry breaking can also be achieved via shape asymmetry \cite{Popescu2011,Michelin2015,Reigh2018} or even spontaneously \cite{Michelin2013}. These effects have also been investigated and verified using Stochastic Rotation Dynamics (SRD) simulations \cite{Kapral2007,Reigh2018}.

If we have more than one species of chemicals, as it is common with the case of catalytic chemical reactions with several reactants and products, the above calculation should be done for all species $k$, and the resulting expression for the slip velocity will be a superposition of all the contributions in the form of
\begin{equation}
{\bfv}_s=\sum_{k} \mu^{(k)}(\bfr_s) \left(\boldsymbol{I}-\bfn \bfn \right) \cdot \nabla C^{(k)}(\bfr_s).\label{vs-mu-k}
\end{equation}
For independent species, $C^{(k)} \sim \alpha^{(k)}/D_k$, and the resulting propulsion velocity will be a superposition of the different contributions. For species that are interlinked through catalytic reactions, the resulting correlations will be reflected in the result. This suggests that the direction of propulsion is in general quite sensitive to the details and can even change for the same system under different conditions.

The surface slip velocity profile will lead to a hydrodynamic flow field generated in the vicinity of these swimmers. In free space in 3D, the flow profile decays as $1/r^3$ for Janus particles that are fore-aft symmetric \cite{Golestanian:2005,Golestanian:2007} whereas a profile that is not fore-aft symmetric will lead to a stronger flow that decays as $1/r^2$ \cite{Jlicher2009}. When the swimmers are in contact with a surface, the force monopole that they experience as a result of this contact will change the velocity profile so that it only decays as $1/r$. In Sec. \ref{sec:exp} below we discuss the measured flow field around the Pt-PS catalytic Janus swimmer.

\section{Stochastic Dynamics of Phoretically Active Particles}		\label{sec:stoch}

We now investigate how phoretic activity modifies the stochastic dynamics of a colloid \cite{Golestanian:2009}. In essence we would like to know how we can quantify the behaviour of such active colloids and identify deviations from Brownian motion. We can describe the dynamics of a spherical colloid and the chemical field around it in the comoving frame of reference by solving the relevant diffusion equation for the concentration profile of the solute particles
\begin{equation}
\partial_t C({\bfr},t)-D \nabla^2 C({\bfr},t)=\alpha(\theta,\phi,t) \delta(r-R),\label{diff-eq-1}
\end{equation}
where $\alpha(\theta,\phi,t)$ is the surface activity function of the sphere. The time dependent solution to this equation can be inserted into ${\bfV}(t)=- \int  \frac{d \Omega}{4 \pi}\;\mu \nabla_{\parallel}C(R,\theta,\phi,t)$ to obtain the instantaneous velocity of the colloid. The axis of symmetry of the colloid, which points to the direction of propulsion is defined by the unit vector ${\bf n}(t)=(\sin \theta_n(t) \cos \phi_n(t),\sin \theta_n(t) \sin \phi_n(t),\cos \theta_n(t))$. The stochastic dynamics of ${\bf n}(t)$ due to rotational diffusion causes the cloud of solute particles to constantly redistribute, which will in turn make the velocity of the active colloid fluctuate. We can represent the axially symmetric activity function in terms of the spherical harmonics as
\begin{equation}
\alpha(\theta,\phi,t)=\sum_{\ell,m} \left(\frac{4 \pi }{2\ell+1}\right) \alpha_{\ell} \; Y^*_{\ell m}(\theta_n(t),\phi_n(t))
Y_{\ell m}(\theta,\phi).\label{alpha-Omega}
\end{equation}
Once we determine the instantaneous velocity, we can calculate the mean-squared displacement via
\begin{math}
\Delta L^2(t)=\int_0^t d t_1 \int_0^t d t_2 \left \langle {\bfV}(t_1) \cdot {\bfV}(t_2) \right \rangle.
\end{math}

Equation (\ref{diff-eq-1}) only gives the average density, and the linear relation between the velocity and the concentration profile suggests that in order to calculate velocity correlations we need to incorporate the density fluctuations as well. To this end, we start from the Langevin equation ${\dot{\bfr}}_i(t)={\bfu}_i(t)$ for the $i$-th particle whose position is described by ${\bfr}_i(t)$ and is subject to a random noise ${\bfu}_i(t)$, which has a Gaussian distribution $P[{\bfu}]=\exp\left[-\frac{1}{4 D}\sum_i \int \dd t \;{\bfu}_i(t)^2\right]$ controlled by the diffusion coefficient. We define a stochastic density ${\hat C}({\bfr},t)=\sum_i \delta^3({\bfr}-{\bfr}_i(t))$, which can be seen to satisfy Eq. (\ref{diff-eq-1}) with a noise term ${\hat Q}({\bfr},t)$ added to the right hand side. Using the distribution $P[{\bfu}]$, we can calculate the moments of the noise term, and show that $\langle {\hat Q}({\bfr},t) \rangle=0$ and $\langle {\hat Q}({\bfr},t) {\hat Q}({\bfr}',t')\rangle=2 D (-\nabla^2) \delta^3({\bfr}-{\bfr}') \delta(t-t') C({\bfr},t)$, where $C({\bfr},t)=\langle \hat{C}({\bfr},t) \rangle$.

The dynamics of the system involves a number of different regimes due to the existence of a number of intrinsic time scales.
The rotational diffusion time, $\tau_{\rm r}=4 \pi \eta R^3/k_{\rm B} T$, controls the changes in the orientation of the sphere. The
characteristic diffusion time of the chemicals around the sphere $\tau_{\rm d}=R^2/D$, where $D={k_{\rm B} T}/{(6 \pi \eta a)}$
depends on the radius of the solute particles $a$. This time scale sets the relaxation time of the redistribution of the particles
around the sphere when it changes orientation. Finally, the hydrodynamic time that controls the crossover between the inertial and the
viscous regimes is given as $\tau_{\rm h}=R^2/\nu$, where $\nu=\eta/\rho$ is the kinematic viscosity of water that involves the mass density
$\rho$. We can write the time scales (for water at room temperature and using a typical value of $a=1 ~\AA$) in the following convenient forms:
$\tau_{\rm h}=10^{-6} ~(R/1 \mu{\rm m})^2$ s, $\tau_{\rm d}=10^{-3} ~(R/1 \mu{\rm m})^2$ s, and $\tau_{\rm r}=3 ~(R/1 \mu{\rm m})^3$ s. This shows
that we have a clear separation of time scale with $\tau_{\rm h} \ll \tau_{\rm d} \ll \tau_{\rm r}$, and thus a number of different dynamical regimes in between these scales. 

There are two independent mechanisms driving stochasticity: (i) density fluctuations, which are relevant for $t \sim \tau_{\rm d}$ and for both symmetric (apolar) and asymmetric (polar) coatings of the colloid, and (ii) rotational diffusion, which is relevant for asymmetric particles when $t \sim \tau_{\rm r}$. The interplay between these mechanisms will lead to a number of different dynamical regimes with distinct features, including anomalous diffusion and memory effects, which we highlight below.

\subsection{Anomalous Diffusion}

In the the intermediate regime where $\tau_{\rm h} < t < \tau_{\rm d}$ a symmetric particle can instantaneously propel itself because of polarization of the cloud of solutes due to density fluctuations. This motion, however, will be decorrelated via density fluctuations themselves, leading to fluctuations without symmetry breaking. We can use a scaling argument to characterize the nature of this anomalous dynamics. To build a scaling relation, we consider $\Delta L^2 \sim v(t)^2 t^2$, and insert $v \sim \mu \nabla C \sim \mu \delta C/R$, to find $\Delta L^2 \sim \mu^2 \langle \delta C(t) \delta C(0) \rangle t^2/R^2$. The density auto-correlation function can be written as $\langle \delta C(t) \delta C(0) \rangle=\langle \delta C^2 \rangle k(t)$, involving the density fluctuations $\langle \delta C^2 \rangle$ and the kernel $k(t)$ that controls the relevant relaxation mode. Here, relaxation is controlled by diffusion, hence $k(t) \sim 1/(D t)^{d/2}$ in $d$-dimensions, and the number fluctuations are controlled by the average
number of particles ($\langle \delta N^2 \rangle \sim N_{\rm ave}$, as inherent to any Poisson process), which yields $\langle \delta C^2 \rangle \sim C_{\rm ave}$. The average density is controlled by the average particle production rate (per unit area) $\alpha_0$ as $C_{\rm ave}\sim (\alpha_0 R^{d-1} t)/R^d$.
Putting these all together, we find that the fluctuations exhibit anomalous diffusion
\begin{equation}
\Delta L^2 \sim\frac{\alpha_0 \mu^2}{R^3 D^{d/2}} \; t^{3-d/2}\hskip 1cm  (t \ll \tau_{\rm d}).\label{MSD-sym-1}
\end{equation}
This expression shows that the active velocity fluctuations are controlled by two mechanisms: particle production (that controls the density fluctuations) and diffusion of the produced particles. The exponent $3-d/2$ indicates superdiffusive behaviour for $d < 4$. At time scales longer than $\tau_{\rm d}$, there is a crossover to diffusive behaviour 
\begin{equation}
\Delta L \sim \frac{\alpha_0 \mu^2}{R^{d-1} D^{2}} \; t \hskip1cm (t \gg \tau_{\rm d}).\label{MSD-sym-2}
\end{equation}

\subsection{Memory Effect}

For a given time dependent orientation trajectory, we can calculate the propulsion velocity of the colloid as a function of time by solving Eq. (\ref{diff-eq-1}) without the noise term. This calculation reveals that the propulsion velocity of the colloid at any instant of time depends on the recent history of the orientation as 
\begin{equation}
{\bf v}(t)=\frac{v_0}{\tau_{\rm d}} \int_{-\infty}^t \dd t' {\cal M}(t-t') \;{\bf n}(t'),\label{vMn-def}
\end{equation}
where $v_0=-\alpha_1 \mu/(3 D)$ is the mean propulsion velocity, and the {\em memory} kernel is given as
\begin{math}
{\cal M}(t)=\frac{2}{\pi} \int_0^\infty d u \frac{u^{3/2}}{(u^2+4)}
e^{-u (t/\tau_{\rm d})},
\end{math}
with asymptotic behaviors ${\cal M}(t) \sim t^{-1/2}$ for $t \ll \tau_{\rm d}$ and ${\cal M}(t) \sim t^{-5/2}$ for $t \gg \tau_{\rm d}$. Note that the propulsion velocity is controlled by the $\ell=1$ term ($\alpha_1$) in the surface activity profile. 

Since the rotational diffusion of the colloid randomizes its orientation over the time scale $\tau_{\rm r}$, the velocity autocorrelation function takes on the form of a convolution between two memory kernels and the orientation autocorrelation function. Consequently, the mean-squared displacement will have three different regimes. We find the asymptotic form of
\begin{equation}
\Delta L^2 \sim v_0^2 t^2  (t \ll \tau_{\rm d} \ll \tau_{\rm r}),\label{MSD-asym-1}
\end{equation}
at short times,
\begin{equation}
\Delta L^2 \simeq v_0^2 t^2-\left(\frac{8}{3\sqrt{\pi}}\right) \frac{v_0^2 \tau_{\rm d}^{3/2}}{\tau_{\rm r}} \; t^{3/2}  \hskip1cm  (\tau_{\rm d} \ll t \ll \tau_{\rm r}),\label{MSD-asym-2}
\end{equation}
at intermediate times, and
\begin{equation}
\Delta L^2 \simeq 2 v_0^2 \tau_{\rm r} \; t  \hskip1cm (\tau_{\rm d} \ll \tau_{\rm r} \ll t),\label{MSD-asym-3}
\end{equation}
at long times, with a smooth crossover between them. For  $\tau_{\rm d} < t < \tau_{\rm r}$ the memory effect that exists for self-propelled asymmetric colloids introduces an anomalous {\em anti-correlation} (i.e. contribution with negative sign) in the velocity autocorrelation function and the mean-squared displacement [Eq. (\ref{MSD-asym-2})]. Such anomalous corrections are reminiscent of the effect of the hydrodynamic long-time tail \cite{Alder1967,Zwanzig1970}. Note, however, that the anomalous $-\gamma t^{3/2}$ correction in Eq. (\ref{MSD-asym-2}) corresponds to much longer time scales and should be more easily observable than the hydrodynamic long-time tail.

\subsection{Effective Diffusivity} 

\begin{figure}[t]
\includegraphics[width=0.6\linewidth]{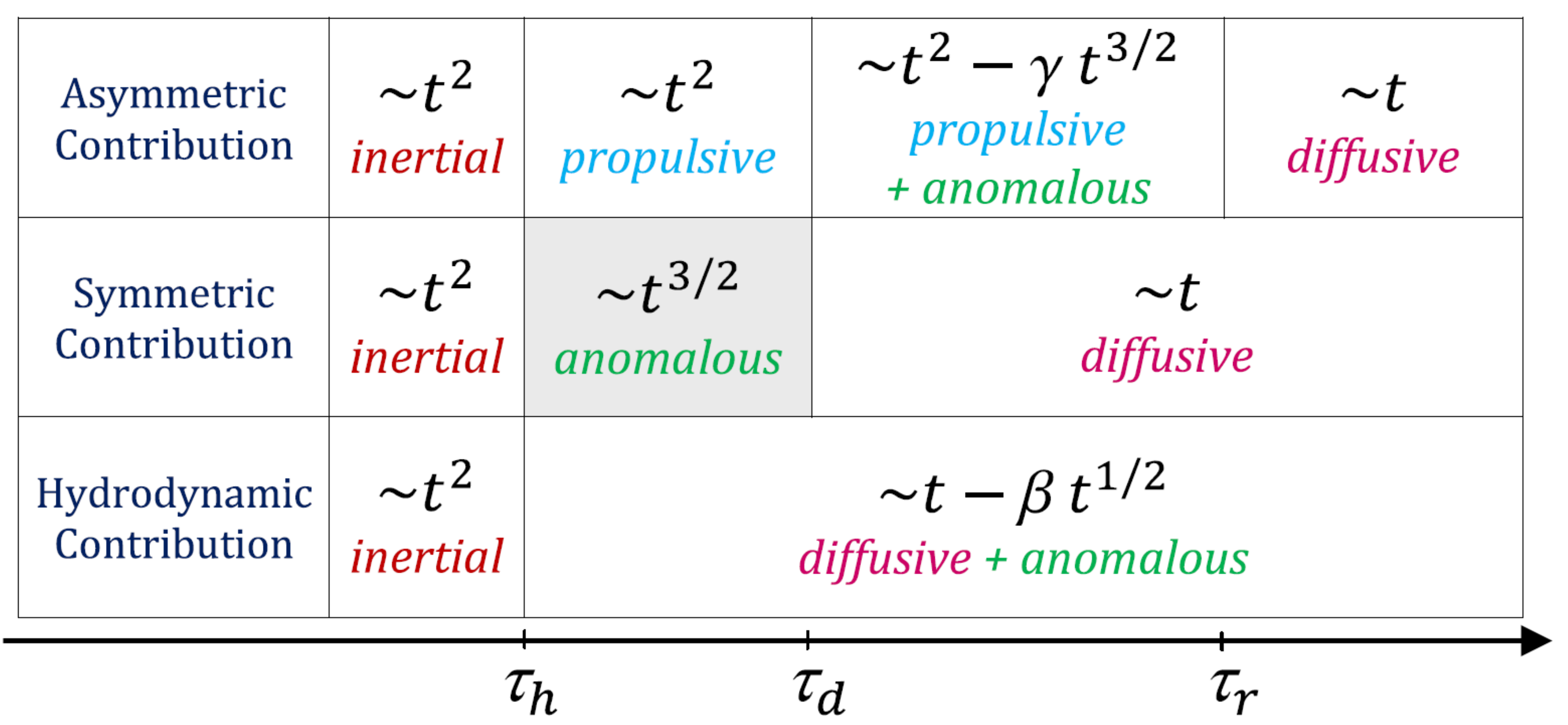}
\caption{Different contributions to the mean-squared displacement of active colloids. The total mean-squared displacement can be obtained by summing all of the contributions for asymmetric colloids, and the bottom two rows (only) for symmetric colloids \cite{Golestanian:2009}.\label{fig:summ}}
\end{figure}

At the longest time scales ($t > \tau_{\rm r}$), all of the contributions are diffusive, leading to a total effective diffusion coefficient
\begin{equation}
D_{\rm eff}=\frac{k_{\rm B} T}{6 \pi \eta R}+\frac{4 \pi \alpha_1^2 \mu^2 \eta R^3}{27 D^2 k_{\rm B} T}+\frac{c_1 \alpha_0 \mu^2}{3 \pi^2 D^2 R^2},\label{Deff}
\end{equation}
where $c_1=1.17180$. The different terms in the above expression exhibit different $R$-dependencies, which causes the asymmetric contribution to be dominant for $R \gtrsim \left[{D k_{\rm B} T}/{(\alpha_1 \mu \eta)}\right]^{1/2}$, while the symmetric contribution takes over when $R \lesssim {\alpha_0 \mu^2 \eta^2}/({D^2 k_{\rm B} T})$. At the shortest time scales, on the other hand, the contribution due to phoretic effects will also be dominated by inertial effects that should lead to ballistic contributions (see Fig. \ref{fig:summ}). Moreover, the different terms depend differently on temperature as well, with the active contributions typically decreasing as temperature is increased contrary to the trend observed in the equilibrium Stokes-Einstein relation.

\section{Experiments on Self-phoresis}		\label{sec:exp}

Spherical Janus particles made from polystyrene (PS) beads that are half-coated with platinum have been shown to self-propel because platinum (Pt) catalyzes the breakdown of hydrogen peroxide into water and oxygen
\beq
{\rm H_2 O_2} \xrightarrow[]{\rm Pt} {\rm H_2 O+\frac{1}{2}O_2},\label{reaction-1}
\eeq
and the continuous flux of the reaction products establishes a steady gradient across the body of the Janus particle \cite{Howse:2007}. Using the Active Brownian Particle model, which was developed for the purpose of analyzing the stochastic trajectories observed from this experiment, it was possible to extract the average propulsion speed of this swimmer as a function of the fuel concentration $C$. It was observed that the speed depends on the fuel concentration according the Michaelis-Menten rule
\beq
V(C)=V_{\infty} \cdot \frac{C}{C+K_M},\label{MM}
\eeq
where $K_M$ is the relevant Michaelis constant. This behaviour is consistent with the catalytic activity that drives the propulsion by setting up the stationary-state gradient, and suggests that the reaction must include a diffusion-limited binding step followed by a reaction-limited step 
\begin{equation}
{\rm H_2 O_2+Pt} \xrightarrow[]{k_1} {\rm Pt(H_2 O_2)}\xrightarrow[]{k_2}{\rm H_2 O+\frac{1}{2}O_2+Pt}.\label{reaction-2}
\end{equation}
The swimming speed was also found to depend on the size of the colloid as $V \sim 1/R$ most of the time, which could be for different reasons, for example a competition between the diffusion- and reaction-limited steps of the catalysis and their interplay with the finite size of the colloid \cite{Ebbens2012}. It is also possible to make self-phoretic active colloids that have spontaneous angular and translational velocities at the same time \cite{Ebbens2010}.

\begin{figure}[t]
\includegraphics[width=0.42\linewidth]{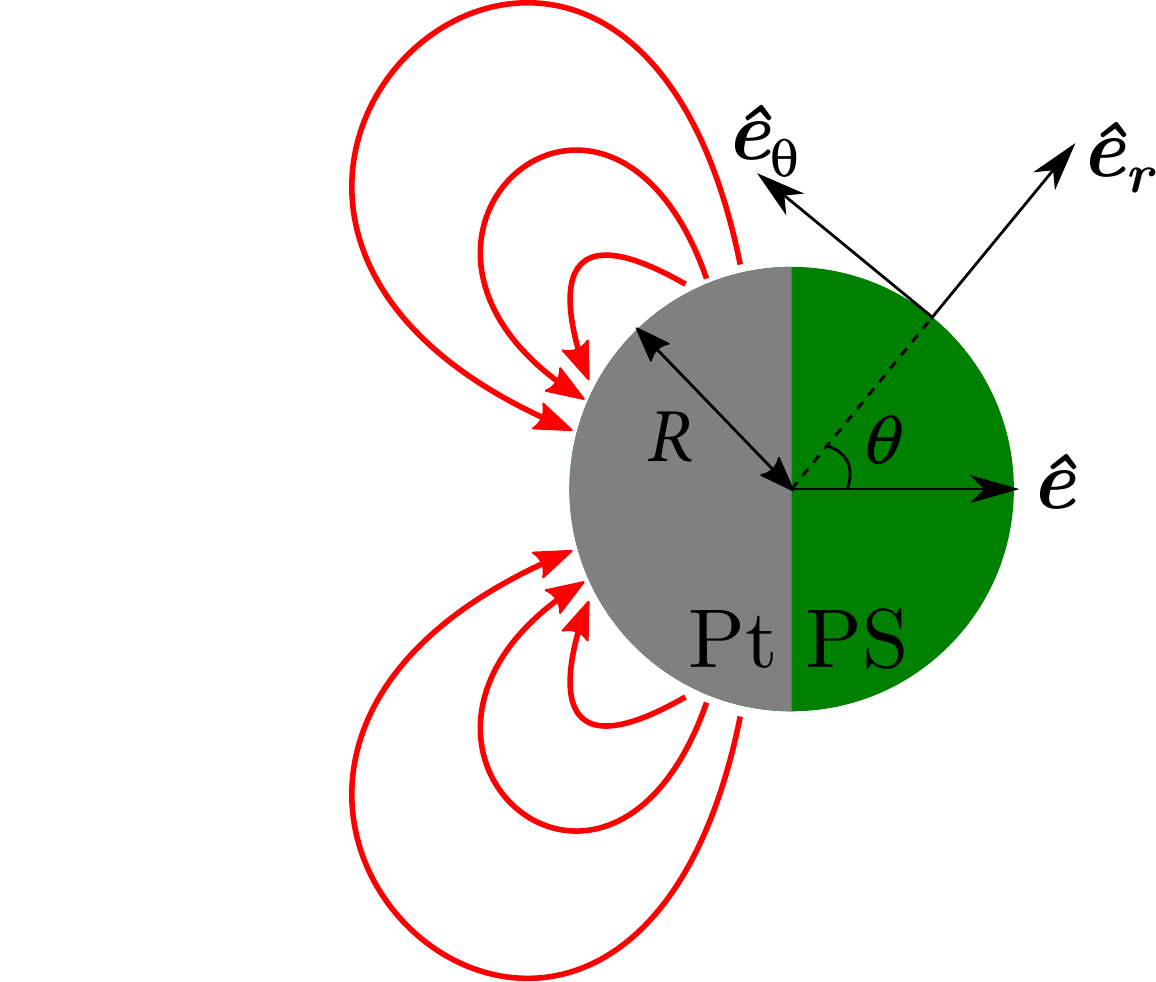}\hskip1cm
\includegraphics[width=0.5\linewidth]{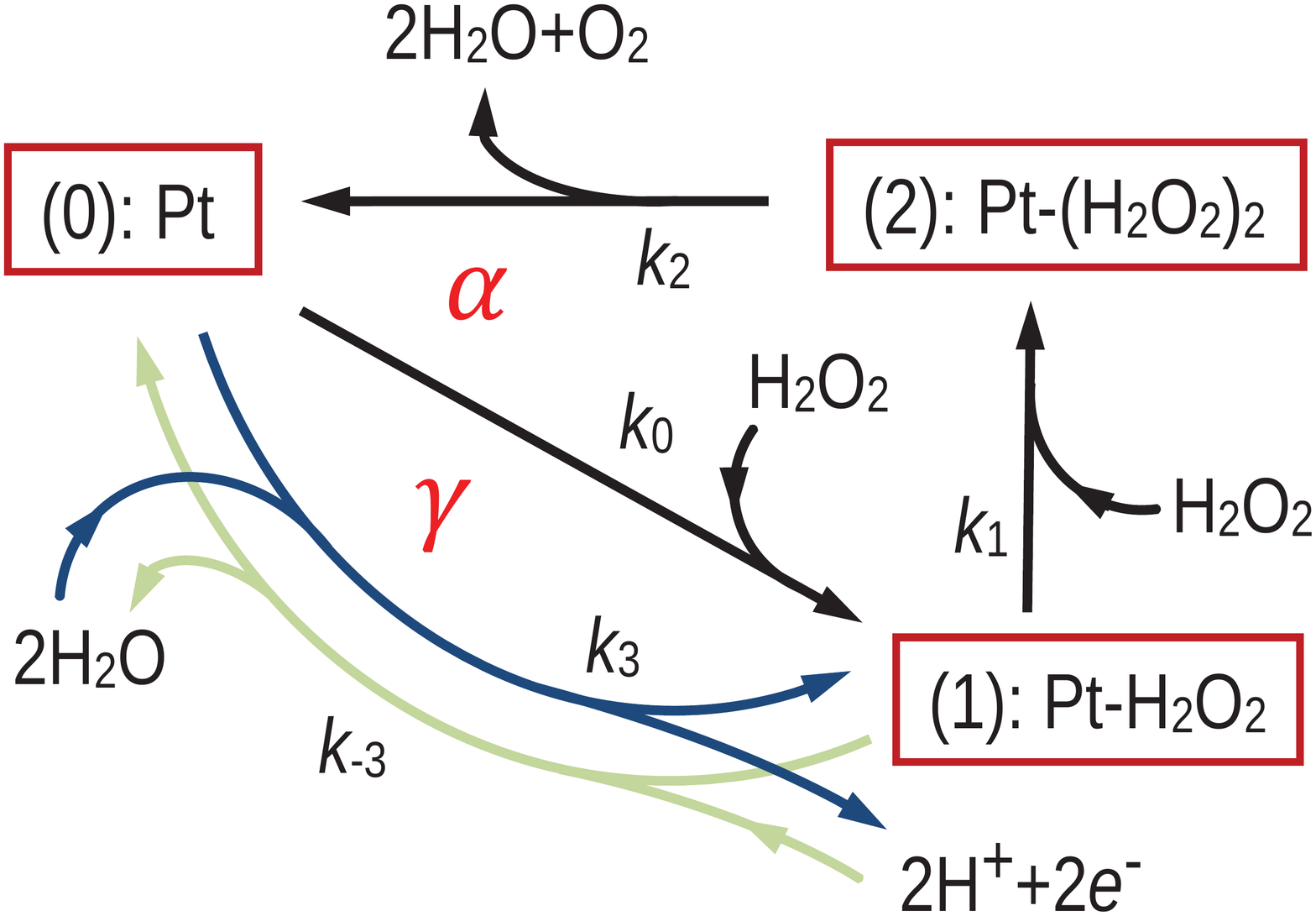}
\caption{Sketch of the Pt-PS Janus particle (left). $\hat{e}$ is the direction of swimming. The red arrows represent the current loops that start at the equator of the particle and end at its pole. The reaction cycle (right), including two sub-cycles with fluxes $\alpha$ and $\gamma$, one of which involves ionic intermediates  \cite{Ebbens2014,Ibrahim2017}.\label{fig:proton-loop}}
\end{figure}

Since the reaction was known to involve only (electrostatically) neutral components, it was a surprise when it was experimentally discovered that adding salt to the solution affects the swimming speed strongly \cite{Ebbens2014,Brown2014}. A possible explanation for this behaviour has been proposed by postulating the existence of ionic intermediates in the catalytic reaction cycle that will give rise to closed loops of current on the platinum coat, as shown in Fig. \ref{fig:proton-loop} \cite{Ebbens2014,Ibrahim2017}. The specific bi-cyclic topology of the reaction has been constructed based on the experimental observations. For example, it has been observed that addition of salt does not significantly affect the rate of consumption of hydrogen peroxide or the rate of production of oxygen, while it strongly affects the swimming speed  \cite{Ebbens2014}. The observations suggest that the Janus particle employs a self-electrophoretic channel in addition to the neutral self-diffusiophoretic channel for its motility.

The conjectured current loops on the Pt hemisphere and the dominance of the resulting self-electrophoretic contribution have a major implication on the distribution of the effective surface slip velocity: the slip velocity will be concentrated on the Pt side, with a profile that can be represented with the following simplified form \cite{das15}
\begin{equation}
\label{slip}
\bfv_{\rm s}|_{r=R} =
\begin{cases}
  v_0(1+\cos\theta)(-\cos\theta) \hat{e}_\theta    & \text{ for $\pi/2<\theta<\pi$}, \\
  0    & \text{otherwise}.
\end{cases}
\end{equation}
This profile has two significant properties: (i) it leads to swimming away from the Pt patch, and (ii) it is not fore-aft symmetric. Interestingly, any surface slip velocity distribution with these two properties should lead to a quenching of the orientation of the swimmer in a direction parallel to the surface due to hydrodynamic interaction. This is indeed observed experimentally \cite{das15}. Interactions between phoretic swimmers and surfaces can lead to a wide variety of different behaviours \cite{Uspal2015,Bayati2019}. In this regards, it has been revealed that the Pt-PS swimmer is a special case. The surface alignment property of the Pt-PS swimmer is not shared by other types, as it arises from its specific form of the slip velocity; usually, other swimmer prototypes do not possess one or both of the above-mentioned necessary properties.   

While the alignment property corroborated the proposed structure of the slip velocity due to the current loops, recent measurements of the complete flow field profile around (swimming and stationary) Pt-PS Janus particles provided a direct visualization of the flow and measurement of the slip velocity profile \cite{Ebbens2018}. As can be seen in Fig. \ref{fig:flow-exp}, the slip velocity is maximum in the middle part of the Pt region, which was in full agreement with the above picture. The measured flow field gave access to the squirmer $B_n$ coefficients, and revealed that for the Pt-PS Janus swimmer $B_2/|B_1|\simeq -2.45 <0$, which is consistent with a {\em pusher}, in terms of the classification for hydrodynamic interactions. However, this experiment has provided a more complete picture with regards to the near-field properties of the hydrodynamic interactions than a simplistic squirmer of pusher type, which can be used to build a more faithful representation of the hydrodynamic interactions. 

\begin{figure}[t]
\includegraphics[width=0.49\linewidth]{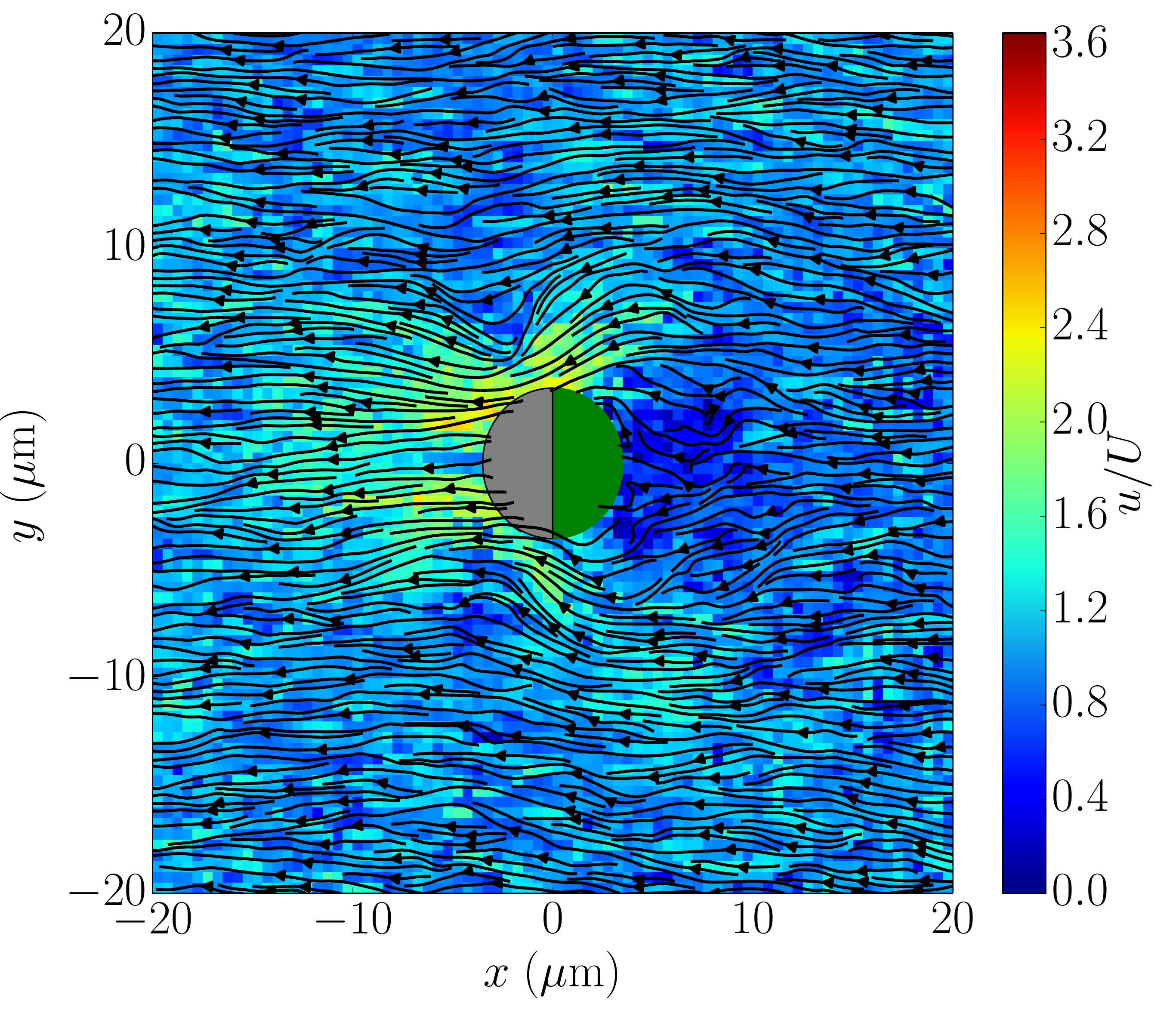}
\includegraphics[width=0.49\linewidth]{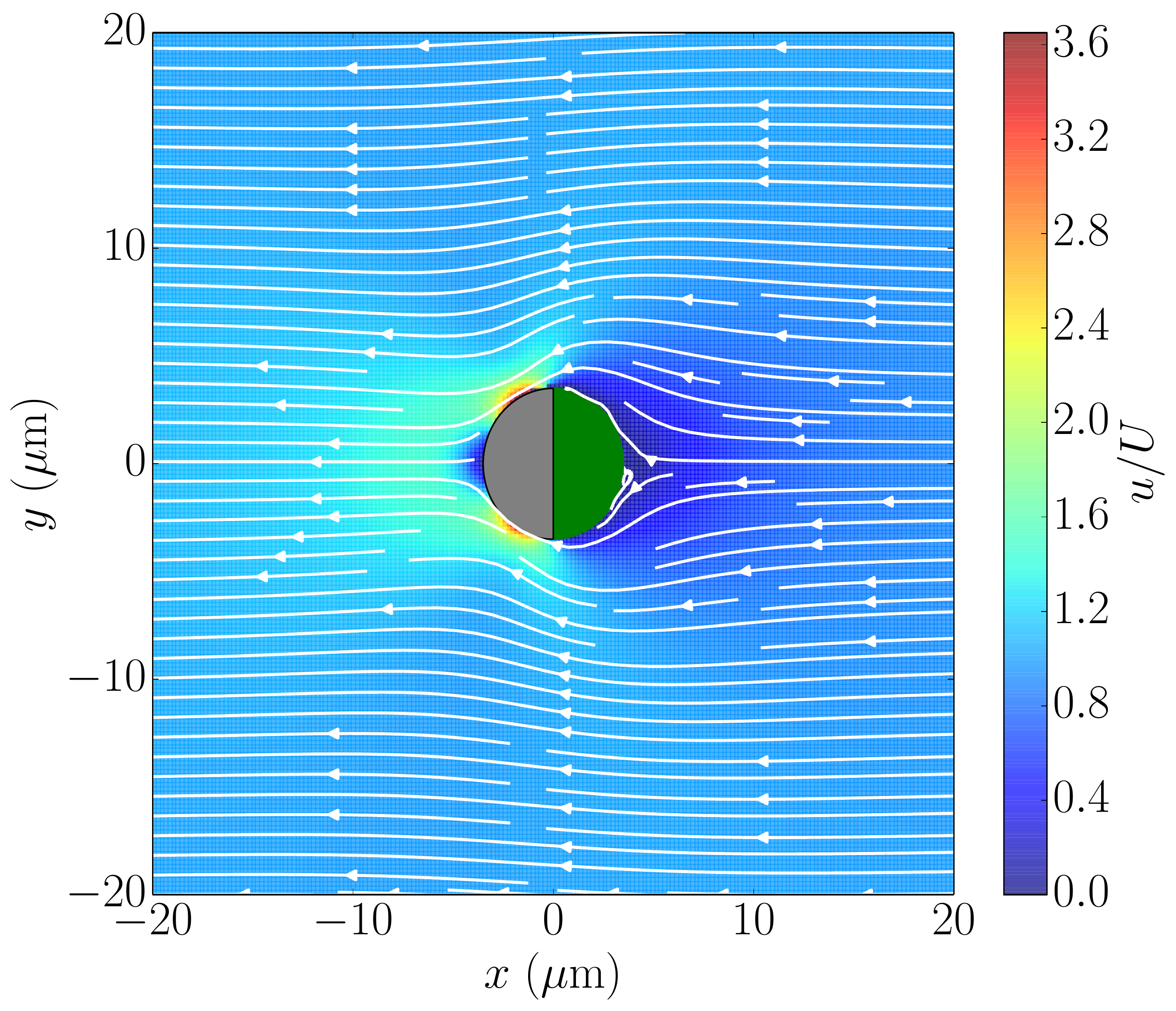}
\caption{Streamlines around freely moving swimming Janus particles obtained experimentally (left) and theoretically (right). The background colors represent the magnitude of the velocity rescaled by the swimming velocity of the Janus particle \cite{Ebbens2018}.
 \label{fig:flow-exp}}
\end{figure}

No other prototype phoretic microswimmer has been experimentally characterized as thoroughly and systematically as the Pt-PS Janus particle.

\section{Apolar Active Colloids: Swarming due to External Steering}			\label{sec:apolar-steer}

Using light as the external source of energy to induce self-thermophoresis is extremely versatile as it can be used to engineer collective swarming behaviour \cite{Cohen:2014oth}. The colloids then take advantage of the natural asymmetries in the system to create non-equilibrium conditions that drive them into a range of collective behaviour. Here a simple system is discussed where colloids that convert light into heat and move in response to self- and collectively generated thermal gradients. The system exhibits self-organization into a moving comet-like swarm with novel non-equilibrium dynamics. Although these active colloids are controlled by viscous hydrodynamics, their collective behaviour shows very dynamic structures with inertial traits. In particular, it exhibits propagation of transverse waves from back to front of the swarm with no dispersion, ejection of hot colloids from the head of the swarm, and persistent circulation flow within the swarm. The rich behaviour of the dynamic comet-like swarm can be controlled by a single external parameter, the intensity of light.

\begin{figure}[t]
\includegraphics[width=0.5\linewidth]{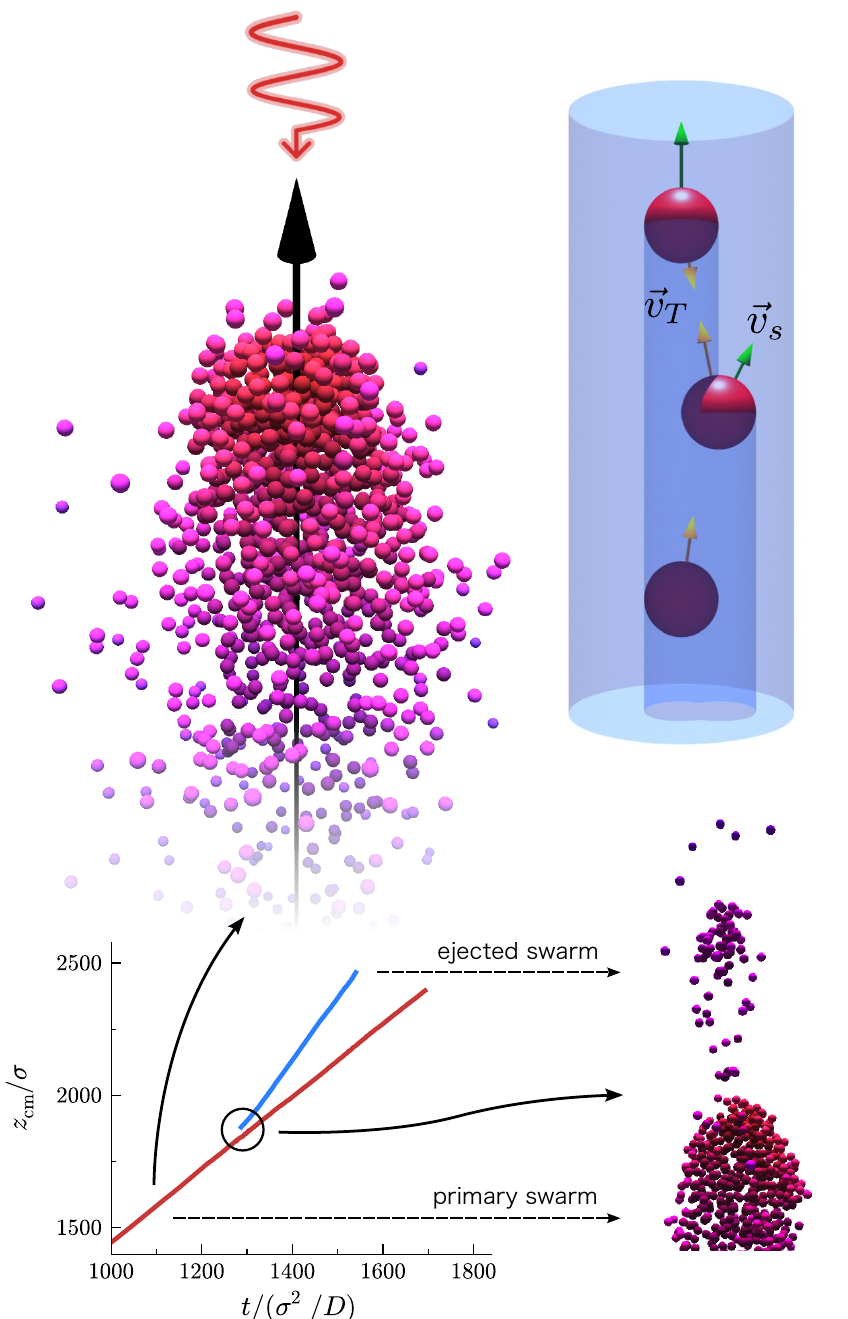}
\caption{Optically driven thermally active colloids self-organize into a moving comet-like swarm under illumination of a uniform intensity plane wave light source (top left). Colloids can cast partial or complete shadows on those below them and propel through production of a collective thermal field, $\bfv_T$ (yellow arrow), and by local self propulsion, $\bfv_{s}$ (green arrow), shown here with the magnitude of $v_T$ magnified by a factor of five for presentation. Fluctuations in the swarm shape facilitates ejection of hot colloids at the swarm tip forming faster moving sub swarms (bottom) \cite{Cohen:2014oth}.
 \label{fig:sch}}
\end{figure}

\begin{figure}[t]

\includegraphics[width=0.9\linewidth]{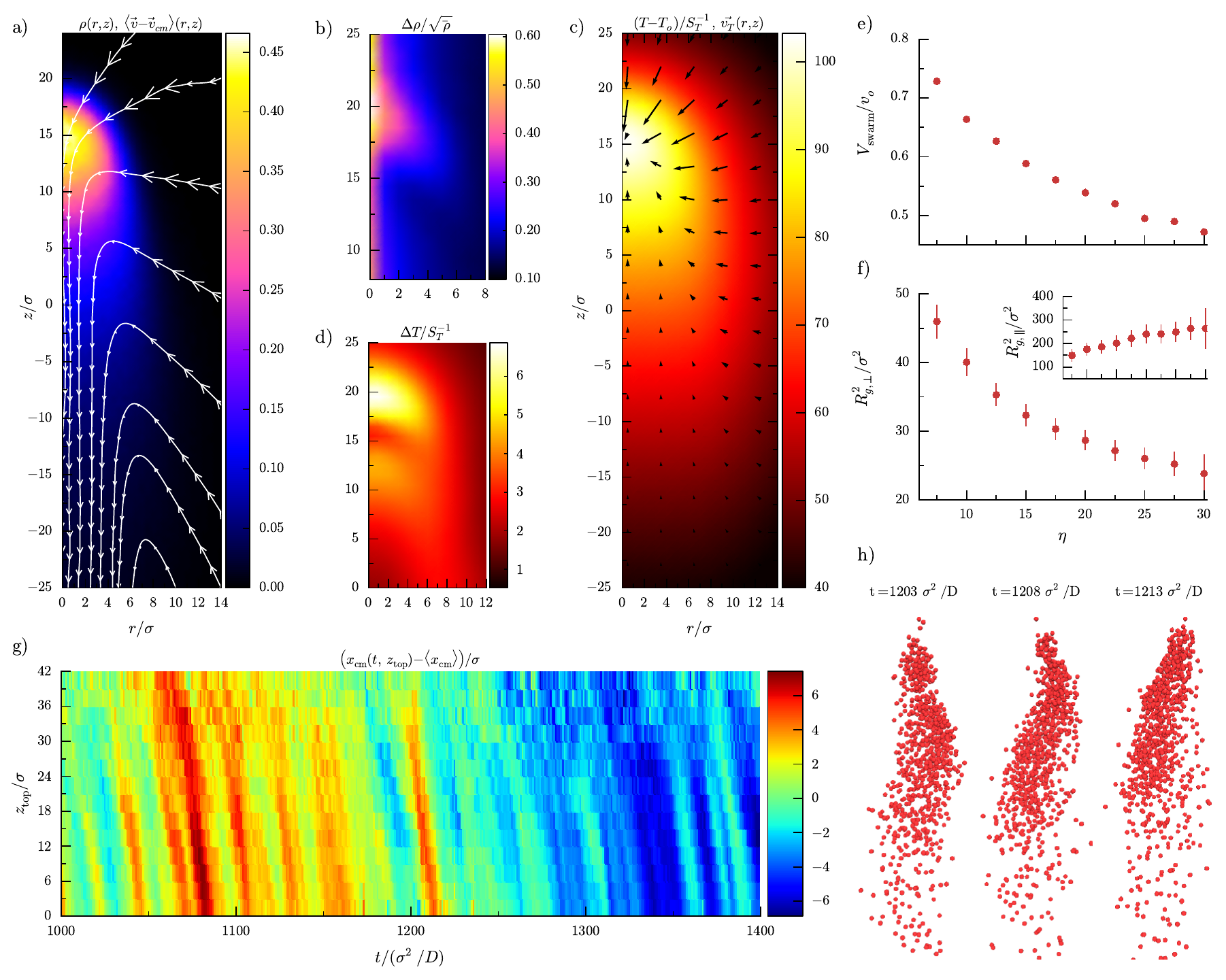}
\caption{Using the moving centre of mass swarm frame of reference fields can be reconstructed using time and ensemble averaging. The swarm fields are axially averaged and presented in cylindrical coordinates $(r,z)$ with $z$ axis pointing towards the light source and $r$ the radial distance. (a) Density field and average colloid velocity relative to centre of mass velocity. The relative colloid velocity is represented by the size of the arrows placed along the streamlines. (b) Density fluctuations normalized by equilibrium fluctuations. (c) Temperature field with phoretic velocity due to temperature gradients. (d) Temperature fluctuations. The velocity and shape of the swarm is affected by the dimensionless coupling strength $\eta$. (e) Swarm velocity normalized to individual colloid drift velocity. (f) Radius of gyration of the swarm in the direction perpendicular to the axis of illumination with error bars show standard deviation over ensemble averaging of $10$ simulations (inset showing radius of gyration in the parallel direction). (g) Kymograph of the centre of mass $x$ location for the section of thickness $\Delta z = 3\sigma$ a distance $z_{\rm top}$ from the swarm top at time $t$ for $\eta = 25$. Colour represents the $x$ value of the centre of mass around the time average within the displayed window. (h) Snapshots of the swarm conformation at three successive times as shown by the kymograph in (g) displaying wave propagation from bottom to top. \cite{Cohen:2014oth}
 \label{fig:kymo}}
\end{figure}

Consider a collection of particles illuminated with a directed light source with uniform intensity $I$. The light intensity received at the surface of each colloid is determined by the distribution of the shadows of the colloids above it, as shown in Fig. \ref{fig:sch}. The light received by each colloid is converted into a heat flux that increases the temperature of the colloid and the surrounding fluid in an anisotropic way. A particle with a clear view of the light source will have an illuminated hot top hemisphere and a dark cold bottom hemisphere. This asymmetric temperature distribution results in the self-propulsion of the colloid via a process known as self-thermophoresis, with a maximum velocity of magnitude $v_o=I |D_T|/(9 \kappa)$, where $D_T$ is the thermophoretic mobility (also known as the thermodiffusion coefficient) and $\kappa$ is the thermal conductivity, which is set to be equal for the colloid and the solvent for simplicity. When $D_T$ is negative, which is allowed as it is an off-diagonal Onsager coefficient and possible via appropriate surface treatment of colloids \cite{Piazza2008}, the self-propulsion will be predominantly towards the light source with a velocity $\bfv_{s}$ (see Fig.  \ref{fig:sch}), leading to an effective attractive artificial phototaxis. Moreover, all colloids (whether illuminated or not) will experience a thermodiffusion drift velocity due the temperature gradient generated by the illuminated colloids, ${\bfv}_T = -D_T {\nabla} T$ (see Fig.  \ref{fig:sch}). The choice of negative $D_T$ causes the colloids to act as both heat sources and heat seekers; a combination that could lead to self-organization and instability, as seen in a diverse range of non-equilibrium phenomena.

The behaviour of the system depends on the intensity of the light source, which we can represent using the dimensionless coupling strength $\eta = \sigma I |S_T|/\kappa$, where $\sigma$ is the diameter of the colloid, $S_T = D_T/D_\mathrm{c}$ is the Soret coefficient, and $D_\mathrm{c}$ is the colloid diffusion coefficient. The system self-organizes into a moving swarm of apparent constant centre-of-mass velocity $V_{\rm swarm}$ with a comet-like structure: a high density head region with the outer most illuminated colloids generating a central hot core, and a relatively more dilute trailing aggregate in the form of a tail as shown in Fig. \ref{fig:sch}. Axially averaged fields are presented in Fig. \ref{fig:kymo} along with their fluctuations. The high density head region forms a hot core, as can be clearly seen in Fig. \ref{fig:kymo}a and  \ref{fig:kymo}c, which pulls the tail of the comet along, and also drives the fluctuations. Density fluctuations are normalized by the local equilibrium expectation values in Fig. \ref{fig:kymo}b such that any deviation from uniformity indicates non-equilibrium density fluctuations. A particularly interesting mode of density fluctuations occurs at the very tip of the head region as a result of the illuminated self-propelled particles (with the strongest $v_{s}$ component) attempting to escape the influence of the thermal attraction (also at its strongest, as shown by the vector field in Fig. \ref{fig:kymo}c), leading to fluctuations in the swarm shape. These particles usually return to the swarm, although spectacular ejection events are also observed at the tip with likelihood increasing with $\eta$; see Fig. \ref{fig:sch} (bottom). Density fluctuations at the swarm tip lead to novel temperature fluctuations due to the transient appearance of heat sources as seen in Fig. \ref{fig:kymo}d.

A circulation can be observed in the average colloid velocity streamlines in the swarm centre-of-mass frame, as shown in Fig. \ref{fig:kymo}a; the colloids that are attracted to the hot core reverse their direction on crossing the shadow boundary. This phenomenon also results from the competition between the strong thermally induced drift velocity towards the core shown in the vector field of Fig. \ref{fig:kymo}c and the propulsion of individual colloids towards the light source. The partially illuminated colloids that are near boundary of the swarm introduce a ``thermal drag'' that slows down the swarm as compared to the external fully illuminated isolated colloids. Figure  \ref{fig:kymo}e shows how this slowing down becomes more prominent as the coupling strength is increased, leading to an effectively sub-linear increase of $V_{\rm swarm}$ with respect to $\eta$. This suggests the following explanation for the observed circulation. A particle in the shadowed tail of the swarm, where the thermal attraction of the core is not strong enough to keep particles within the bulk of the swarm, may be left behind but remain in the shadow. At some point this inactive colloid will diffuse out of the shadow to become active again propelling towards the source. As it moves faster individually than in the swarm it may catch up and find itself attracted back to the hot core creating a circulation. Alternatively, the colloid may escape the influence of thermal attraction and propel past the swarm. The average shape of the swarm is also affected by the value of $\eta$ in line with the above picture, as shown in Fig. \ref{fig:kymo}f. The radius of gyration perpendicular (parallel) to the axis of illumination becomes smaller (larger) as $\eta$ is increased, resulting in an increased aspect ratio.

Increasing the coupling strength will also make the swarm more dynamic. Transverse waves of colloids can be observed (Fig. \ref{fig:kymo}g and Fig. \ref{fig:kymo}h) propagating from tail to head in randomly selected azimuthal directions, with pronunciation increasing with higher $\eta$. The waves appear to be randomly initiated at the back of the swarm, and propagate with a constant speed (that increases with $\eta$) towards the front without any dispersion, as can be seen from the kymograph displayed in Fig. \ref{fig:kymo}g. The existence of these waves is a result of the competition between the transverse (xy) components of the self-propulsion and thermal drift velocities, as shown in Fig.  \ref{fig:sch}. The undulations arise from the colloids in the tail region diffusing out of the shadow, aided by the effect of partial illumination upon crossing the shadow boundary that further drives their motion away from the swarm. The colloids then become thermally active and attract higher up colloids out of the shadow, thereby initiating a propagating wave along the swarm length.

\section{Mixtures of Apolar Active Colloids: Active Molecules}\label{sec:molecules}

After understanding how nonequilibrium phoretic activity affects the dynamics of single colloidal particles, we can now study the nonequilibrium interactions between these particles. The simplest case to consider is the interaction between two different types of apolar (symmetric) active colloids \cite{SotoGolestanianPRL:2014}.

Consider a concentration field $C(\bfr)$ that satisfies the stationary diffusion equation $\nabla^2 C=0$, subject to the boundary condition of production or consumption on the surfaces of colloidal particles. For activity $\alpha$, the boundary condition is $-D \partial _r C|_{r=R}=\alpha$, which leads to a contribution to the concentration field of the form $C(\bfr)=\frac{\alpha R^2}{D r}$. At a distance $\bfr$ away from the colloid that acts as a source or a sink for the chemical, a colloidal particle will experience  a drift velocity of $\bfv=-\mu \nabla C=-\frac{\mu \alpha}{D} \nabla\left(\frac{1}{r}\right)$. Therefore, our system is governed by a dissipative equivalent of gravitational or Coulomb interactions with a $1/r$ potential. There is, however, a peculiar feature when we consider a mixture of, say, type-A and type-B active colloids. In this case, the symmetry between action and reaction will be, in general, broken. This can be observed from the drift velocities of A and B particles:
\begin{eqnarray}
\bfv_{\rm A}&=&-\mu_{\rm A} \nabla C|_{\rm A}=-\mu_{\rm A} \alpha_{\rm B} \nabla_{\rm A}\left(\frac{R^2}{D |\bfr_{\rm B}-\bfr_{\rm A}|}\right)=-\mu_{\rm A} \alpha_{\rm B} \cdot \frac{R^2}{D} \cdot \frac{\bfr_{\rm B}-\bfr_{\rm A}}{|\bfr_{\rm B}-\bfr_{\rm A}|^3}, \label{eq:vA}\\
\bfv_{\rm B}&=&-\mu_{\rm B} \nabla C|_{\rm B}=-\mu_{\rm B} \alpha_{\rm A} \nabla_{\rm B}\left(\frac{R^2}{D |\bfr_{\rm B}-\bfr_{\rm A}|}\right)=+\mu_{\rm B} \alpha_{\rm A} \cdot \frac{R^2}{D} \cdot \frac{\bfr_{\rm B}-\bfr_{\rm A}}{|\bfr_{\rm B}-\bfr_{\rm A}|^3}. \label{eq:vB}
\end{eqnarray}
Unless we have a specially fine-tuned system, we have $\mu_{\rm A} \alpha_{\rm B} \neq \mu_{\rm B} \alpha_{\rm A}$, and therefore, the action-reaction symmetry is broken. We can understand this property as a generalization of electrostatics or gravity in which for every particle the charge or mass that creates the field is different from the charge or mass that responds to the field (created by others).

To examine the behaviour of a mixture in the dilute limit, we start by considering two colloids. Assuming an additional equilibrium interaction potential $U(|\bfr_{\rm B}-\bfr_{\rm A}|)$, which can arise from excluded volume interaction for example, the particles will experience an additional contribution to the drift velocity
\begin{equation}
\bfv_{\rm A}^{\rm eq}=- D_{\rm c} \beta  \nabla_{\rm A} U=-\bfv_{\rm B}^{\rm eq},
\end{equation}
where $D_{\rm c}$ is the diffusion coefficient of the colloids. We can change coordinates from $\bfr_{\rm A}$ and $\bfr_{\rm B}$ to the relative and centre of mass coordinates, $\bfr=\bfr_{\rm B}-\bfr_{\rm A}$ and $\bfr_{\rm CM}=\frac{1}{2}\left(\bfr_{\rm A}+\bfr_{\rm B}\right)$. For the velocities, we obtain
\begin{eqnarray}
\bfv&=&\bfv_{\rm B}-\bfv_{\rm A}=\left(\mu_{\rm B} \alpha_{\rm A}+\mu_{\rm A} \alpha_{\rm B}\right) \cdot \frac{R^2}{D} \cdot\frac{\bfr}{r^3}-2 D_{\rm c} \beta  \nabla U, \\
{\bm V}&=&\frac{1}{2}\left(\bfv_{\rm A}+\bfv_{\rm B}\right)=\frac{1}{2}\left(\mu_{\rm B} \alpha_{\rm A}-\mu_{\rm A} \alpha_{\rm B}\right) \cdot \frac{R^2}{D} \cdot\frac{\bfr}{r^3}.
\end{eqnarray}
Equilibration in the relative distance yields
\begin{equation}
\cP(r)={\cal A} r^{d-1} \exp\left\{-\left(\mu_{\rm B} \alpha_{\rm A}+\mu_{\rm A} \alpha_{\rm B}\right) \cdot \frac{R^2}{D D_{\rm c} r}-2  \beta  U\right\},
\end{equation}
where ${\cal A}$ is a normalization constant. Invoking the analogy to electrolytes, we can define a generalized Bjerrum length as $\ell_{\rm B}=|\mu_{\rm B} \alpha_{\rm A}+\mu_{\rm A} \alpha_{\rm B}| R^2/(D D_{\rm c})$, which represents the distance at which ``energy'' and ``entropy'' are comparable. Using the Bjerrum length, the stationary distribution for the relative distance is $\cP(r) \sim r^{d-1} \exp\left({\frac{\ell_{\rm B}}{r}-2 \beta U}\right)$. Since the centre of mass speed is determined by $r$ as $V=\frac{1}{2}\left|\mu_{\rm B} \alpha_{\rm A}-\mu_{\rm A} \alpha_{\rm B}\right| \cdot \frac{R^2}{D r^2}$, we can deduce the distribution of swimming speed as follows
\begin{equation}
\cP(V)={\cal B} \cdot \frac{e^{\sqrt{V/V_0}}}{V^{1+d/2}}
\end{equation}
where $V_0 \equiv \frac{|\mu_{\rm B} \alpha_{\rm A}-\mu_{\rm A} \alpha_{\rm B}| D D_{\rm c}^2}{2 |\mu_{\rm B} \alpha_{\rm A}+\mu_{\rm A} \alpha_{\rm B}|^2 R^2}$ and ${\cal B}$ is a normalization constant. The distribution is cut off at $V_{\rm max}=\frac{1}{8D}\left|\mu_{\rm B} \alpha_{\rm A}-\mu_{\rm A} \alpha_{\rm B}\right|$.

\subsection{Designing the Configurations of Active Molecules}

When the phoretic interaction strengths are sufficiently strong, the active colloids will form clusters. We can examine the configurations of different clusters and determine which ones have the potential to form stable configurations, which can be regarded as stable active molecules. Let us consider a cluster formed with one A particle and two B particles. We can parametrize reflection-symmetric configurations using the angle $\phi$ defined in Fig. \ref{fig:AB2}. We can write down the following expressions for various contributions to the velocities of the three particles
\begin{eqnarray}
\bfv_{\rm A}&=&\frac{\mu_{\rm A} \alpha_{\rm B}}{D} \bfn_1+\frac{\mu_{\rm A} \alpha_{\rm B}}{D} \bfn_2+v^{\rm eq} \bfn_1+v^{\rm eq} \bfn_2,\\
\bfv_{{\rm B}_1}&=&-\frac{\mu_{\rm B} \alpha_{\rm A}}{D} \bfn_1-v^{\rm eq} \bfn_1-\frac{\mu_{\rm B} \alpha_{\rm B}}{D} \cdot \frac{1}{(2 \cos \phi)^2} \bfe_x,\\
\bfv_{{\rm B}_2}&=&-\frac{\mu_{\rm B} \alpha_{\rm A}}{D} \bfn_2-v^{\rm eq} \bfn_2+\frac{\mu_{\rm B} \alpha_{\rm B}}{D} \cdot \frac{1}{(2 \cos \phi)^2} \bfe_x.
\end{eqnarray}
The unit vectors $\bfn_1$ and $\bfn_2$ are defined in Fig. \ref{fig:AB2}. We can invoke the constraint of no penetration between the particles via $\bfn_i \cdot \left(\bfv_{{\rm B}_i}-\bfv_{\rm A}\right)=0$ and obtain the equilibrium contribution of the velocities as
\begin{equation}
v^{\rm eq}=\frac{\left(\frac{\mu_{\rm B} \alpha_{\rm B}}{D}\right)}{2-\cos 2\phi}\left[-\left(\frac{\alpha_{\rm A}}{ \alpha_{\rm B}}+\frac{\mu_{\rm A}}{\mu_{\rm B}}\right)+\left(\frac{\mu_{\rm A}}{\mu_{\rm B}}\right) \cos 2 \phi-\frac{1}{4 \cos\phi}\right].
\end{equation}
Using a kinematic definition $R \dot\phi=-\bfn_i \times \left(\bfv_{{\rm B}_i}-\bfv_{\rm A}\right) \cdot \bfe_z$, we find the following dynamical equation for the configuration of the molecule
\begin{equation}
\dot \phi=\frac{2}{R}\left(\frac{\mu_{\rm B} \alpha_{\rm B}}{D}\right) \frac{\sin \phi}{\left(2-\cos 2 \phi\right)}\left[\left(\frac{\mu_{\rm A}}{\mu_{\rm B}}-\frac{\alpha_{\rm A}}{ \alpha_{\rm B}}\right) \cos \phi-\frac{3}{8 \cos^2\phi}\right].\label{eq:phidot}
\end{equation}
Figure \ref{fig:AB2} shows the behaviour of this dynamical system for different values of the tuning parameter. When $\frac{\mu_{\rm A}}{\mu_{\rm B}}-\frac{\alpha_{\rm A}}{ \alpha_{\rm B}} < \frac{3}{8}$ the dynamical system in Eq. (\ref{eq:phidot}) has only one stable fixed point at $\phi=0$, which corresponds to a linear B--A--B conformation for the ${\rm AB}_2$ molecule; since the conformation is symmetric, the molecule is not self-propelled. At  $\frac{\mu_{\rm A}}{\mu_{\rm B}}-\frac{\alpha_{\rm A}}{ \alpha_{\rm B}} = \frac{3}{8}$ the dynamical system exhibits a supercritical pitchfork bifurcation, and for $\frac{\mu_{\rm A}}{\mu_{\rm B}}-\frac{\alpha_{\rm A}}{ \alpha_{\rm B}} > \frac{3}{8}$ two stable fixed points appear at $\phi=\pm \phi_s$, defined via $\cos \phi_s=\left(\frac{3/8}{{\mu_{\rm A}}/{\mu_{\rm B}}-{\alpha_{\rm A}}/{ \alpha_{\rm B}}}\right)^{1/3}$, while the $\phi=0$ fixed point becomes unstable. Due to the symmetry breaking in the conformation, the ${\rm AB}_2$ molecule will now be self-propelled, with a speed that can be determined from the above equations in terms of the parameters.

\begin{figure}[t]
\includegraphics[width=0.3\linewidth]{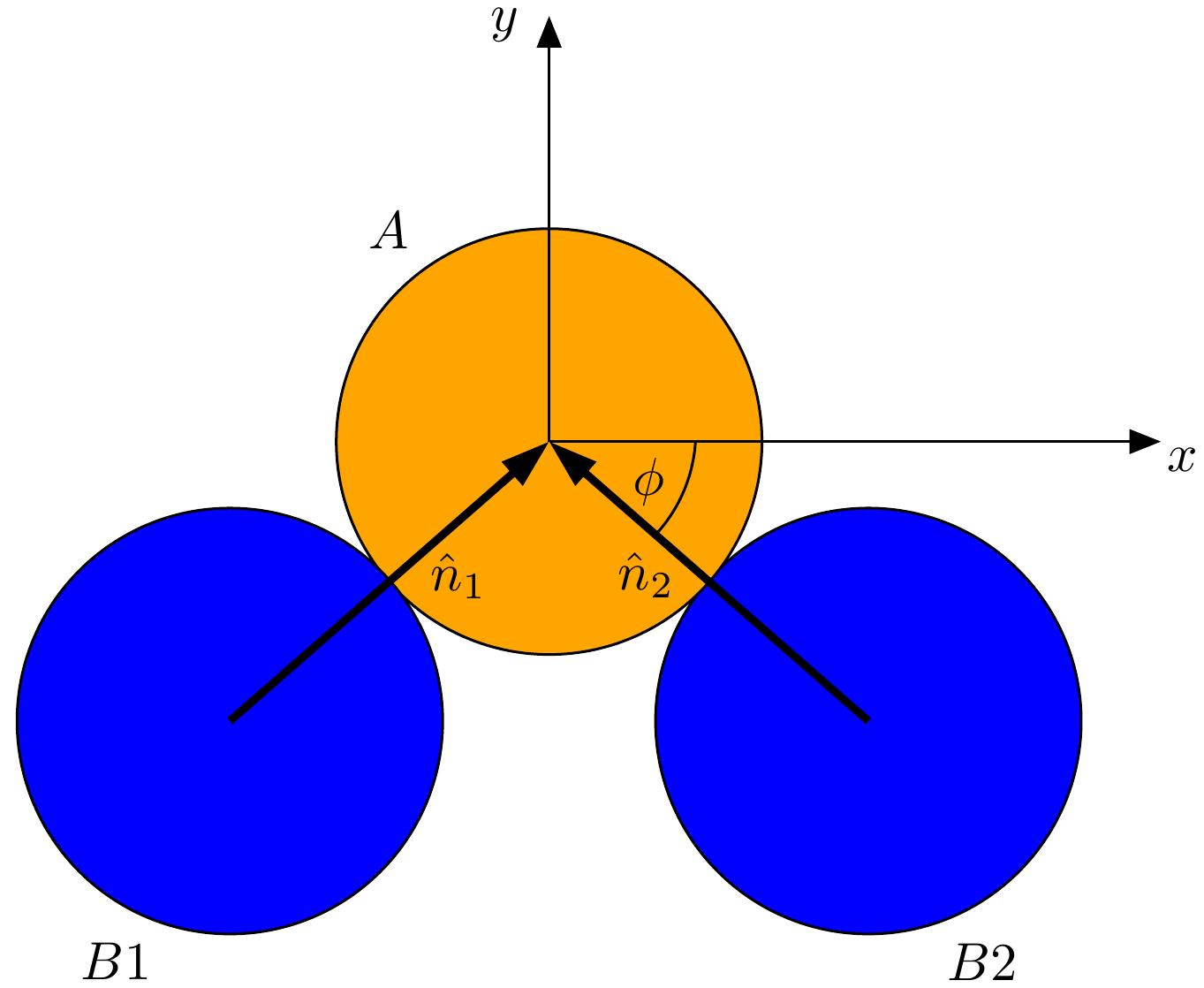}\hskip1cm
\includegraphics[width=0.5\linewidth]{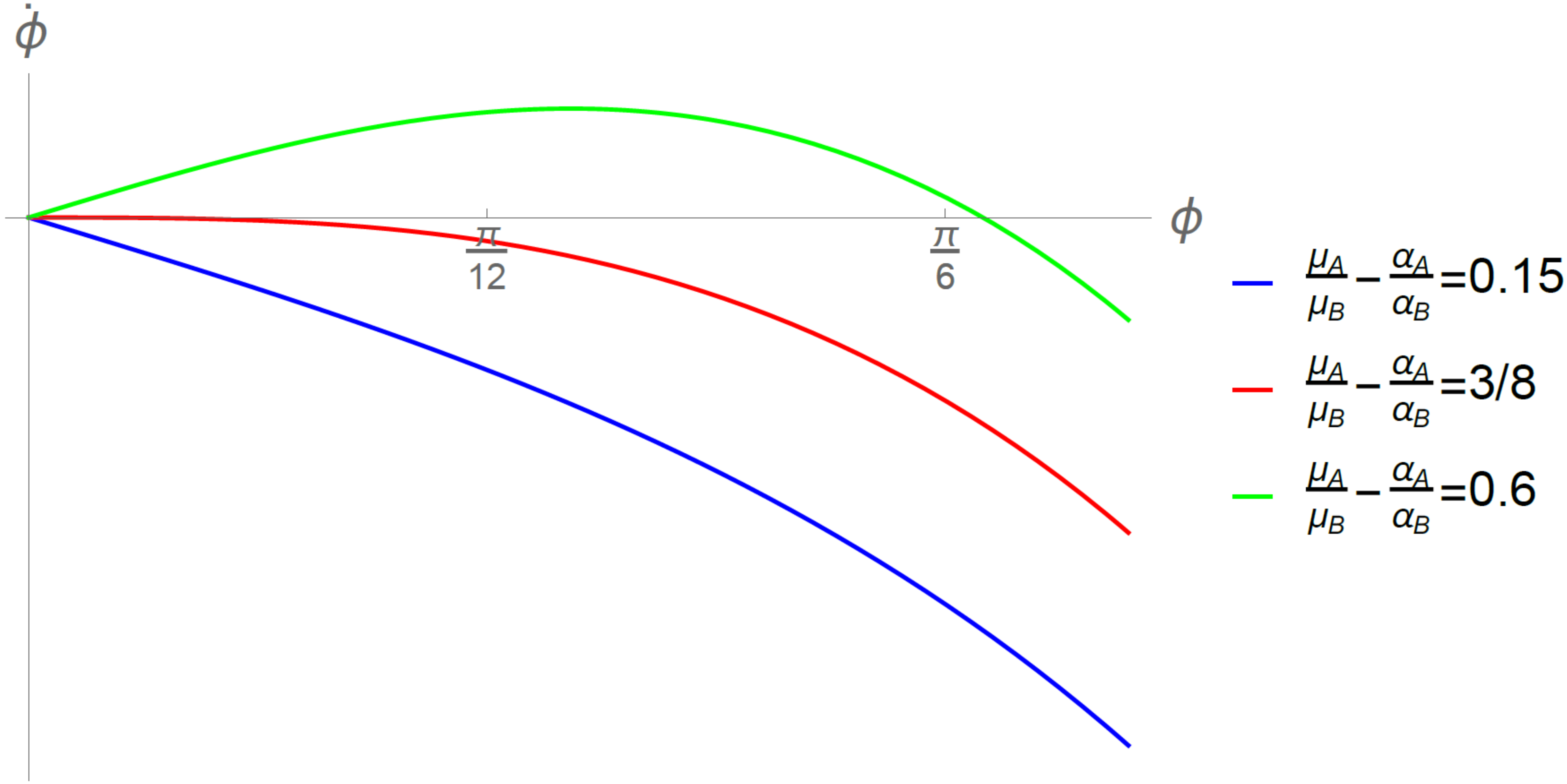}
\caption{Stability of $AB_2$ molecule. The angle $\phi$ parametrizes the deviation from a linear chain of the $AB_2$ molecule (left). Plots of Eq. (\ref{eq:phidot}) for different values of the parameter $\frac{\mu_{\rm A}}{\mu_{\rm B}}-\frac{\alpha_{\rm A}}{ \alpha_{\rm B}}$.
 \label{fig:AB2}}
\end{figure}

This calculation demonstrates how it is possible to design stable shapes for various active colloidal molecules using the parameters of the system, namely the values of the activities and the mobilities.

\subsection{Dynamic Function}

For sufficiently large molecules, it is possible to have cases where the conformations of the molecule change dynamically, and consequently this will be translated to the manifested non-equilibrium function exhibited by those molecules. To have such dynamical changes, one possibility is to have multiple stable fixed points and noise-induced transitions between them, presumably across barrier. An interesting case with such behaviour is the ${\rm AB}_3$ molecule, which has a stable Y-isomer with no self-propulsion, and a stable T-isomer with self-propulsion; stochastic switching between them leads to an emergent run-and-tumble behaviour in a system with a continuous configuration space. Another possibility is the existence of an oscillatory conformation. When such conformations are symmetric, such as the case for ${\rm A}_4{\rm B}_8$, the molecule will exhibit spontaneous oscillations without self-propulsion. In asymmetric cases, such ${\rm A}_5{\rm B}_8$, the oscillations can lead to self-propulsion, in a way that is reminiscent of the swimming of sperm \cite{SotoGolestanianPRE:2015}.
 
 \subsection{From Structure to Function: A New Non-equilibrium Paradigm}

The framework described above can be generalized to cases where different parameters such as size, surface chemistry, and surface activity are tuned in order to achieve desired clusters and molecules. With such capabilities, the framework provides a paradigm in which we can design certain {\it structures}---i.e. 3D geometry and conformation---that will exhibit certain non-equilibrium {\it functions} entirely due to their shape. The function can be derived from symmetry properties of the conformations. For example, axially symmetric molecules will exhibit an intrinsic (self-propulsion) translational velocity, whereas non-axially symmetric molecules will have intrinsic angular velocity or spin. If the molecules are ``too symmetric'' they might not exhibit any mechanical function and can be categorized as inert. Sufficiently large complexes can spontaneously break time-translation invariance and exhibit oscillations. The paradigm has similarities to the way proteins are designed from sequences to shapes to biological function.

\section{Mixtures of Apolar Active Colloids: Stability of Suspensions}		\label{sec:apolar-susp}

As described in the previous section, two different apolar chemically active colloids interact with each other through the chemical fields that they themselves produce, and a key feature of these interactions is that they are in general non-reciprocal; see Eqs. (\ref{eq:vA}) and (\ref{eq:vB}). It is therefore pertinent to investigate the phase behaviour of mixtures of several species of active colloids  \cite{Agudo-Canalejo2019}. Brownian dynamics simulations of a dilute suspention of many colloids belonging to different species and interacting through Eqs. (\ref{eq:vA}) and (\ref{eq:vB}) show a variety of phase separation phenomena. For binary mixtures, the simulations reveal that, while in a large region of the parameter space the mixtures remain homogeneous, the homogeneous state can also become unstable leading to a great variety of phase separation phenomena; see Figs.~\ref{intro}(c--e). Here, phase separation is used in the sense of macroscopic (system-spanning) separation typically into a single large cluster [occasionally into two; see Fig.~\ref{intro}(c)] that coexists with a dilute (or empty) phase. The phase separation process may lead to aggregation of the two species into a single mixed cluster, or to separation of the two into either two distinct clusters or into a cluster of a given stoichiometry and a dilute phase. The resulting configurations are qualitatively distinct for mixtures of one chemical-producer and one chemical-consumer species, as opposed to mixtures of two producer (or consumer) species; compare Fig.~\ref{intro}(c) and Fig.~\ref{intro}(e). While the typical steady-state configurations are static, for mixtures of producer and consumer species it is observed that static clusters can undergo a shape-instability that breaks their symmetry, leading to a self-propelling cluster. Randomly-generated highly-polydisperse mixtures of up to 20 species also show homogeneous as well as phase-separated states [Fig.~\ref{intro}(f)].

This variety of phase separation phenomena can be understood within a continuum theory of the mixture. Let us consider a system consisting of $M$ different species of chemically-interacting particles, with concentrations $\rho_i(\rr,t)$ for $i=1,...,M$; and a messenger chemical with concentration $C(\rr,t)$. The concentration of species $i$ is described by 
\beq
\partial_t \rho_i(\rr,t) - \nabla \cdot [\Dc \nabla \rho_i + (\mu_i \nabla C) \rho_i ]=0,
\eeq
which includes a diffusive term with diffusion coefficient $\Dc$, which for simplicity is taken to be equal for all types, as well as the phoretic drift term with mobility $\mu_i$ which is positive or negative if the particle is repelled or attracted to the chemical, respectively; see Figs.~\ref{intro}(a) and \ref{intro}(b). The concentration of the chemical is described by 
\beq
\partial_t C(\rr,t) - D \nabla^2 C = \sum_{i} \alpha_i \rho_i,
\eeq
where the right hand side represents production or consumption of the chemical by all particle species. Within this continuum theory for the mixture, we can study the stability of the homogeneous state, and show that under certain conditions the system undergoes macroscopic phase separation.

\begin{figure}
\includegraphics[width=1\linewidth]{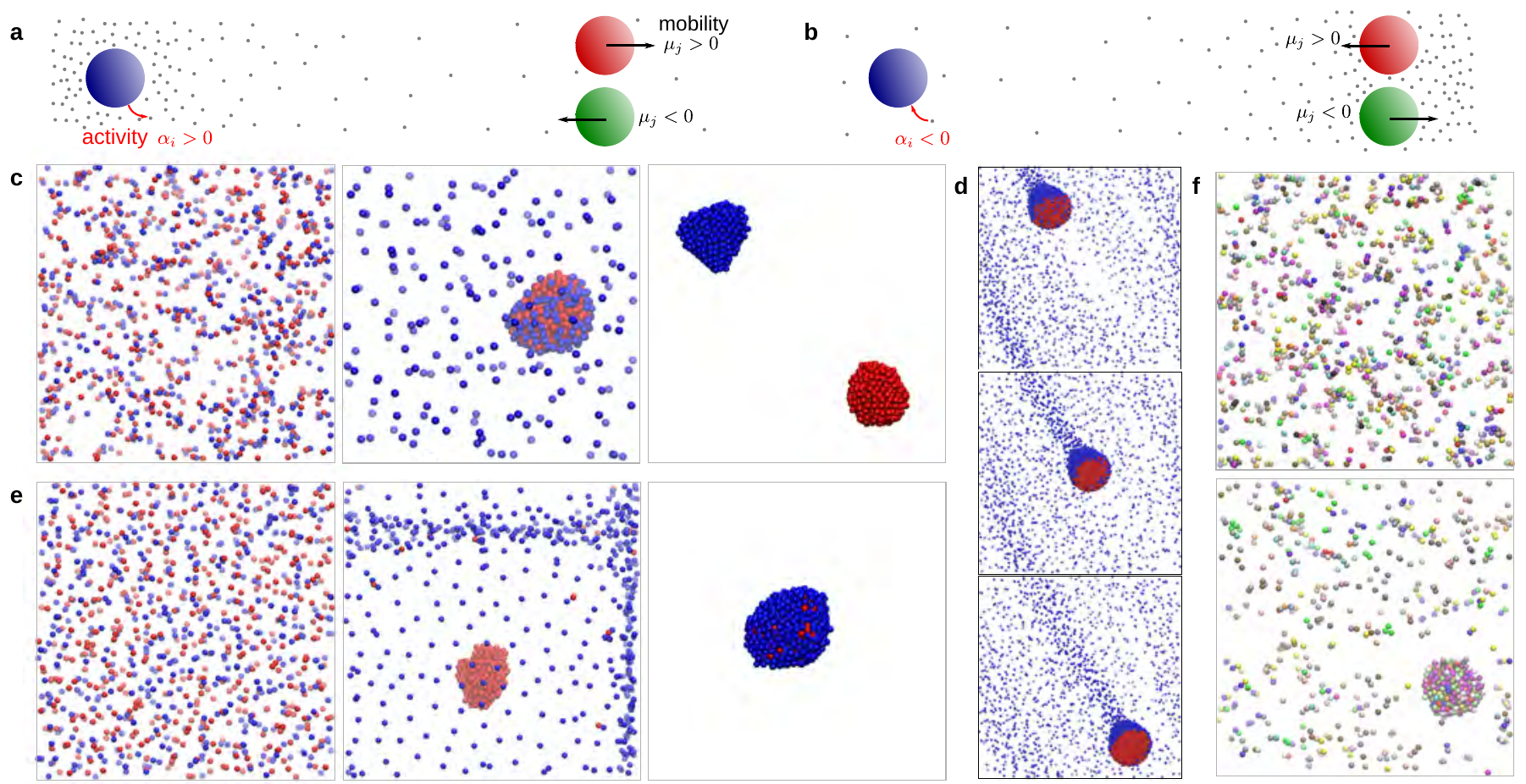}
\caption{Active phase separation phenomena in mixtures of chemically-interacting particles. (a) Producer particles repel (attract) particles with positive (negative) mobility, while (b) the opposite is true for consumer particles. (c) Binary mixtures of producer (blue) and consumer (red) species show, from left to right, homogeneous states with association of particles into small molecules, see Section \ref{sec:molecules}, aggregation into a static dense phase that coexists with a dilute phase, and separation into two static collapsed clusters. (d) The static aggregate [(c), centre] can undergo symmetry breaking to form a self propelled macroscopic cluster. (e) Binary mixtures of producer species (blue and red) show homogeneous states without molecule formation, separation into a static dense phase and a dilute phase that are pushed away from each other, and aggregation into a static collapsed cluster. (f) Randomly-generated highly polydisperse mixtures (20 different species) can remain homogeneous or undergo macroscopic phase separation. \label{intro}}
\end{figure}

\subsection{Linear Stability Analysis of the Homogeneous Mixture}

We consider small deviations from the homogeneous state, so that the colloid density is described by $\rho_i(\rr,t) = \rho_{0i} + \delta \rho_i(\rr,t)$. The net catalytic activity of a mixture is defined as $A \equiv \sum_{i} \alpha_i \rho_{0i}$, where we note that $A$ represents activity in the homogeneous state, while locally we have $\sum_{i} \alpha_i \rho_{i} = A + \sum_{i} \alpha_i \delta \rho_{i}$. The chemical concentration can be separated into a (time-dependent) uniform value and the deviations from this uniform value in response to nonuniformities of the colloid distribution, so that $C(\rr,t) = C_0 + A t + \delta C (\rr, t)$. Introducing this into the evolution equation for $C(\rr,t)$ we obtain an equation for the deviations $\delta C (\rr,t)$ given by
\begin{equation}
\partial_t \delta C(\rr,t) - D \nabla^2 \delta C = \sum_{i} \alpha_i \delta \rho_i.
\label{deltac}
\end{equation}
Because the small chemical diffuses much faster than the large colloids, the deviations $\delta C(\rr,t)$ of the chemical concentration from the uniform value $C_0+At$ can be assumed to reach a steady state instantaneously for each configuration of the colloids, so that from Eq. (\ref{deltac}) we obtain 
\beq
- D \nabla^2 \delta C =  \sum_{i} \alpha_i \delta \rho_i. 
\eeq
Introducing this into the evolution equation for $\rho_i(\rr,t)$, and staying only to linear order in $\delta \rho_i(\rr,t)$, we obtain
\begin{equation}
\partial_t \delta \rho_i(\rr,t) = \Dc \nabla^2 \delta \rho_i - \frac{\mu_i \rho_{0i}}{D} \sum_{j} \alpha_j \delta \rho_j.
\label{deltarho}
\end{equation}

The linearized system of equations [Eqs. (\ref{deltarho})] with $i=1,...,M$ describes the evolution of the deviations of the colloid density around the homogeneous state. This result is valid at all times for mixtures with net positive or zero production $A \geq 0$; and for mixtures with net consumption $A<0$ as long as the chemical concentration is still large enough that the consumption of chemical by the colloids can be considered to be taking place in the saturated regime, i.e. at a rate $\alpha_i<0$ independent of the local chemical concentration. If $K$ is the (largest) equilibrium constant of the consumption reaction at the surface of the colloids, then this approach is valid as long as $C_0 -|A|t \gg K$, i.e. for sufficiently short experiments with $t \ll (C_0 - K) / |A|$. 

The stability analysis of Eqs. (\ref{deltarho}) is done most conveniently by defining the new variables $U_i \equiv \alpha_i \delta \rho_i$ and the parameters $\gamma_i \equiv \mu_i \alpha_i \rho_{0i}/D$. The system of equations (\ref{deltarho}) can be rewritten as
\begin{equation}
[\partial_t - \Dc \nabla^2 + \gamma_i ] U_i + \gamma_i \sum_{j \neq i} U_j = 0
\label{u1}
\end{equation}
the solution of which is given by a sum of Fourier modes of the form $U_i (\rr,t) = U_{\qq i} \ex{i \qq \cdot \rr} \ex{\lambda t}$. Introducing this into (\ref{u1}) finally results in the eigenvalue problem
\begin{equation}
[\lambda + \Dc \qq^2 + \gamma_i ] U_{\qq i} + \gamma_i \sum_{j \neq i} U_{\qq j} = 0
\label{u2}
\end{equation}
for the growth rate $\lambda$ of the perturbation modes with wavenumber $\qq$.

By defining $\tilde{\lambda} \equiv - (\lambda + \Dc \qq^2)$, the eigenvalue problem (\ref{u2}) is equivalent to finding the eigenvalues of a $M \times M$ matrix with $M$ identical rows each given by $[\gamma_1~\gamma_2~...~\gamma_M]$. Such a matrix has rank 1 and therefore at least $M-1$ of its eigenvalues are equal to zero, $\tilde{\lambda}_-=0$. Because the trace of a matrix is equal to the sum of its eigenvalues, the remaining eigenvalue is equal to the trace of the matrix, so that $\tilde{\lambda}_+=\sum_i \gamma_i$.

Transforming from $\tilde{\lambda}$ back to $\lambda$, we finally find $M-1$ identical eigenvalues $\lambda_- = -\Dc \qq^2$, and one eigenvalue $\lambda_+ = -\Dc \qq^2 - \sum_i \gamma_i$. The latter eigenvalue can become positive, indicating an instability. 
When rewritten in the original variables, we find that the homogeneous state becomes unstable towards a spatially-inhomogeneous state when the following condition holds
\begin{equation}
\sum_i \mu_i \alpha_i \rho_{0i} < 0.
\label{instNsame}
\end{equation}
The instability corresponds to macroscopic phase separation, in the sense that it occurs for perturbations of infinite wave length, specifically for perturbations with wave number 
\beq
\qq^2 < -  (D \Dc)^{-1} \sum_i \mu_i \alpha_i \rho_{0i}, 
\eeq
with those having infinite wave length $\qq \to 0$ being the first and most unstable. Importantly, the stability analysis also tells us about the stoichiometry of the different particle species at the onset of growth of the perturbation, which follows
\begin{equation}
(\delta \rho_1,\delta \rho_2,...,\delta \rho_M) = \left(1,\frac{\mu_2 \rho_{02}}{\mu_1 \rho_{01}},...,\frac{\mu_M \rho_{0M}}{\mu_1 \rho_{01}}\right)\delta \rho_1.
\label{stoichNsame}
\end{equation}

If only a single particle species is present ($M=1$), the instability criterion (\ref{instNsame}) describes the well-known Keller-Segel instability \cite{Keller1970}, which simply says that the homogeneous state is stable for particles that repel each other ($\mu_1 \alpha_1>0$), whereas particles that attract each other ($\mu_1 \alpha_1<0$) tend to aggregate, with the end state being a featureless macroscopic cluster containing all particles. In contrast, as soon as we have mixtures of more than one species, the combination of the instability criterion (\ref{instNsame}) and the stoichiometric relation (\ref{stoichNsame}) predicts a wealth of new phase separation phenomena.

 \begin{figure}
 \includegraphics[width=1\linewidth]{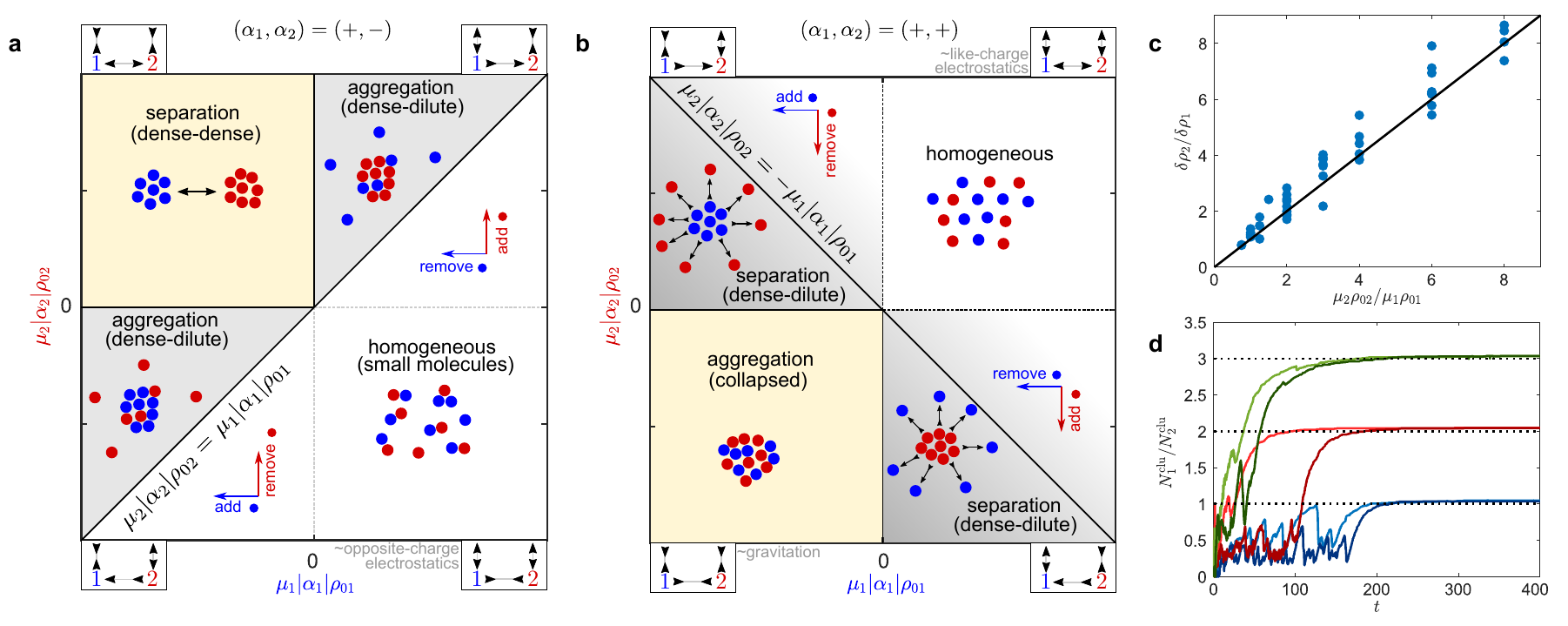}
 \caption{Stability diagrams and stoichiometry for binary mixtures. (a) Stability diagram for mixtures of one producer and one consumer species [cf.~Fig.~\ref{intro}(c)], and (b) for mixtures of two producer species [cf.~Fig.~\ref{intro}(e)]. In (a) and (b) the boxed legends attached to each quadrant symbolize the ``interaction network'' representing the sign of interactions between each species in the system. Phase separation [aggregation in (a), separation in (b)] can be triggered by addition or removal of particles (density changes) only when interactions between the two species are intrinsically non-reciprocal. (c) Stoichiometry at the onset of the instability, obtained from 44 simulations (blue circles) compared to the stability analysis prediction [Eq.~(\ref{stoichNsame})]. (d) Time evolution of the stoichiometry of the biggest cluster arising from aggregation of $(\alpha_1,\alpha_2)=(+,-)$ mixtures, demonstrating that the long time stoichiometry is predicted by the neutrality rule [Eq.~(\ref{neutr})] and is independent of the species' mobility (blue: $\tilde{\alpha_2}=-1$, $\tilde{\mu_2}=8$ and $12$; red: $\tilde{\alpha_2}=-2$, $\tilde{\mu_2}=4$ and $8$; green: $\tilde{\alpha_2}=-3$, $\tilde{\mu_2}=3$ and $5$; in all cases $N_1=800$, $N_2=200$, $\tilde{\alpha}_1=\tilde{\mu}_1=1$). \label{phasediag}}
 \end{figure}

\subsection{Phase Separation in Binary Mixtures}

For binary mixtures ($M=2$), the instability condition (\ref{instNsame}) becomes $\mu_1 \alpha_1 \rho_{01} + \mu_2 \alpha_2 \rho_{02} < 0$, and the stoichiometric constraint (\ref{stoichNsame}) implies that when $\mu_1$ and $\mu_2$ have equal or opposite sign, the instability will lead respectively to aggregation or separation of the two species. Combining these criteria we can construct a stability diagram for the binary mixture, although we must distinguish between two qualitatively-different kinds of mixtures: those of one producer and one consumer species, see Fig~\ref{phasediag}(a) where we can choose $(\alpha_1,\alpha_2)=(+,-)$ without loss of generality; and those of two producer species, see Fig~\ref{phasediag}(b). The case of two consumer species is related to the latter by the symmetry $(\mu_1,\mu_2) \to -(\mu_1,\mu_2)$. In this way, the parameter space for each type of mixture can be divided into regions leading to homogeneous, aggregated, or separated states, which correspond directly to those observed in simulations; compare Figs.~\ref{phasediag}(a) and \ref{phasediag}(b) to Figs.~\ref{intro}(c) and \ref{intro}(e). We note, however, that while for $(\alpha_1,\alpha_2)=(+,-)$ mixtures the simulations are always seen to match the predicted phase behaviour, for $(\alpha_1,\alpha_2)=(+,+)$ mixtures once can observe separation in the simulations even when the continuum theory predicts the homogeneous state to be linearly stable, although proceeding much more slowly, indicating that in this region separation may be occurring through a nucleation-and-growth process controlled by fluctuations. This is denoted as the shaded gray region extending past the instability line in Fig~\ref{phasediag}(b).

The wide variety of phase separation phenomena arising in these mixtures is intimately related to the active, non-reciprocal character of the chemical interactions. In particular, it is useful to consider the sign of both inter-species as well as intra-species interactions. In the stability diagrams in Figs.~\ref{phasediag}(a) and\ref{phasediag}(b), one finds that each quadrant corresponds to a distinct ``interaction network'' between species, as depicted in the boxed legends attached to every quadrant (as an example, the top-right interaction network in Fig. \ref{phasediag}(a) can be read as ``1 is attracted to 2, 2 is repelled from 1, 1 is repelled from 1, and 2 is attracted to 2''). Note that only three regions in the parameter space have passive analogs: (i) The bottom-right of \ref{phasediag}(a) corresponds to electrostatics with opposite charges, where equals repel and opposites attract, allowing for the formation of small active molecules as studied in Section \ref{sec:molecules}. (ii) The top-right of Fig. \ref{phasediag}(b) corresponds to electrostatics with like charges, where all interactions are repulsive leading to a homogeneous state. (iii) The bottom-left of Fig. \ref{phasediag}(b) corresponds to gravitation, where all interactions are attractive. The top-left of Fig. \ref{phasediag}(a) can be thought of as the opposite of electrostatics (or as gravitation including a negative mass species), where equals attract and opposites repel. The remaining four quadrants involve intrinsically non-reciprocal interactions where one species chases after the other: in Fig. \ref{phasediag}(a), a self-repelling species chases after a self-attracting species; whereas in Fig. \ref{phasediag}(b), a self-attracting species chases after a self-repelling species. Importantly, the most non-trivial instances of phase separation, which are also those that can be triggered simply by density changes (e.g. by addition or removal of particles), occur in regions with such chasing interactions, which are in turn a direct signature of non-equilibrium activity. 

 \begin{figure}[t]
 \includegraphics[width=0.7\linewidth]{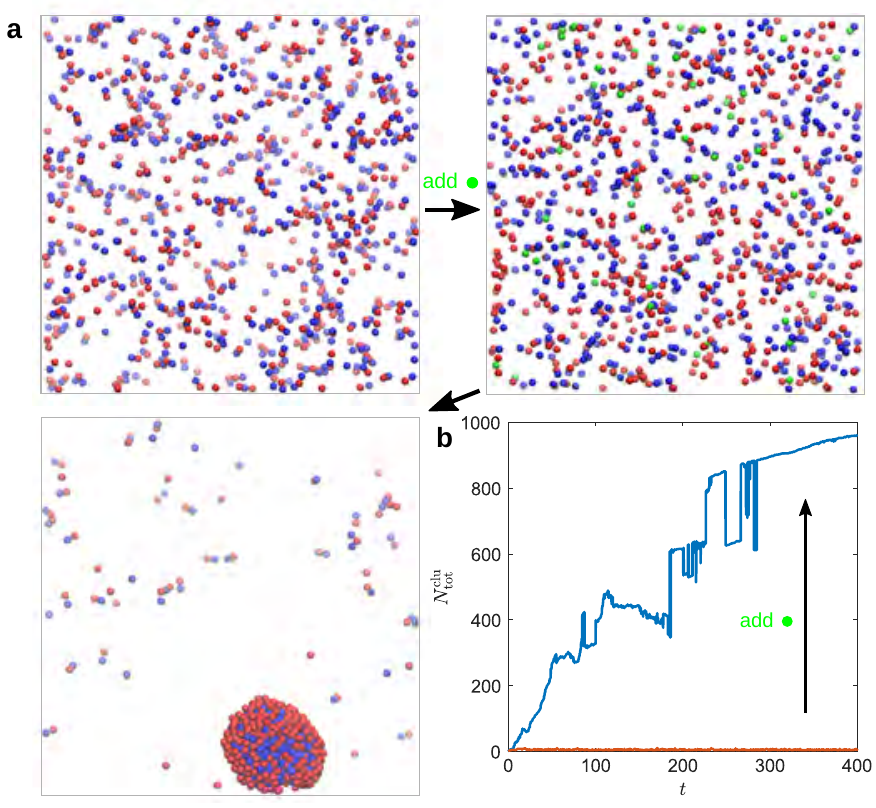}
 \caption{Phase separation induced by a small amount of an active ``doping agent''. (a) Simulation snapshots showing macroscopic aggregation of a previously homogeneous mixture ($N_1=N_2=500$, $\tilde{\alpha}_1=\tilde{\mu}_1=1$, $\tilde{\alpha}_2 = -1$, $\tilde{\mu}_2=1/2$) after addition of 5 \% of a third species ($N_3=50$, $\tilde{\alpha}_3=-5$, $\tilde{\mu}_3=2$). (b) Time evolution of the size of the largest cluster (total number of particles), in the absence and presence of the third species. \label{dope}}
 \end{figure}

Fourier analysis of the Brownian dynamics simulations (44 simulations with varying $N_i$, $\alpha_i$, and $\mu_i$) agrees quantitatively with the theoretical prediction (\ref{stoichNsame}) for the stoichiometry at the onset of the instability; see Fig.~\ref{phasediag}(c). However, this initial value is not representative of the long-time stoichiometry of the phases. For $(\alpha_1,\alpha_2)=(+,+)$ mixtures, shown in Figs.~\ref{intro}(e) and \ref{phasediag}(b), we always observe final configurations with either complete aggregation or separation of the two species. For $(\alpha_1,\alpha_2)=(+,-)$ mixtures, shown in Figs.~\ref{intro}(c) and \ref{phasediag}(a), we always observe complete separation, but in this case aggregation leads to a cluster with non-trivial stoichiometry [Fig.~\ref{intro}(c), centre]. Phenomenologically, we observe that cluster formation in this case proceeds by fast initial aggregation of the particles of the self-attractive species ($\alpha_i \mu_i<0$) followed by slower recruitment of particles of the self-repelling species ($\alpha_i \mu_i>0$) until the cluster is chemically ``neutral'', in the sense that its net consumption or production of chemicals vanishes, namely
\begin{equation}
\alpha_1 N_1^\mathrm{clu}+\alpha_2 N_2^\mathrm{clu} = 0,
\label{neutr}
\end{equation}
where $N_i^\mathrm{clu}$ is the number of particles of species $i$ in the cluster. The long-time stoichiometry of the clusters thus depends on the activity of the species, but it is independent of their mobility; see Fig.~\ref{phasediag}(d). An intuitive explanation for this observation can be provided as follows: once the cluster becomes neutral, the remaining self-repelling particles will no longer ``sense'' its presence and stay in a dilute phase. At high values of activity and mobility for the self-attractive species, however, these static neutral clusters can become unstable \emph{via} shape-symmetry breaking towards a self-propelled asymmetric cluster [Fig.~\ref{intro}(d)], which also involves the ``shedding'' of some of the self-repelling particles. Such self-propelled clusters are possible only thanks to the existence of non-reciprocal interactions.

\subsection{Beyond Binary Mixtures}

Going beyond binary mixtures ($M>2$), the phase separation phenomenology becomes even more complex due to the increasing number of parameter combinations, leading to a large variety of possible interaction networks between the different species. The instability condition (\ref{instNsame}), however, remains extremely useful. Figure~\ref{dope} demonstrates as an example how a small amount of a highly active ``dopant'' third species can be added to an otherwise homogeneous binary mixture in order to trigger macroscopic phase separation of the whole mixture on demand. Moreover, the instability condition (\ref{instNsame}) can also be used to predict whether highly polydisperse mixtures will phase separate or remain homogeneous, see Fig.~\ref{intro}(f).

\section{Polar Active Colloids: Moment Expansion}		\label{sec:polar-mom}

The description of the collective behaviour of polar active colloids is considerably more complicated than apolar particles due to the additional complexity that arises from the coupling between polarity and motion. Here we develop a systematic framework that can accommodate this complexity in terms of a hierarchical expansion in terms of the moments of the distribution in the Fokker-Planck equation \cite{Golestanian:2012,saha+golestanian14}.

\subsection{From Trajectories to Hydrodynamic Equations}\label{sec:traj}

We consider a collections of $N$ polar spherical particles of radius $R$ and describe the configuration of a particle labeled $i$ with position $\bfr_i(t)$ and orientation $\bfn_i(t)$. The particle experiences deterministic translational velocity $\bfv_i$ and angular velocity ${\bm \omega}_i$, as well as noise characterized by $D_\mathrm{c}$ and $D_\mathrm{r}$, which are the translational and rotational diffusion coefficients. In a medium with uniform temperature $T$, we have $D_\mathrm{c}=\frac{k_{\rm B} T}{6 \pi \eta R}$ and $D_\mathrm{r}=\frac{k_{\rm B} T}{8 \pi \eta R^3}$, where $\eta$ is the viscosity of water. The resulting general Langevin equations for the translational and rotational degrees of freedom are as follows
\beqa
&&\frac{\dd \;}{\dd t}{\bfr}_i(t) =\bfv_i+\sqrt{2 D_\mathrm{c}}\,{\bm \xi}_i, \\
&&\frac{\dd \;}{\dd t} {\bfn}_i(t)={\bm \omega}_i+\sqrt{2 D_\mathrm{r}}\,{\bm \eta}_i \times {\bf n}_i,
\eeqa
where ${\bm \xi}_i$ and ${\bm \eta}_i$ are Gaussian-distributed white noise terms of unit strength. From the stochastic trajectories, we can define the probability distribution
$$
{\cal P}({\bfr},{\bfn},t) \equiv \left\langle \sum_{i=1}^N \delta\left({\bfr}-{\bfr}_i(t)\right) \delta\left({\bfn}-{\bfn}_i(t)\right)\right \rangle,
$$
and recast the Langevin equations, which are governing equations for the trajectories, into an evolution equation for the probability distribution. Defining the rotational gradient operator ${\bm {\mathcal R}} \equiv {\bfn} \times \partial_{\bfn}$, which has the properties $ {\cal R}_\alpha n_\beta=-\epsilon_{\alpha \beta \gamma} n_{\gamma}$ and ${\bf {\cal R}}^2 n_\beta=-2 n_\beta$, allows us to construct the translational and the rotational fluxes as follows
\beqa
&&\bfJ =\bfv(\bfr,\bfn) \;{\cal P}({\bfr},{\bfn})-D_\mathrm{c} {\bm \nabla} {\cal P}({\bfr},{\bfn}),\\
&&\bfJ_r={\bm \omega}(\bfr,\bfn) \;{\cal P}({\bfr},{\bfn})-D_\mathrm{r} {\bm {\mathcal R} {\cal P}}({\bfr},{\bfn}).
\eeqa
Then, we can write the Fokker-Planck equation as a conservation law
\begin{equation}
\partial_t {\cal P}+\nabla \cdot \bfJ+{\bm {\mathcal R}}  \cdot \bfJ_r=0. \label{eq:FP}
\end{equation}

To describe the collective behaviour of active particles with phoretic interactions plus translational self-propulsion, we choose the following forms for the velocities
\beqa
&&\bfv(\bfr,\bfn)=v_0 \bfn- \mu  {\bm \nabla} \psi,\\
&&{\bm \omega}(\bfr,\bfn)=\chi \bfn \times {\bm \nabla} \psi,
\eeqa
where $v_0$ is the self-propulsion speed and $\psi$ represents a thermodynamic potential such as solute concentration (diffusiophoresis), electrostatic potential (electrophoresis), or temperature (thermophoresis). The resulting Fokker-Planck equation reads
\beq
\partial_t {\cal P}+\nabla \cdot \left[\left(v_0 \bfn- \mu  {\bm \nabla} \psi\right){\cal P}-D_\mathrm{c} {\bm \nabla} {\cal P}\right]+{\bm {\mathcal R}}  \cdot \left[\left(\chi \bfn \times {\bm \nabla} \psi\right){\cal P}-D_\mathrm{r} {\bm {\mathcal R}} {\cal P}\right]=0. \label{eq:FP-2}
\eeq
Equation (\ref{eq:FP-2}) is rather complex as it mixes orientation and position. A systematic approximation framework called moment expansion helps us to tackle this complication. The method builds on the orientation moments of the distribution, namely, the density $\rho({\bfr})=\int_{{\bfn}} {\cal P}({\bfr},{\bfn})$, the polarization field ${\bfp}({\bfr})=\int_{{\bfn}} {\bfn} \; {\cal P}({\bfr},{\bfn})$, the nematic order parameter ${\bfQ}({\bfr})=\int_{{\bfn}} \left[{\bfn} {\bfn}-\frac{1}{3} {\bfI}\right] {\cal P}({\bfr},{\bfn})$ etc, and a hierarchy of equations derived from Eq. (\ref{eq:FP-2}), which connect them.

The governing equation for the zeroth moment of orientation is obtain by integrating Eq. (\ref{eq:FP-2}) over $\bfn$. This gives
\begin{equation}
\partial_t \rho+v_0 {\bm \nabla} \cdot {\bfp}-\mu {\bm \nabla} \cdot \bigl[\left({\bm \nabla} \psi\right) \rho  \bigr]-D_\mathrm{c} {\bm \nabla}^2 \rho=0, \label{eq:rho-1}
\end{equation}
which has a source term in the form of $-v_0 \nabla \cdot {\bf p}$ due to the self-propulsion of the colloids. Performing $\int_{{\bf n}} {\bf n} \times$
Eq. (\ref{eq:FP-2}), we can obtain an equation for the polarization field as
\begin{equation}
\partial_t {\bfp}+\frac{v_0}{3} {\bm \nabla} \rho+v_0 {\bm \nabla} \cdot {\bfQ}-\mu \partial_\alpha \bigl[\left(\partial_\alpha \psi\right) {\bfp}\bigr]-D_\mathrm{c} {\bm \nabla}^2 {\bfp}+\chi \bfQ \cdot {\bm \nabla} \psi- \frac{2}{3} \chi \rho  {\bm \nabla} \psi +2 D_\mathrm{r} {\bfp}=0. \label{eq:p-1}
\end{equation}
Continuing this process will produce the interconnected hierarchy of equations for the moments. To make further progress, we truncate the hierarchy so that we can deal with a finite number of equations. For sufficiently dilute solutions (i.e. when $\rho R^3 \ll 1$) and in the absence of any external mechanisms that can lead to alignment, such as external fields or boundaries \cite{pala10,holger}, we can ignore the nematic order and set ${\bfQ} \simeq 0$. This yields
\begin{equation}
\partial_t {\bfp}+\frac{v_0}{3} {\bm \nabla} \rho-\mu \partial_\alpha \bigl[\left(\partial_\alpha \psi\right) {\bfp}\bigr]-D_\mathrm{c} {\bm \nabla}^2 {\bfp}- \frac{2}{3} \chi \rho  {\bm \nabla} \psi +2 D_\mathrm{r} {\bfp}=0. \label{eq:p-2}
\end{equation}

\subsection{Self-consistent Field Equations}

To complete the description of the system, we need to specify how the field $\psi$ is generated by the phoretically active particles. A generic governing equation for the field can be written as
\beq
\xcancel{\partial_t \psi}-K  {\bm \nabla}^2 \psi=\text{sources and sinks},\label{eq:psi-1}
\eeq
where $K$ represents the solute diffusion coefficient (diffusiophoresis) or the heat conductivity (thermophoresis) etc. The time derivative term is ignored because we are interested in the long time limit and assume that solute, heat, etc diffuse much faster than the colloids. Assuming a surface activity coverage $\alpha_i(\Omega_i)$ for the $i$th colloid, we can describe the right hand side of Eq. (\ref{eq:psi-1}) as follows
\beqa
-K  {\bm \nabla}^2 \psi &=& \sum_i \int \dd S_i \; \alpha_i(\Omega_i) \; \delta(\bfr-\bfr_i-R {\hat \bfR_i}) \nonumber \\
&=&  \sum_i \int \dd S_i \; \alpha_i(\Omega_i) \bigl[\delta(\bfr-\bfr_i) - R {\hat \bfR_i} \cdot {\bm \nabla} \delta(\bfr-\bfr_i)+ \cdots \bigr] \nonumber \\
&=& R^2 \sum_i \left[\int \dd \Omega_i \; \alpha_i(\Omega_i)\right] - R^3 \sum_i \left[\int \dd \Omega_i \; \alpha_i(\Omega_i) {\hat \bfR_i}\right] \cdot {\bm \nabla} \delta(\bfr-\bfr_i)+\cdots.\label{eq:sources-1}
\eeqa
Assuming all colloids are the same and using the expansion of Eq. (\ref{alpha-Omega}), we find $\int \dd \Omega \alpha(\Omega)=4 \pi \alpha_0$ and $\int \dd \Omega \alpha(\Omega) {\hat \bfR}=\frac{4 \pi}{3} \alpha_1 \bfn$. Therefore, Eq. (\ref{eq:sources-1}) reads
\beq
-K  {\bm \nabla}^2 \psi =4 \pi R^2 \alpha_0  \sum_i \delta(\bfr-\bfr_i)- \frac{4 \pi}{3} R^3 \alpha_1  {\bm \nabla} \cdot \left[\sum_i \bfn_i\delta(\bfr-\bfr_i)\right]+ \cdots.\label{eq:psi-2}
\eeq
The above form of the equation for $\psi$ highlights its stochastic nature. To proceed, we implement a mean-field approximation and replace $\psi$ by its average over the trajectories, which yields
\beq
-K  {\bm \nabla}^2 \psi =4 \pi R^2 \alpha_0  \rho- \frac{4 \pi}{3} R^3 \alpha_1  {\bm \nabla} \cdot \bfp+ \cdots.\label{eq:psi-3}
\eeq
This equation now complements Eqs (\ref{eq:rho-1}) and (\ref{eq:p-2}) for a complete approximate description of the system.

\subsection{Behaviour at Long Times and Large Length Scales}

We are interested in the behaviour of the system at time scales sufficiently longer than the rotational diffusion time of the colloids $D_{\rm r}^{-1}$ and lengths much larger than the size of the colloid $R$. A description of this regime can achieved by ignoring a number of terms in Eq. (\ref{eq:p-2}) as follows
\begin{equation}
\xcancel{\partial_t {\bfp}}+\frac{v_0}{3} {\bm \nabla} \rho-\mu \left({\bm \nabla}^2 \psi \right){\bfp}-\xcancel{\mu \left(\partial_\alpha \psi\right) \left(\partial_\alpha {\bfp}\right)}-\xcancel{D_\mathrm{c} {\bm \nabla}^2 {\bfp}}- \frac{2}{3} \chi \rho  {\bm \nabla} \psi +2 D_\mathrm{r} {\bfp}=0. \label{eq:p-3}
\end{equation}
Note that the $4^{\rm th}$ term is similar to the $3^{\rm rd}$ term, but it adds a tensorial structure to the equation. We have ignored it here for simplicity. Inserting an approximate form of Eq. (\ref{eq:psi-3}), namely ${\bm \nabla}^2 \psi \simeq -4 \pi R^2 \alpha_0 \rho/K$, in Eq. (\ref{eq:p-3}), we obtain
\begin{equation}
\left[(d-1) D_\mathrm{r}+S_d R^{d-1} \;\frac{\alpha_0 \mu}{K} \;\rho\right] \bfp=-\frac{v_0}{d} {\bm \nabla}\rho+\frac{d-1}{d} \;\chi \rho {\bm \nabla} \psi,\label{eq:p-vs-rho-1}
\end{equation}
where we have written the coefficients explicitly in terms of the dimensionality of space $d$. 
Here $S_d=2 \pi^{d/2}/\Gamma(d/2)$ is the surface area of the unit sphere embedded in $d$ dimensions. From Eq. (\ref{eq:p-vs-rho-1}), we can find an explicit expression for the polarization in terms of the density and the field, which reads
\begin{equation}
\bfp=\frac{\left(-\frac{v_0}{3} {\bm \nabla}\rho+\frac{2}{3} \chi \rho {\bm \nabla} \psi\right)}{\left(2  D_\mathrm{r}+\frac{4 \pi R^{2}\alpha_0 \mu}{K} \;\rho\right)},\label{eq:p-vs-rho-2}
\end{equation}
in $d=3$. Note that phoretic interaction renormalizes the rotational diffusion of the colloids. Setting $\rho \simeq \rho_0={\rm const}$ in the denominator of Eq. (\ref{eq:p-vs-rho-2}) and inserting the resulting form for $\bfp$ back into Eq. (\ref{eq:rho-1}), we obtain the following equation for the density field
\beq
\partial_t \rho+{\bm \nabla} \cdot \bfJ_{\rm eff}=0, \label{eq:rho-2}
\eeq
where the effective flux is defined as
\beq
\bfJ_{\rm eff}=-D_{\rm eff} {\bm \nabla}\rho-\mu_{\rm eff} ({\bm \nabla} \psi) \rho,\label{eq:J-eff-1}
\eeq
in terms of the effective diffusion coefficient
\beq
D_{\rm eff}=D_\mathrm{c}+\frac{v_0^2}{6 (D_\mathrm{r}+2 \pi R^2 \alpha_0 \mu \rho_0/K)},\label{eq:D-eff-1}
\eeq
and the effective phoretic mobility
\beq
\mu_{\rm eff}=\mu-\frac{v_0 \chi}{3 (D_\mathrm{r}+2 \pi R^2 \alpha_0 \mu \rho_0/K)}.\label{eq:mu-eff-1}
\eeq
We thus find that self-propulsion leads to an enhancement of the translational diffusion of the colloid on time scales longer than the rotational diffusion, while the combination of phoretic alignment and self-propulsion leads to a renormalization of the translational phoretic mobility at long times. 

In stationary state, Eq. (\ref{eq:rho-2}) is satisfied if $\bfJ_{\rm eff}=0$, which yields
\begin{equation}
\rho({\bfr})=\rho_0 \;\exp\left[-\frac{\mu_{\rm eff}}{D_{\rm eff}} \psi(\bfr)\right].\label{eq:rho-psi}
\end{equation}
This is a generalized Boltzmann distribution, which will allow us to use analogies to equilibrium theories of electrolytes.

\subsubsection{Stationary State Polarization}

We can insert the stationary distribution into Eq. (\ref{eq:p-vs-rho-2}) to find a direct relationship between the polarization and the density gradient as follows
\beq
\bfp=-\Gamma {\bm \nabla}\rho,\label{eq:p-rho}
\eeq
where the response coefficient is given as
\beq
\Gamma=\frac{\frac{D_\mathrm{c}}{v_0} \chi+\mu/2}{3 \frac{D_\mathrm{r}}{v_0} \left(1+\frac{2 \pi R^2 \alpha_0 \mu \rho_0}{K D_\mathrm{r}}\right) \mu-\chi},\label{eq:Gamma-def}
\eeq
in terms of $\mu$ and $\chi$, which can both be either positive or negative. The parameters can thus be tuned such that $\Gamma >0$, in which case polarization tends to stabilize accumulation of particles via a tendency for the particles to swim away from high density regions. When $\Gamma <0$, on the other hand, the particles tend to be aligned with the concentration gradient and the particles tend to swim towards already crowded regions, hence instigating an instability. Therefore, the alignment or polarization tendencies of the system as controlled by $\chi$ will determine the phase behaviour of the system in competition with the translational or positional tendencies that are controlled by $\mu$ (see Fig. \ref{inst}). 

\begin{figure}[t]
\includegraphics[width=0.4\linewidth]{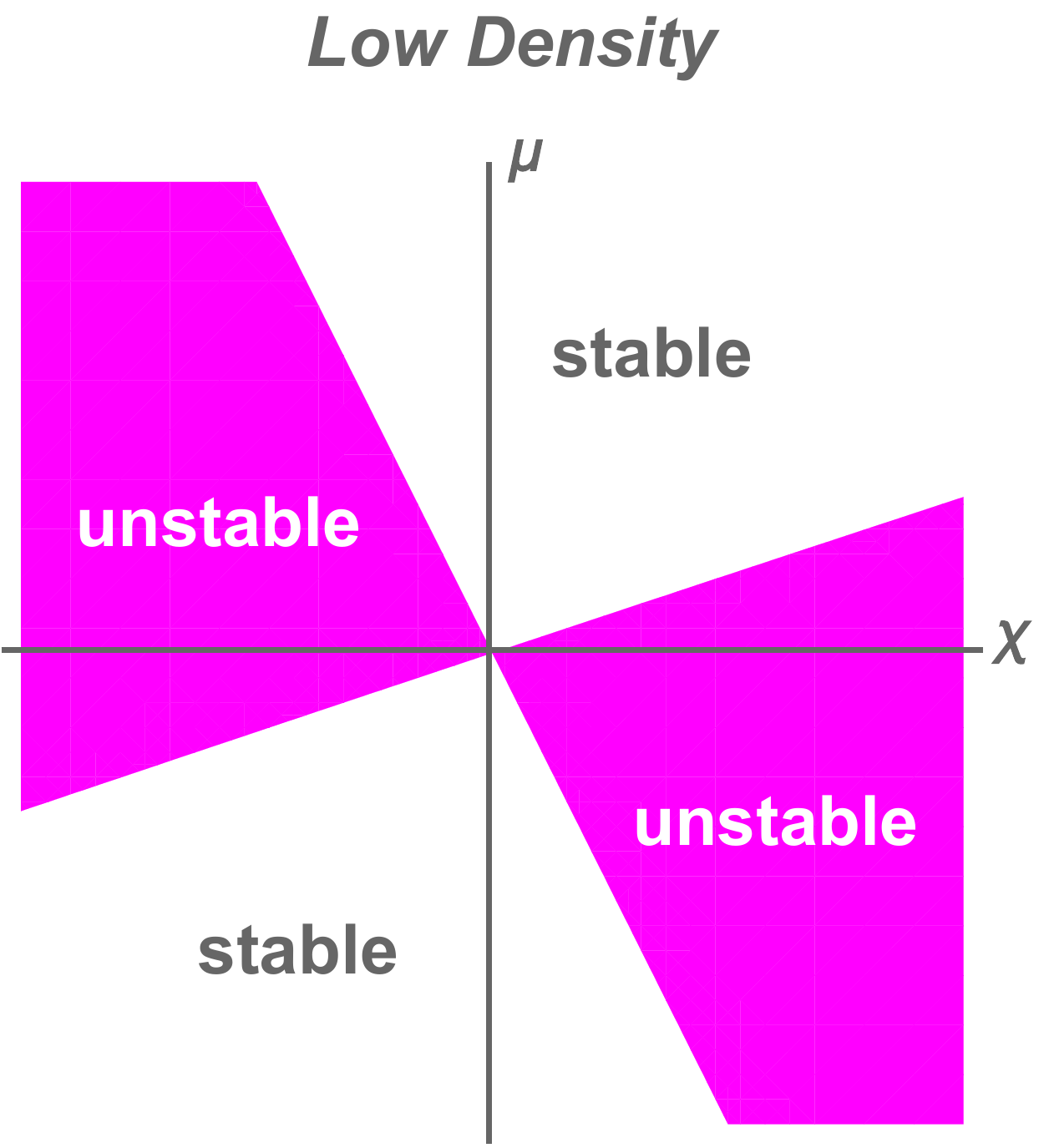}\hskip1.5cm
\includegraphics[width=0.4\linewidth]{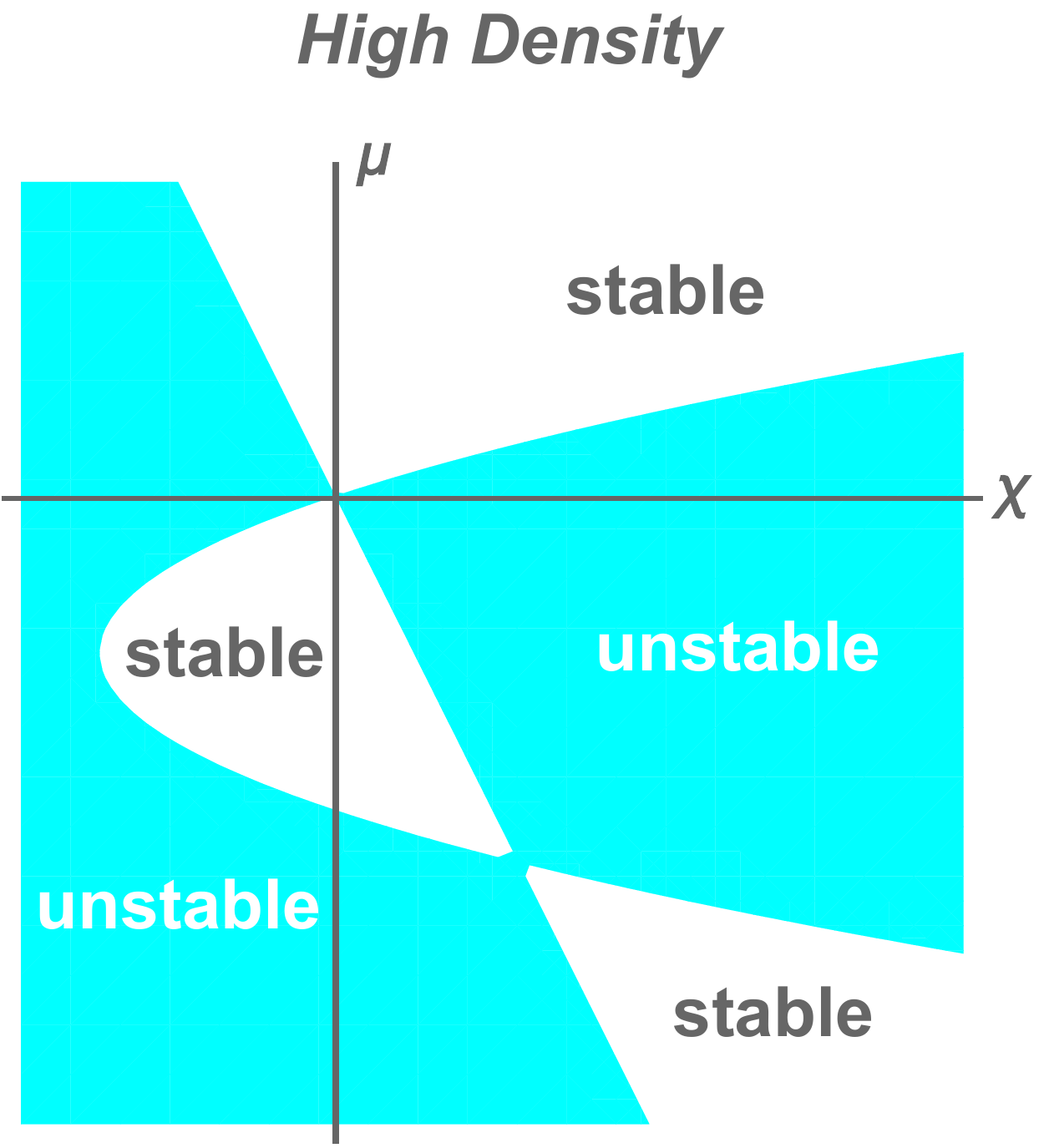}
\caption{Stability of the suspension of polar phoretic active colloids as determined by the condition $\Gamma <0$, which ensures that the colloids will point in the direction of higher concentrations and actively swim towards the denser region. 
\label{inst}}
\end{figure}

\subsubsection{Generalized Poisson-Boltzmann Equation}

Going back to Eq. (\ref{eq:psi-3}), we can now eliminate the polarization by inserting its explicit form from Eq. (\ref{eq:p-vs-rho-2}). This yields
\beq
- {\bm \nabla} \cdot \bigl(K_{\rm eff} {\bm \nabla} \psi\bigr) =4 \pi R^2 \alpha_0  \rho+ \xcancel{\frac{2 \pi}{9} \frac{R^2 \alpha_1 v_0}{(D_\mathrm{r}+2 \pi R^2 \alpha_0 \mu \rho_0/K)} {\bm \nabla}^2\rho}+ \cdots,\label{eq:psi-4}
\eeq
where the $K$ coefficient is renormalized due to the polarization of the Janus particles as follows
\beq
K_{\rm eff}=K-\frac{R^3 \alpha_1 \chi \rho}{(D_\mathrm{r}+2 \pi R^2 \alpha_0 \mu \rho_0/K)}.\label{eq:K-eff-1}
\eeq
This phenomenon is analogous to the emergence of the polarization field inside dielectric material, which is accounted for by an effective dielectric constant that reduces or screens the field. We can simplify further and assume a constant density profile in Eq. (\ref{eq:K-eff-1}), and therefore treat $K_{\rm eff}$ as a constant.
Putting Eq. (\ref{eq:rho-psi}) into this simplified form of Eq. (\ref{eq:psi-4}), we find
\begin{equation}
-K_{\rm eff} {\bm \nabla}^2 \psi=4 \pi R^2 \alpha_0 \rho_0  \;\exp\left[-\frac{\mu_{\rm eff}}{D_{\rm eff}} \psi(\bfr)\right].\label{eq:PB-psi-1}
\end{equation}
This equation is reminiscent of the Poisson-Boltzmann equation for electrolytes, which should be solved subject to the normalization constraint
\beq
N=\int \dd^d \bfr \;\rho(\bfr)=\rho_0 \int \dd^d \bfr \;\exp\left[-\frac{\mu_{\rm eff}}{D_{\rm eff}} \psi(\bfr)\right].\label{eq:Normal-1}
\eeq 
We can identify two distinct classes described by the above equations:
\begin{itemize}
\item {\bf Electrostatic}, in which like charges predominantly repel. This corresponds to $\alpha_0 \mu_{\rm eff} >0$.
\item {\bf Gravitational}, in which like charges predominantly attract. This corresponds to $\alpha_0 \mu_{\rm eff} <0$.
\end{itemize}

\begin{figure}[b]
\includegraphics[width=0.9\linewidth]{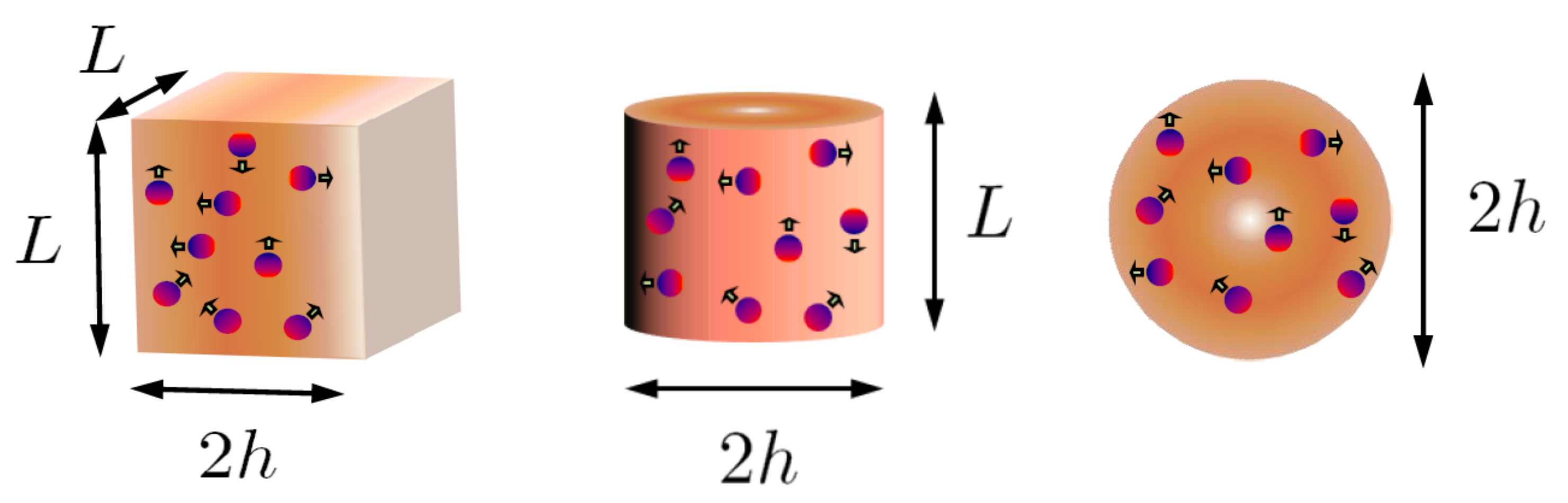}
\caption{Different geometries for the electrostatic and gravitational cases in analogy with cases where Poisson-Boltzmann equation has been studied.
\label{hot}}
\end{figure}

We can define a dimensionless field as
\beq
\Psi \equiv \frac{\mu_{\rm eff} \psi}{D_{\rm eff}} \cdot {\rm sgn}(\alpha_0 \mu_{\rm eff}),
\eeq
and a characteristic Bjerrum length scale
\begin{equation}
\ell\equiv \frac{R^2 |\alpha_0 \mu_{\rm eff}|}{K_{\rm eff} D_{\rm eff}},\label{eq:Bjerrum}
\end{equation}
as well as a corresponding Debye length $\kappa^{-1}$ defined via
\beq
\kappa^2=4 \pi \ell \rho_0.
\eeq
Then our Poisson-Boltzmann equation reads
\begin{equation}
-\nabla^2 \Psi=\kappa^2 e^{\mp\Psi},\label{eq:PB-T-2}
\end{equation}
where the sign choice is $\mp=-{\rm sgn}(\alpha_0 \mu_{\rm eff})$. Equation (\ref{eq:PB-T-2})
is subject to the constraint $N=\rho_0 \int \dd^d{\bfr} \;e^{\mp\Psi}$.

In analogy with studies of Poisson-Boltzmann equation, we can look for exact solutions of the above equation under confinement, by applying the following boundary condition
\beq
-\partial_{\perp} \Psi |_{S}=4 \pi \ell N/A,
\eeq
which we obtain by invoking Gauss theorem, as well as symmetry considerations. We will now consider a number of different geometries as shown in Fig. \ref{hot}.

\begin{figure}[b]
\includegraphics[width=0.4\linewidth]{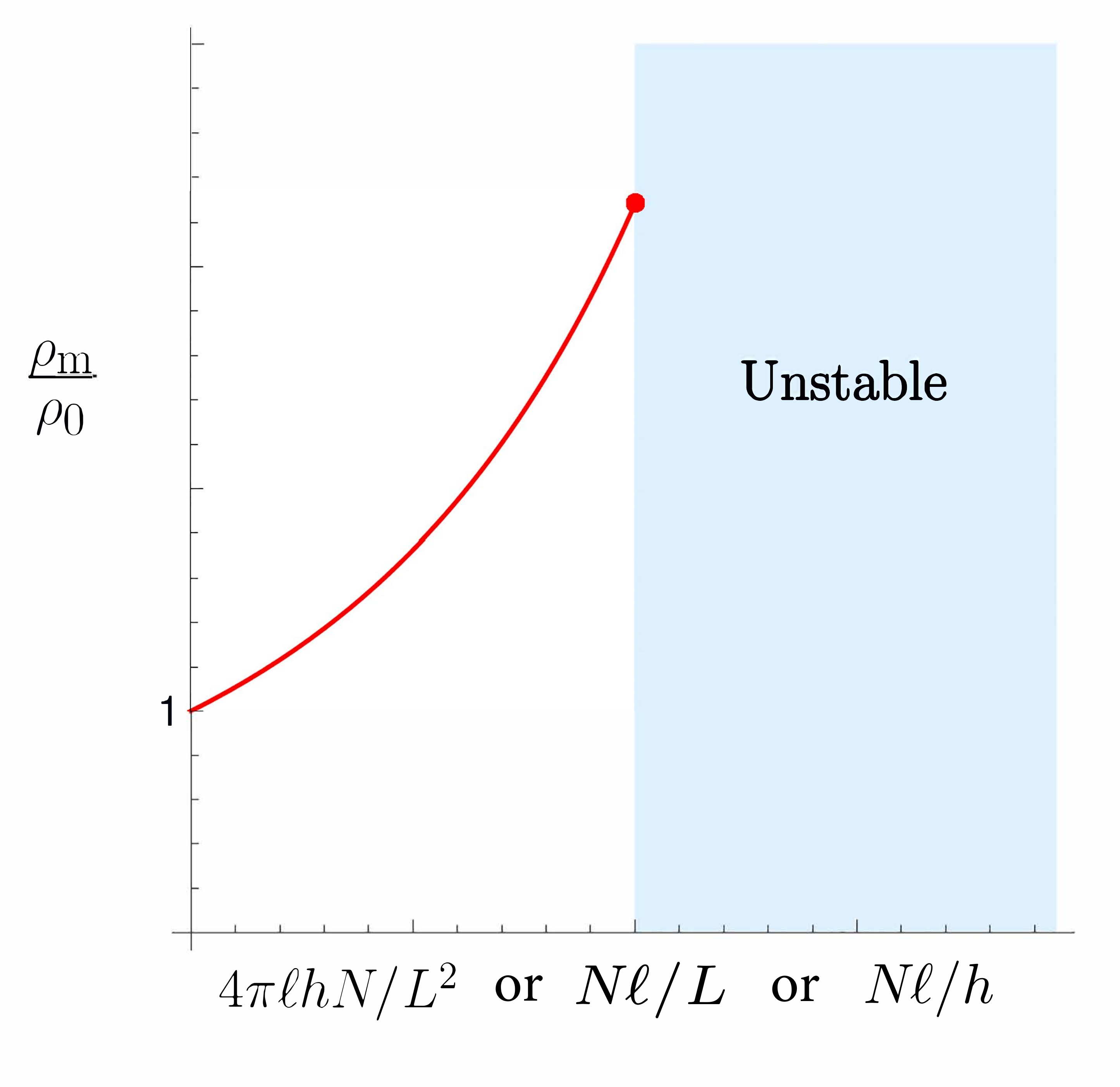}
\caption{The density of colloids in the middle of the confined space relative to the density at the edge as a function of the dimensionless phoretic coupling constant for the gravitational case. 
\label{gravity}}
\end{figure}

In the electrostatic case, we can obtain exact solutions in cases with 1D and 2D confinement \cite{Levin2002}. When the colloids are confined between two plates of lateral size $L$ and distance $2 h$, the exact density profile is found as
\begin{equation}
\rho(x)=\frac{\rho_0}{\left[1+\frac{2 \pi^2 \ell^2}{k^2} \left(\frac{N}{L^2}\right)^2\right] \cos^2 \left(\frac{k x}{\sqrt{2}}\right)},\label{eq:sol-1D-rho-1}
\end{equation}
where $\rho_0$ is the concentration at the edge of the confining wall, and $k$ satisfies the following transcendental equation
\beq
\left(\frac{k h}{\sqrt{2}}\right) \tan\left(\frac{k h}{\sqrt{2}}\right)=\pi \ell h \left(\frac{N}{L^2}\right),\label{eq:trans-1}
\eeq
The profile of Eq. (\ref{eq:sol-1D-rho-1}) describes an accumulation of the colloids near the confining boundary that is analogous to the phenomenon of {\em counterion condensation} \cite{Levin2002}, and a resulting depletion zone in the central
region of the system. In the strong coupling limit when $N \ell h/L^2 \gg 1$, we can obtain an approximate solution to Eq. (\ref{eq:trans-1}) as $k h \simeq \frac{\pi}{\sqrt{2}} \left[1-\frac{1}{\pi \left({N \ell h}/{L^2}\right)}\right]$.
In this limit, the ratio between the density of the colloids in the middle and at the edge can be found as $\frac{\rho_{\rm m}}{\rho_0} \simeq \frac{1}{4} \;(N \ell h/L^2)^{-2}$, which shows a significant depletion effect. Note that the depletion becomes stronger as $h$ is increased, when other parameters are kept fixed. The length scale $1/[2 \pi \ell (N/L^2)]$ is equivalent to the Gouy-Chapman length in the electrostatic analogy \cite{Levin2002}.

For an active colloidal solution confined in a cylindrical cage of length $L$ and width $2 h$, the density profile is obtained as
\begin{equation}
\rho(r)=\frac{\rho_0}{\left[1+\frac{1}{2}\left(\frac{N \ell}{L}\right)\right]^2
\left[1-\frac{1}{8} k^2 r^2\right]^2},\label{eq:sol-2D-rho-1}
\end{equation}
where $k$ is given by following closed-form expression
\beq
k h=\sqrt{\frac{8 (N \ell/L)}{2+(N \ell/L)}}.
\eeq
In this geometry, the strong coupling limit corresponds to $N \ell/L \gg 1$, in which case we obtain a measure of depletion as follows $\frac{\rho_{\rm m}}{\rho_0} \simeq 4 \;(N \ell/L)^{-2}$, which is independent of the confinement size in this geometry. The ratio $\ell N/L$ is analogous to the so-called Manning-Oosawa parameter for highly charged rodlike polyelectrolytes \cite{Levin2002}. The same type of profile is obtained when the colloidal solution is confined in 3D to a spherical cage
of diameter $2 h$, where in the strong coupling limit defined via $N \ell/h \gg 1$, we have $\frac{\rho_{\rm m}}{\rho_0} \simeq 21.4 \;(N \ell/h)^{-2}$. Here, the depletion is inversely related to the size of the cage, i.e. it decreases for larger cages.

In the gravitational case, we observe accumulation of the colloidal particles at the centre of the confined region, which contrasts from the electrostatic case, while the potential profile $\Psi(\bfr)$ is still peaked at the centre as in the electrostatic case. 
The relevant coupling constants denoted as $g(d)$ are, $g(1)=N \ell h/L^2$ (1D), $g(2)=N \ell/L$ (2D), $g(3)=N\ell/h$ (3D), as discussed above. As $g(d)$ increases, the ratio $\rho_{\rm m}/\rho_0$ increases as well, signalling accumulation at the centre. This structure, which is still a relatively dilute as of particles that are free to diffuse within the confined region, is analogous to a ``colloidal star'', in the gravitational analogy. This dilute structure is stable up to a critical point $g_c(d)$ beyond which a stable (stationary-state) solution no longer exists; see Fig. \ref{gravity}. For example, in 1D the onset of instability occurs at $g_c(1)=1/2\pi$, at which $(\rho_{\rm m}/\rho_0)_c=3.29$, while similar thresholds hold for the 2D and 3D confinement cases \cite{LanLif-fluid}. The instability occurs because the particles that act as sources for the potential attract each other and result in a suspension that becomes increasingly denser and more attractive. In this case, the flux at the outer boundary of the system cannot balance the field generated inside the confined region, which leads to an uncontrolled buildup of thermodynamic energy associated with the potential $\Psi$. This state of the system can be called a ``colloidal supernova'' in our gravitational analogy. However, we should bear in mind that the analogy is not exact as the colloidal system operates in the dissipative regime, in contrast with the inertial
and conserved dynamics of the gravitational system.

\begin{figure}[t]
\includegraphics[width=0.50\linewidth]{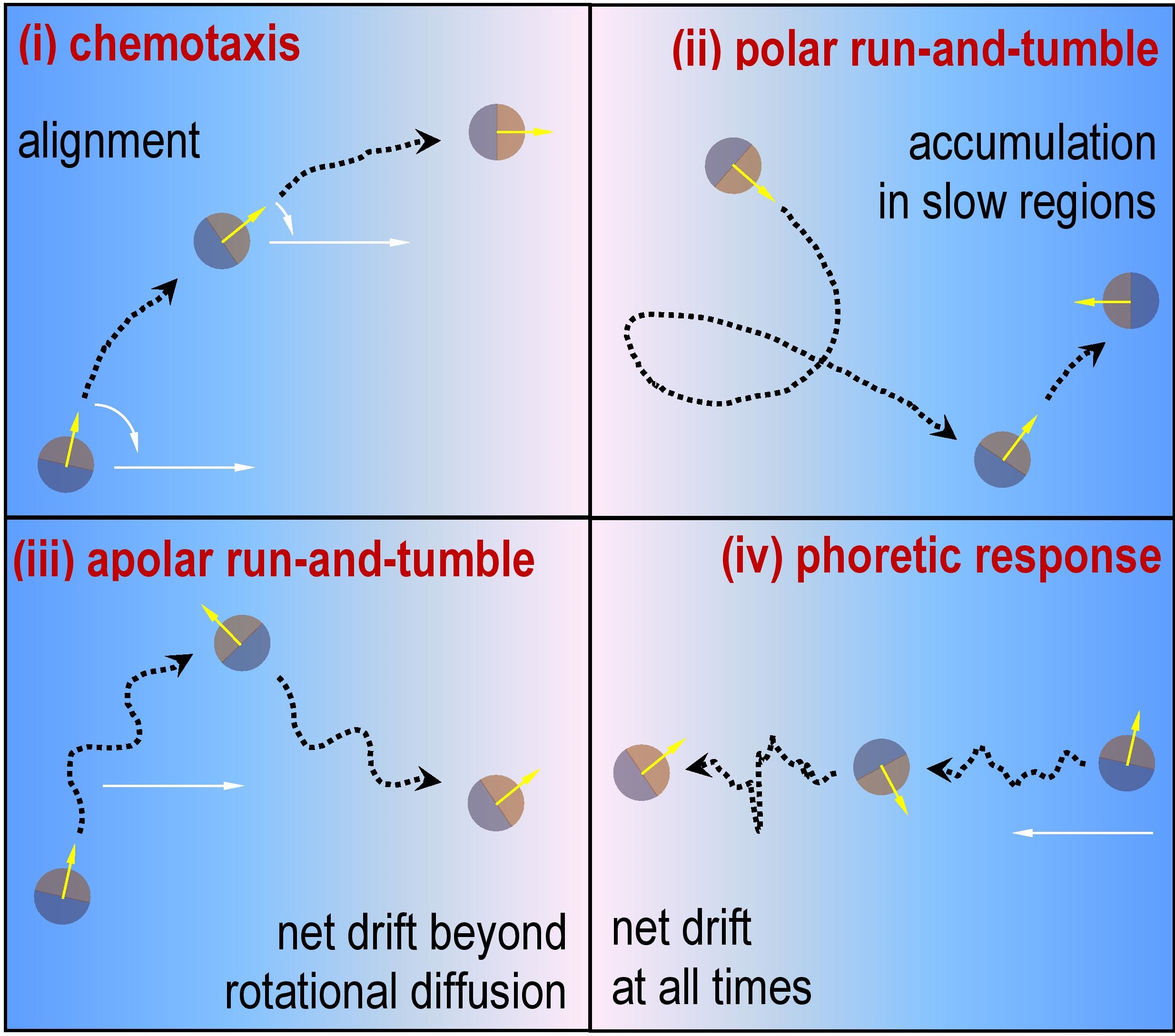}
\caption{Four different mechanisms or channels through which a Janus particle responds to gradients in substrate concentration \cite{saha+golestanian14}. 
\label{fig:chemotaxis}}
\end{figure}

\subsubsection{Additional Generalizations}

In our simplified description, we have so far ignored a number of important features that will affect the dynamics of catalytically active Janus particles. Typically, the catalytic activity on Janus particles involves reactions that convert {\it substrates} (as reactants are called in the biochemistry literature) into {\it products}, i.e. S$\to$P. This means that there a number of chemical species in the solution, each producing their own gradients and contributing to phoretic transport with their corresponding mobilities, as shown in Eq. (\ref{vs-mu-k}). 

Moreover, the mobilities will in general be tensors for Janus particles, allowing in general different drift velocities along the polar axis of the particle and perpendicular to it. Additionally, the Janus structure will introduce an alignment response to a concentration gradient due to the angular velocity given in Eq. (\ref{ASS-2}). Finally, the variations in the local concentration of the substrate molecule that fuels the propulsion will also modulate the swimming velocity $v_0$. Putting together all these contributions, we find that a single Janus particle can respond to variations in substrate concentrate via four different channels or mechanisms as summarized in Fig. \ref{fig:chemotaxis}.

Taking these effects into consideration will give us a complex phase diagram that includes a range of collective dynamical regimes including clustering, pattern formation, aster condensation, plasma oscillations, and spontaneous oscillations \cite{saha+golestanian14}. The existence of such a range of different regimes can be traced back to different possibilities provided by the positional and orientational interactions, as discussed in Sec. \ref{sec:traj} above. When $\mu <0$ and $\chi >0$, the particles are translationally attracted to one another while they would also tend to orient towards each other and swim towards one another; this is a clear cut case of collapse instability. If, on the other hand, $\mu >0$ and $\chi >0$, they repel each while they tend to orient towards each other and swim to one another; this is a frustrated case, which can lead to oscillations and pattern formation. Similar observations have been reported from studies using Brownian dynamics simulations \cite{Stark2014,Stark2018}. Enhanced density fluctuations and clustering that can arise from phoretic instabilities as discussed above have been observed experimentally in suspensions of catalytic Janus swimmers \cite{pala10,Palacci2013}.

\section{Polar Active Colloids: Scattering and Orbiting}		\label{sec:polar-scat}

The existence of different modes of chemotactic coupling to position and orientation has interesting implications on how two active colloids interact with one another \cite{Saha2019}.

\begin{figure}[t]
\begin{center}
\includegraphics[width=0.80\linewidth]{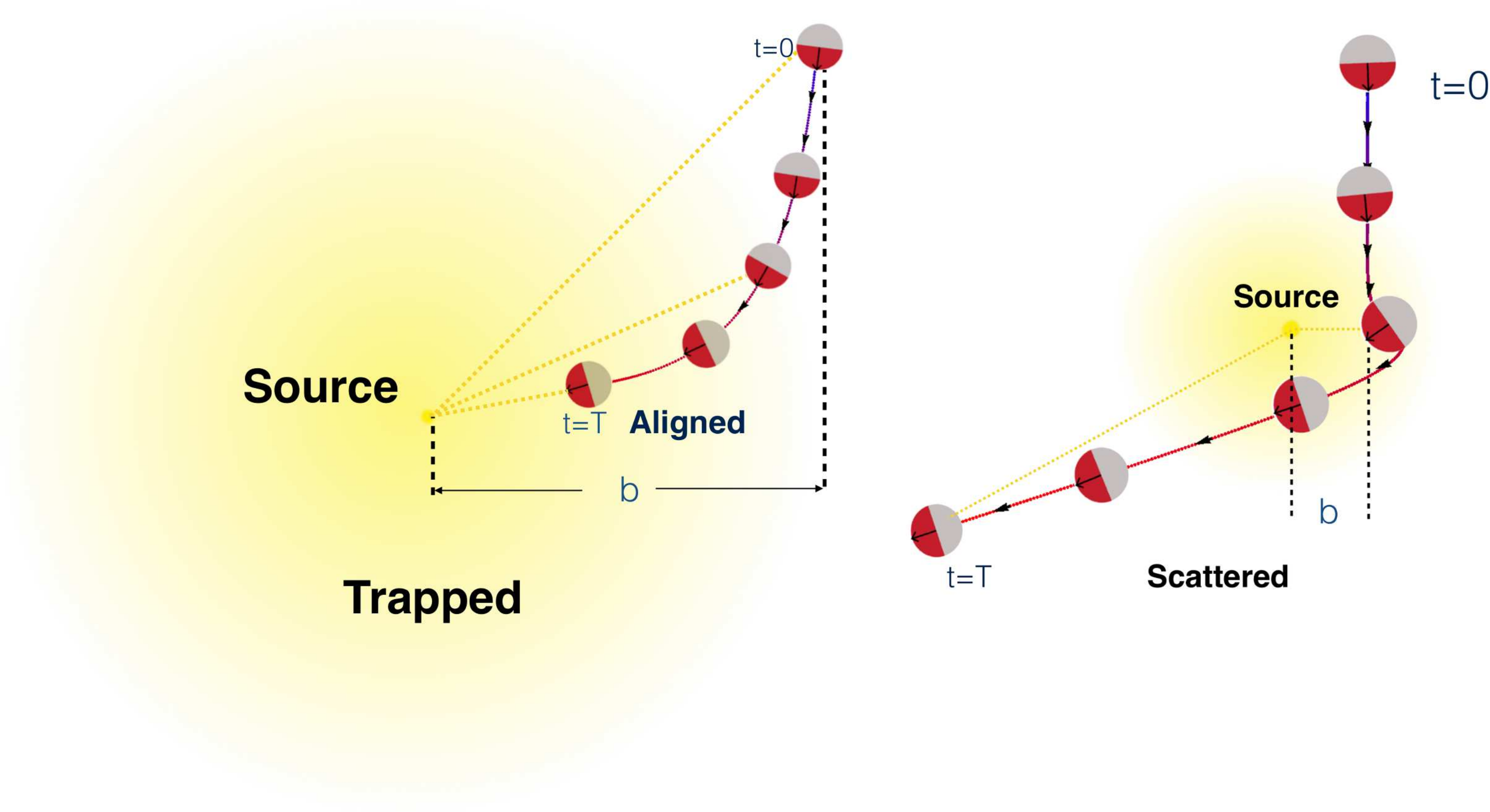}         
		    \caption{{Typical trajectories of an active colloid in an isotropic source of fuel, illustrating the role of initial conditions and active rotation of the polar axis. The swimmer can be captured by a source of chemicals or interact with it briefly before scattering, depending on initial conditions. If it approaches the source with an impact parameter $b$ lower than a threshold, a bound state is formed when self-propulsion and phoretic interactions balance each other. The swimmer on the left is trapped when the polarity of the aligns with the local chemical gradient marked with dotted orange lines. The swimmer on the right is scattered as it cannot be trapped. (Figure by Suropriya Saha).
}}
\label{fig:scattering}
\end{center}
\end{figure}

When a polar active colloids interacts with an apolar source of chemical, two different types of behaviour can emerge (see Fig. \ref{fig:scattering}). An active colloid that tends to align with the local gradient of an externally imposed chemical field can be trapped by a source of fuel.  A trapped swimmer either comes to rest at a fixed distance from the source or executes periodic orbits. By tuning initial condition, it is possible to transition to a state where the swimmer interacts with the source for a short period before running away, thereby undergoing scattering. 

Two interacting chemotactic active colloids, which can rotate their polar axis to align with an external chemical gradient, form new bound states by cancellation of velocities rather than by minimization of a free energy. The interactions are dynamical in origin, resulting from an interplay of self propulsion and gradient-seeking mechanisms, and are thus non-central and non-reciprocal.  Bound states are formed where the distance between their centres and relative orientation of their polarity remains fixed or traces a periodic cycle. These states fall in two broad classes: (i) active dimers, where the centre of mass translates linearly and (ii) orbits, where the centre of mass moves in a closed orbit. A necessary condition is that the chemotactic alignment response of at least one colloid in the pair must be positive. Similarly to the case of a single swimmer near a source, they can unbind and scatter when the surface activity is changed. The fixed points underlying the bound states correspond to the case when exactly one of the two colloids is stationary, and show that the transition happens through bifurcations. These findings are robust upon the introduction of hydrodynamic interactions and (relevant) thermal fluctuations \cite{Saha2019}. Similar effects have been studied for a system of active colloids under confinement \cite{Kanso2019}.

\section{Non-equilibrium Dynamics of Active Enzymes}		\label{sec:enz}

Enzymes are molecular machines that catalyze chemical reactions. The appropriate description of a chemical reaction is a Kramers (escape) process in the reaction space in which the system aims to go from an initial higher energy state, which corresponds to the substrate to a final lower energy state, corresponding to the product, by overcoming an energy barrier. A schematic description of how enzymes work can be constructed as follows (see Fig. \ref{fig:enzyme-schem}). Consider a chemical reaction S$\to$P as an activated process along a specific reaction coordinate, with a barrier that is considerably larger than $\kT$; this transition will happen very slowly on its own. An enzyme can speed up this reaction if upon binding to the substrate it can effectively lower the barrier, or perhaps more accurately, open up a new trajectory with a lower barrier that was not accessible before. Therefore, enzymes are drivers of non-equilibrium activity at the right time at the right place.

For the reaction path described in Fig. \ref{fig:enzyme-schem}, the overall rate of product formation follows the so-called Michaelis-Menten rule
\beq
\frac{\dd}{\dd t} P=k E_0,\label{eq:MM0}
\eeq
where $E_0$ is the bulk concentration of enzymes, and $k$ is the effective catalytic reaction for a single enzyme, given as
\beq
k=k_{\rm cat} \; \frac{S}{K+S},\label{eq:MM}
\eeq
with $K$ being the Michaelis constant.

\begin{figure}[t]
\includegraphics[width=0.50\linewidth,angle =-90]{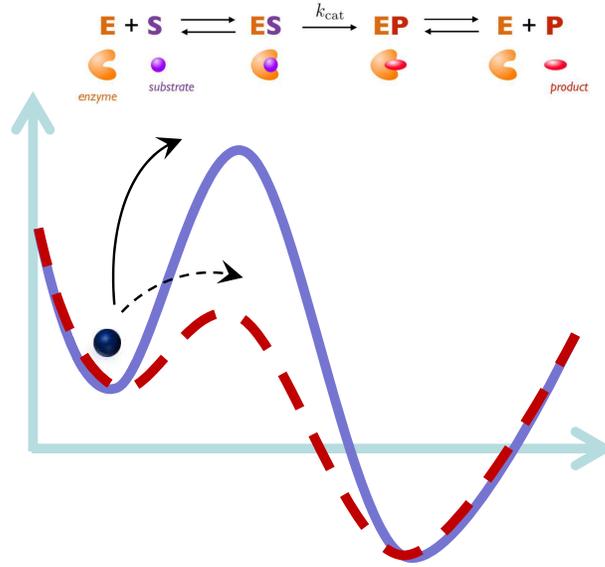}
\caption{The interaction of an enzyme and its substrate effectively lowers the energy barrier along the relevant reaction coordinate and speed up a chemical reaction, which would normally happen, but at a much slower rate. Solid line represents a free reaction and dashed lined represents a reaction that is facilitated by an enzyme.
\label{fig:enzyme-schem}}
\end{figure}

\subsection{Enhanced Diffusion of Enzymes}

There have been a number of experimental reports on the effect of catalytic activity on the diffusion of enzymes \cite{mudd10,seng13,seng14,ried15}. Typically, the enzymes have been found to undergo diffusion with an effective diffusion coefficient that depends on the substrate concentration, which can be approximately described via
\beq
D_{\rm eff}(S)=D_0+\Delta D \; \frac{S}{K+S},\label{eq:DeffMM}
\eeq
where $D_0\equiv D(S=0)$, and $\Delta D/D_0$ is often of the order of a fraction of one (ten percent or so) \cite{mudd10,seng13,seng14,ried15}.

There are several mechanisms that can contribute to enhanced diffusion of enzymes with varying degrees of significance \cite{Golestanian2015,Agudo-Canalejo2018}: 

{ \bf (i) self-phoresis}, due to self-generated chemical gradients or temperature gradients if the reactions are exothermic. This contribution can typically yield $\Delta D/D_0 \sim 10^{-16}$ for fast enzymes. 

{\bf (ii) boost in kinetic energy}, as caused by the release of the energy of the reaction to the centre of mass translational degrees of freedom by equipartition. This mechanism leads to an effective diffusion coefficient given by  
\beq
D_{\rm eff}=D_0 \left[1+\frac{2}{3} \frac{\gamma Q}{\kT} \; k \tau_{\rm b}\right],\label{eq:Boost}
\eeq
where $\gamma$ represents the fraction of the released energy of reaction $Q$ that is transferred to the centre of mass, and $\tau_{\rm b}$ is the time scale characterizing the decay of the inertial boost. Using an estimate of $\gamma=10^{-4}$ based on the number of degrees of freedom in a typical enzyme, as well as $k_{\rm cat}=10^5$ s$^{-1}$ and $Q=40 \kT$ for a fast exothermic enzyme like catalase, we find $\Delta D/D_0 \sim 10^{-9}$.

{\bf (iii) stochastic swimming} due to cyclic stochastic conformational changes associated with the catalytic activity of the enzyme \cite{Golestanian:2008b,Golestanian2009b,Najafi2010,Bai2015}. We can use a simple bead-spring model to estimate this effect. Let us consider two spherical beads of radius $R$ that are attached to each by a linker, that can undergo stochastic elongations with amplitude $b$. The amplitude of these deformations is typically much smaller than the size of the enzyme, e.g. when they arise from mechanochemical coupling of electrostatic nature \cite{Golestanian:2010} (analogous to phosphorylation) or structural changes due to ligand binding \cite{Sakaue:2010}. However, it is possible that the local heat release could disturb the relatively more fragile tertiary structure of the folded protein for a short while, leading to large amplitudes; thereby suggesting $b \lesssim R$ .

To calculate the contribution of such conformational changes to effective diffusion coefficient, we use a simple model in which the conformational change is described by one degree of freedom $L(t)$ representing elongation of the structure along an axis defined by a unit vector ${\bfn}(t)$. To achieve directed swimming, we need at least two degrees of freedom to incorporate the coherence needed for breaking the time-reversal symmetry at a stochastic level \cite{Najafi2004}, and we know that realistic conformational changes must involve many degrees of freedom. The randomization of the orientation, described via $\left\langle {\bfn}(t) \cdot {\bfn}(t')\right\rangle=e^{-2 D_{\rm r} |t-t'|}$, will turn the directed motion into enhanced diffusion over the time scales longer than $1/D_{\rm r}$. Since the same can be achieved through reciprocal conformational changes described by one compact degree of freedom, we will adopt this simpler form. The stochastic motion of the enzyme can be described by the Langevin equation 
\beq
{\bfv}(t) \simeq g \left(\frac{\dd}{\dd t} L(t)\right) \, {\bfn}(t)+\sqrt{2 D_0}\, \boldsymbol{\xi}(t),
\eeq
where $g$ is a numerical pre-factor that depends on the geometry of the enzyme, $\boldsymbol{\xi}(t)$ is a Gaussian white noise of unit strength, and $D_0$ is the intrinsic translational diffusion coefficient.

We can describe the combined mechanochemical cycle using a two-step process, which takes the enzyme from its free state through the reaction that is accompanied by the deformation with rate $k$, and a relaxation back to it native state with rate $k_{\rm r}$. This is a simplification of a more realistic model with three states (free, substrate-bound, and reacted-deformed) and $k$ is to be understood as the combined catalytic rate that has the Michaelis-Menten form as defined above. Therefore, we can describe $L(t)$ as a telegraph process and calculate the elongation speed auto-correlation function as 
\beq
\left\langle \frac{\dd}{\dd t} L(t) \cdot \frac{\dd}{\dd t'} L(t')\right\rangle=2 b^2 \left(\frac{k k_{\rm r}}{k+k_{\rm r}}\right) \left[\delta(t-t')-\frac{1}{2} (k+k_{\rm r})e^{-(k+k_{\rm r}) |t-t'|}\right]. 
\eeq
By combining this with the orientation auto-correlation, we can calculate the effective diffusion coefficient of the enzyme, which gives the following correction
\beq
D_{\rm eff}=D_0+\frac{1}{3} g^2 b^2 \left(\frac{k k_{\rm r}}{k+k_{\rm r}}\right) \frac{2 D_{\rm r}}{2 D_{\rm r}+k+k_{\rm r}},\label{eq:DeltaD-swimming-1}
\eeq
Even for the fastest enzymes, we typically have $k_{\rm r} \approx D_{\rm r} \gg k$. Using an upper bound of $b \sim R$, we can approximate Eq. (\ref{eq:DeltaD-swimming-1}) as $\Delta D \approx k R^2$. For catalase, we obtain $\Delta D \approx 1\,\mu{\rm m}^2 {\rm s}^{-1}$, which gives an upper bound of $\Delta D/D_0 \approx 10^{-2}$. This is one order of magnitude smaller than the observed values.

{\bf (iv) collective heating} caused by treating enzymes as mobile sources of heat for exothermic reactions \cite{Golestanian2015}. This effect will cause the global temperature of the system to raise, leading to an enhancement of the fluctuations as well as simultaneously decreasing the viscosity of the solution. The combined effect can lead to $\Delta D/D_0 \sim 10^{-2}-10^{-1}$ depending on the size of the container. The nonlinearity in the heat conduction equation arising from the dependence of the heat source on temperature can lead to the possibility of the formation of shock waves and fronts, which could contribute to observation of enhanced diffusion after time-averaging.

{\bf (v) modified equilibrium} due to the changes in the hydrodynamic couplings between the different modules of an enzyme while undergoing thermal fluctuations, caused by binding and unbinding of chemicals \cite{Illien2017a,Illien2017b,AdelekeLarodo2019}. This effect originates from the observation that the centre of mass diffusion of compound asymmetric objects depends on their configuration because their internal degrees of freedom are coupled to the centre of mass translation. This phenomenon can be studied using a simple asymmetric dumbbell model, which represent the modular structure of a generic enzyme; see Fig. \ref{fig:dumbbell}(a).

\begin{figure}
\centering
\includegraphics[width=0.8\textwidth]{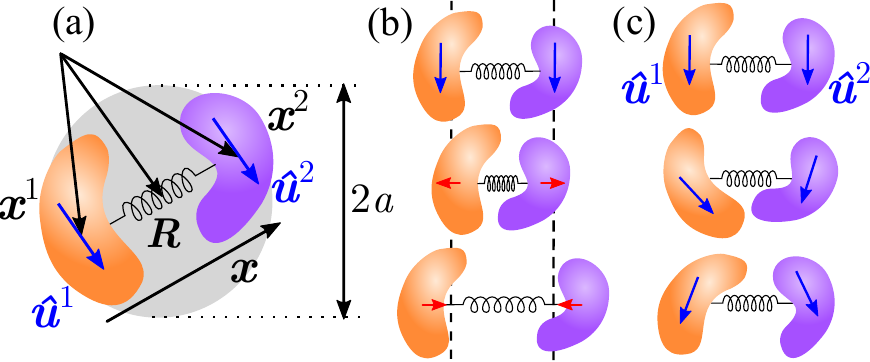}
\caption{ (a) The asymmetric dumbbell model: $a$ is the typical size of the protein, which is made of two subunits with orientations $\uu^1$ and $\uu^2$, and located at positions $\xx^1$ and $\xx^2$. $\RR$ denotes the center of mass of the protein and $\xx$ its elongation. (b) The dumbbell fluctuates around the equilibrium position of the interaction potential. The red arrows represent the forces experienced by the subunits when the dumbbell is contracted or extended. (c) The orientations of the subunits fluctuate around an equilibrium configuration.\label{fig:dumbbell}
}
\end{figure}
		
We begin our analysis with the Smoluchowski equation for a pair of Brownian particles interacting with the potential $U$. The probability $P(\xx^1 ,\xx^2 ,\uu^1, \uu^2; t)$ of finding subunit $\alpha$ at position $\xx^\alpha$ and with orientation $\uu^\alpha$ at time $t$ has the following evolution equation
\begin{eqnarray}
\partial_t P &=& \sum_{\alpha,\beta=1,2} \Big\{\nabla_\alpha \cdot  \mathbf{M}_{\mbox{\scriptsize TT}}^{\alpha\beta} \cdot \left[(\nabla_\beta U)P + \kB T \nabla_\beta P\right]+\nabla_\alpha \cdot \mathbf{M}_{\mbox{\scriptsize TR}}^{\alpha\beta} \cdot\left[(\rotop^\beta  U) P + \kB T \rotop^\beta P \right] \nonumber\\
&+&  \rotop^\alpha \cdot \mathbf{M}_{\mbox{\scriptsize RT}}^{\alpha \beta}  \cdot \left[(\nabla_\beta U) P + \kB T \nabla_\beta P\right]+\rotop^\alpha \cdot \mathbf{M}_{\mbox{\scriptsize RR}}^{\alpha \beta}  \cdot \left[(\rotop^\beta U) P + \kB T \rotop^\beta P\right] \Big\},
\label{smoluchowski}
\end{eqnarray}
where $\mathbf{M}_{\mbox{\scriptsize AB}}^{\alpha\beta}$ are elements of a mobility matrix which couples the interactions between the translational (T) and rotational (R) modes of the subunits. We now use the centre of mass and elongation coordinates, $\RR=(\xx^1+\xx^2)/2$ and $\xx=\xx^2-\xx^1$, to find the Smoluchowski equation for $P$
\begin{eqnarray}
\partial_t P & = & \frac{\kT}{4} \nabla_{\RR} \cdot \mathbf{M} \cdot \nabla_{\RR} P+ \frac{1}{2}\nabla_{\RR} \cdot \boldsymbol{\Gamma} \cdot (\nabla_{\xx}U)P +\frac{\kT}{2}( \nabla_{\RR} \cdot \boldsymbol{\Gamma} \cdot \nabla_{\xx} P +  \nabla_{\xx} \cdot \boldsymbol{\Gamma} \cdot \nabla_{\RR} P) \nonumber\\
&& + \nabla_{\xx} \cdot \mathbf{W} \cdot [\kT\nabla_{\xx} P + (\nabla_{\xx}U)P]  + \sum_{\alpha,\beta=1,2} \rotop^\alpha\cdot \mathbf{M}_\text{RR}^{\alpha \beta} \cdot [\kT \rotop^\beta P+(\rotop^\beta U) P]  \nonumber \\
&+& \sum_{\alpha=1,2} \left\{ \nabla_\alpha \cdot \mathbf{M}_\text{TR}^{\alpha\beta} \cdot  [  (\rotop^\beta  U) P + \kT \rotop^\beta P ] + \rotop^\alpha \cdot \mathbf{M}_\text{RT}^{\alpha \beta}  \cdot [(\nabla_\beta U) P + \kT \nabla_\beta P]  \right\},
\label{FP_in_R_and_x}
\end{eqnarray}
where the new mobility coefficients are mixtures of the previous ones \cite{Illien2017b,AdelekeLarodo2019}. Using a generic form for the interaction potential $U=V_0(x) + V_{01}(x) \nn\cdot \uu^1 + V_{02}(x) \nn\cdot \uu^2  +   V_{12}(x) \uu^1 \cdot \uu^2+ \cdots$, we can average Eq. (\ref{FP_in_R_and_x}) over $x$ and obtain a reduced equation for the orientational degrees of freedom, involving the definitions $\mathcal{P} = \int \dd x\,  x^2 P$ and $\aver{\cdot} = \frac{1}{\mathcal{P}} \int \dd x \, x^2 \cdot P$. The averaging procedure is motivated by the observation of a clear separation of the time-scales in the problem. The dumbbell possesses three time-scales, each describing the relaxation time of one of its degrees of freedom. The slowest of the three is the relaxation time of the centre-of-mass coordinate $\RR$. Given a potential that can be Taylor expanded around the minimum, the quadratic term gives the effective spring constant of the potential ($k$) and the time-scale for $x$ to return to its equilibrium value, $\tau_{\rm s}=\xi/k$, where $\xi$ is the friction coefficient of the enzyme. The rotational diffusion time of the enzyme $\tau_{\rm r}$, which determines the rate of the loss of memory of the orientation, is of the order $\xi a^2/\kB T$. The ratio of the two times $\tau_{\rm s}/\tau_{\rm r}$ goes as $\kB T/ka^2\sim\delta x/a$, which is a the relative deformation of the enzyme due to thermal fluctuations and is therefore bounded by unity. With this consideration, we can average over the separation of the subunits assuming $\nn$, $\uu^1$ and $\uu^2$ to be fixed. 

We define the lowest order moments of the average distribution with respect to the three unit vectors $\nn$, $\uu^1$ and $\uu^2$ as $\rho \equiv \int_{\nn,\uu^1,\uu^2}\mathcal{P}$, $\pp \equiv \int_{\nn,\uu^1,\uu^2}\nn\mathcal{P}$ and $\pp^\alpha \equiv \int_{\nn,\uu^1,\uu^2}\uu^\alpha\mathcal{P}$ and obtain the respective evolution equations in the moment expansion of the resulting equation \cite{Golestanian:2012}. To the lowest order, this yields
\begin{eqnarray}
\partial_t \rho & = & \frac{\kT}{4} \aver{m_0}  \nabla^2_{\RR} \rho + \kT \aver{\frac{\gamma_0}{x}} \nabla_{\RR} \cdot \pp + \frac{1}{3}\bigg[\bigg<\frac{\gamma_0 V_{01}}{x}\bigg>\nabla_{\RR} \cdot \pp^1 + \bigg<\frac{\gamma_0 V_{02}}{x}\bigg>\nabla_{\RR} \cdot \pp^2  \bigg],\label{momexp_rho}\\
\partial_t p_i & = & -\frac{\kT}{3} \aver{\frac{\gamma_0}{x}} \partial_{R_i}  \rho -  2 \kT \aver{\frac{w_0}{x^2}} p_i -\frac{2}{3}  \sum_{\alpha=1,2} \bigg<\frac{w_0 V_{0\alpha}}{x^2}\bigg> p^\alpha_{i},\label{momexp_p}\\ 
\partial_t p^\alpha_{i} & = & - 2\kT\langle \psi_0^{(\alpha)} \rangle  p_i^{\alpha} + \frac{1}{9}\bigg<\frac{\gamma_0 V_{0\alpha}}{x}\bigg>   \partial_{R_{i}} \rho -\frac{2}{3} [\langle\psi_{0}^{(\alpha)} V_{12}\rangle p^\beta_{i} + \langle\psi_{0}^{(\alpha)} V_{0\alpha}\rangle p_{i} - \langle M_{\text{RR}\,\,0}^{\alpha\beta} V_{12}\rangle p^\beta_{i}] \label{momexp_palpha},
\end{eqnarray}
where the coefficients are the the corresponding amplitudes of the mobility tensors in a harmonic expansion \cite{Illien2017b,AdelekeLarodo2019}. A closed equation for the density $\rho$ can be now obtained by taking the stationary limits of these equations, which yields
\beq
\partial_t\rho(\RR;t) = D_\text{eff} \nabla_{\RR}^2 \rho, 
\eeq
where the effective diffusion has the form
\beq
D_\text{eff}=D_\text{ave}-\delta D_\text{fluc}=\frac{\kB T}{4} \aver{m_0}-\frac{\kB T}{6} \frac{\aver{\gamma_0/x}^2}{\aver{w_0/x^2}} \left[1+ \text{corrections}\right]. \label{eq:Deff}
\eeq
This result highlights the fact that due to the coupling between the internal degrees of freedom and the centre of mass translation, the effective diffusion coefficient is composed of an average contribution that relates to the average mobility of the parts that make the enzyme, and a fluctuation--induced correction, which is universally negative. To obtain a physical intuition for the result, consider the force dipole created by the dumbbell on the fluid; see Fig. \ref{fig:dumbbell}(b). If this fluctuating dipole has a nonvanishing average, the asymmetry of the dumbbell couples the dipole to a net drift velocity. Considering, for instance, the force experienced by the second subunit over the timescale when the radial coordinate has equilibrated but the orientation of the dumbbell is frozen, $\aver{\FF_2} = -\aver{U'}\nn$, we notice that a nonzero average dipole will result provided $\aver{U'} \neq 0$. Assuming that the average at the relevant timescale is taken with the weight $\ex{-U/\kB T}$, it is easy to see that $\aver{U'}$ does not vanish in any dimension other than one (even though $U\rq{}|_{x=x_{\rm eq}}=0$) due to the entropic contributions to sampling of the configuration space when using radial coordinates. The resulting contribution to the diffusion coefficient is proportional to $- \aver{U'}^2$, which highlights a similarity to dispersion forces \cite{London1937}. The fluctuations can also explore the orientational degrees of freedom of the sub-units, as shown on Fig. \ref{fig:dumbbell}(c), leading to the corrections highlighted in Eq. (\ref{eq:Deff}). 

The above result is obtained in the longest time limit, as set out by the moment expansion technique. Using a path integral description, it is possible to show that the diffusion coefficient of an enzyme at short time scales is given by the average contribution only, and that the crossover to the long time behaviour---that includes the fluctuation--induced reduction of the diffusion coefficient---happens at the expected time scales \cite{Illien2017b}. 

The above formulation lends itself naturally to a new mechanism contributing to enhanced diffusion of catalytically active enzymes. Since the diffusion coefficient has a component that depends on fluctuations, any process that modifies the fluctuations will change the diffusion coefficient. In particular, substrate binding can generically supress elongation and orientation fluctuations of an enzyme, and thereby reduce the negative fluctuation--induced correction in Eq. (\ref{eq:Deff}), leading to an enhanced diffusion as given by Eq. (\ref{eq:DeffMM}). The Michaelis-Menten form of the substrate concentration dependence originates from the probability of substrate binding in stationary state, and not from the catalytic rate of reaction, which happens to have the same form [see Eq. (\ref{eq:MM})], because it is also proportional to the probability of substrate binding. Note that while this calculation was performed only using the equilibrium components of the forces between the sub-units, the nonequilibrium forces that have been dealt with for the stochastic swimming scenario can also be taken into consideration within this more elaborate formalism.

The existence of multiple mechanisms highlights that it is perhaps misleading to think about enhanced diffusion of catalytically active enzymes as one phenomenon with universal features.

\subsection{Chemotaxis of Enzymes}

\begin{figure}
\centering
\includegraphics[width=1\linewidth]{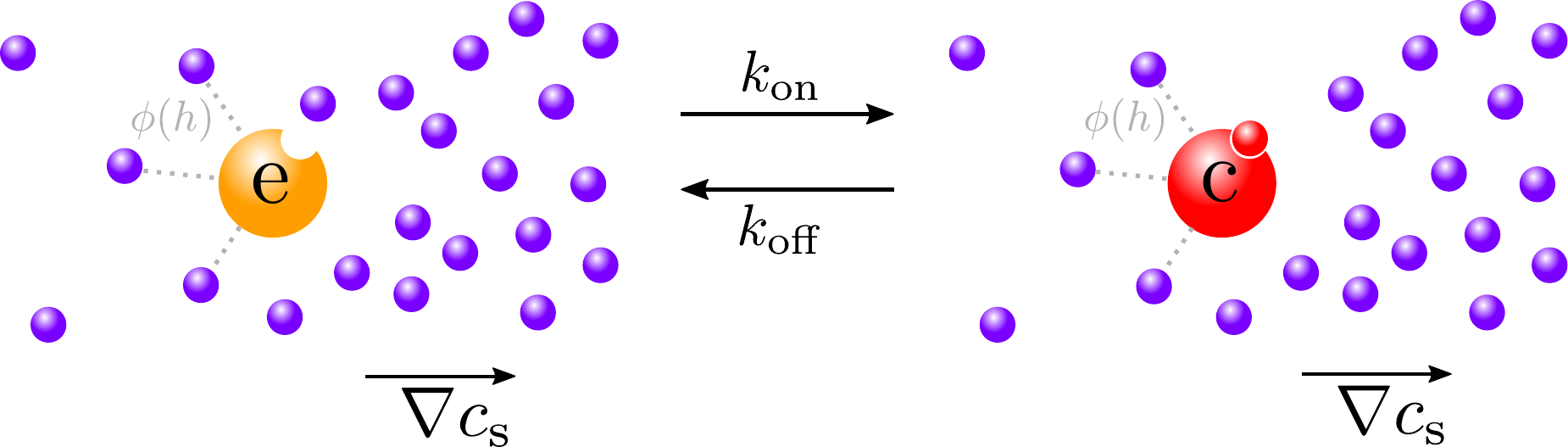}
\caption{Microscopic model for chemotaxis. The free enzyme (yellow) is in a gradient of substrate molecules (purple), with concentration $c_\mathrm{s}(\RR)$. The enzyme can bind one-on-one to a substrate molecule to form a complex (red), with binding rate $k_\mathrm{on}$ and unbinding rate $k_\mathrm{off}$, and also interacts with all other substrate molecules around it through the non-specific potential $\phi(h)$.}
   \label{fig:chemotaxismodel}
\end{figure}

In recent years, it has been observed experimentally that the non-equilibrium chemical activity of enzymes leads to intrinsically non-equilibrium behaviour and interactions already at the microscopic scale, and in particular chemotaxis towards \cite{dey14,guha2017} and away \cite{jee17} from the substrate source. Similarly to the case of enhanced diffusion, it can be argued that several mechanisms contribute to enzyme chemotaxis and that the different tendencies observed in the experiments can be explained as a competition between these contributions \cite{agud18,Agudo-Canalejo2018}. In particular, phoresis tends to direct enzymes towards their substrates, due to the predominantly attractive interaction between them, whereas enhanced diffusion tends to have the opposite effect as the enzymes will evacuate the region with higher substrate concentration more quickly due to enhanced diffusive activity.

We can use a minimal model that takes into account the following ingredients: specific binding between the substrate and the enzyme, nonspecific interactions such as electrostatic, steric and van den Waals interactions, as well as hydrodynamics interactions; see Fig. \ref{fig:chemotaxismodel}. To set up the model, we start from the full $N$-particle Fokker-Planck equation of the system of enzyme and substrate molecules and integrate out the substrate degrees of freedom to arrive at the following effective one-particle description for the free enzyme and complex probability distributions
\begin{eqnarray}
\partial_t C_{\ee}(\RR;t) & = &  \nabla_{\RR}\cdot \left[  D_{\ee} \nabla_{\RR}C_{\ee}  -  \boldsymbol{v_{\ee}}(\RR) C_{\ee} \right]  - k_\mathrm{on} C_{\ee} C_{\sss} + k_\mathrm{off} C_{\cc} \label{eq:coupled1} \\
\partial_t C_{\cc}(\RR;t) & = &  \nabla_{\RR}\cdot \left[  D_{\cc} \nabla_{\RR}C_{\cc} -   \boldsymbol{v_{\cc}}(\RR) C_{\cc} \right]  + k_\mathrm{on} C_{\ee} C_{\sss} - k_\mathrm{off} C_{\cc} \label{eq:coupled2}
\end{eqnarray}
where $C_{\sss}(\RR)$ is the concentration of substrate molecules. Equations (\ref{eq:coupled1}--\ref{eq:coupled2}) capture the effects of three important physical mechanisms. First, the free enzyme can turn into a complex and vice versa through the \emph{binding and unbinding} of a substrate molecule, with rates $k_\mathrm{on}$ and $k_\mathrm{off}$. Second, the free enzyme and complex \emph{diffuse} with diffusion coefficients respectively given by $D_{\ee} = \kT \boldsymbol{\mu}^{\ee \ee}$ and $D_{\cc} = \kT \boldsymbol{\mu}^{\cc \cc}$. Finally, the combination of non-specific and hydrodynamic interactions between the free enzyme or complex and the substrate molecules leads to a \emph{diffusiophoretic drift} of the free enzyme and complex with velocities respectively given by
\begin{eqnarray}
\boldsymbol{v_{\ee}}(\RR) & \approx & -\frac{\kT}{\eta} \left[ \int_0^\infty \mathrm{d}h h \left(1-\ex{- \phi^{\ee \sss}(h)/\kT}\right) \right] \nabla_{\RR} C_{\sss} \equiv -\frac{\kT}{\eta} \lambda_{\ee}^2 \nabla_{\RR} C_{\sss} \label{eq:v1}  \\
\boldsymbol{v_{\cc}}(\RR) & \approx & -\frac{\kT}{\eta} \left[ \int_0^\infty \mathrm{d}h h \left(1-\ex{- \phi^{\cc \sss}(h)/\kT}\right) \right] \nabla_{\RR} C_{\sss} \equiv -\frac{\kT}{\eta} \lambda_{\cc}^2 \nabla_{\RR} C_{\sss} \label{eq:v2}
\end{eqnarray}
where $\eta$ is the viscosity of the fluid. Equations (\ref{eq:v1}--\ref{eq:v2}) are approximate forms of the diffusiophoretic velocity valid for the typical case in which the range of the non-specific interactions is shorter than the size of the enzyme or the complex. We have defined here the Derjaguin length $\lambda_\alpha$ as before. Remember that, in our convention, $\lambda_\alpha^2$ may be positive or negative, with positive (negative) values corresponding to overall repulsive (attractive) interactions that lead to a depletion (an accumulation) of substrate molecules in the proximity of the enzyme.  Because enzyme--substrate interactions are generally attractive, we expect $\lambda_\alpha^2 < 0$, implying that for typical enzymes the phoretic velocity is directed towards higher concentrations of the substrate.

Equations (\ref{eq:coupled1}--\ref{eq:coupled2}) already contain all the ingredients necessary to describe enzyme--substrate interactions. Nevertheless, in order to describe chemotaxis we are actually interested in the \emph{total} concentration of enzyme, both free and bound, given by
\begin{equation}
C_{\ee}^{\tot}(\RR;t) = C_{\ee}(\RR;t) + C_{\cc}(\RR;t) \label{eq:rhotot}
\end{equation}
which corresponds to what is measured in the experiments with fluorescently tagged enzymes (free enzyme and complex cannot be distinguished). Furthermore, the typical timescale of diffusion and phoretic drift is much longer than the typical timescale of binding and unbinding. We can therefore assume that the enzyme is locally and instantaneously at binding equilibrium with the substrate, so that at any position $\RR$ we will have
\begin{equation}
k_\mathrm{on} C_{\ee}(\RR;t) C_{\sss}(\RR;t) \approx k_\mathrm{off} C_{\cc}(\RR;t) \label{eq:fasteq}
\end{equation}
at time $t$. Combining Eq. (\ref{eq:rhotot}) and Eq. (\ref{eq:fasteq}), we find the typical Michaelis-Menten kinetics for the free enzyme and the complex
\begin{equation}
C_{\ee} = \frac{K}{K + C_{\sss}}C_{\ee}^{\tot}~~\text{and}~~C_{\cc} = \frac{C_{\sss}}{K + C_{\sss}}C_{\ee}^{\tot} \label{eq:MichM}
\end{equation}
where we have defined the Michaelis constant $K \equiv k_\mathrm{off}/k_\mathrm{on}$.

Adding together Equations (\ref{eq:coupled1}) and (\ref{eq:coupled2}), and using (\ref{eq:MichM}), we finally obtain an expression for the time evolution of the total enzyme concentration
\begin{equation}
\partial_t C_{\ee}^{\tot}(\RR;t) = \nabla_{\RR} \cdot \left\{ D(\RR)\cdot\nabla_{\RR}C_{\ee}^{\tot} - [\boldsymbol{V}_\mathrm{ph}(\RR) + \boldsymbol{V}_\text{bi}(\RR)] C_{\ee}^{\tot} \right\}
\label{eq:evol}
\end{equation}
with the space-dependent diffusion coefficient
\begin{eqnarray}
D(\RR) &=& D_{\ee} + (D_{\cc} - D_{\ee})   \frac{C_{\sss}(\RR)}{K + C_{\sss}(\RR)},
\end{eqnarray}
a drift velocity arising from phoretic effects
\begin{eqnarray}
\boldsymbol{V}_\mathrm{ph}(\RR) &=& \boldsymbol{v_{\ee}}(\RR) + [\boldsymbol{v_{\cc}}(\RR)-\boldsymbol{v_{\ee}}(\RR)]  \frac{C_{\sss}(\RR)}{K + C_{\sss}(\RR)},
\label{eq:Vph}
\end{eqnarray}
as well as a drift velocity arising from the changes in diffusion coefficient due to substrate binding and unbinding
\begin{eqnarray}
\boldsymbol{V}_\mathrm{bi}(\RR) &=& - (D_{\cc} - D_{\ee}) \nabla_{\RR} \left(  \frac{C_{\sss}(\RR)}{K + C_{\sss}(\RR)} \right)= - \nabla_{\RR} D(\RR).
\label{eq:Vbi}
\end{eqnarray}
The drift velocity arising from diffusiophoresis $\boldsymbol{V}_\mathrm{ph}(\RR)$ corresponds to an `average' of the phoretic velocities of the free enzyme and the complex. With increasing substrate concentration, a smooth Michaelis-Menten-like crossover is observed between the velocity of the free enzyme $\boldsymbol{v_{\ee}}$ and that of the complex $\boldsymbol{v_{\cc}}$. In principle, this velocity may be directed towards or away from higher concentrations of substrate, depending on the details of the non-specific interactions. Because enzyme--substrate interactions are generally attractive, however, we expect that the typical phoretic velocity for enzymes will be directed towards higher substrate concentrations. The drift velocity $\boldsymbol{V}_\mathrm{bi}(\RR)$ is a direct consequence of the changes in the diffusion coefficient of the enzyme due to binding and unbinding of the substrate. This drift velocity is directed towards higher concentrations of substrate in the case of inhibited diffusion with $D_{\cc}<D_{\ee}$, and towards lower concentrations of substrate in the case of enhanced diffusion with $D_{\cc}>D_{\ee}$. In the absence of phoresis with $\boldsymbol{V}_\mathrm{ph}(\RR)=0$, Equation (\ref{eq:evol}) can then be written as $\partial_t C_{\ee}^{\tot}(\RR;t) = \nabla_{\RR}^2[D(\RR) C_{\ee}^{\tot}]$, and we would thus expect $C_{\ee}^{\tot}(\RR) \propto 1/D(\RR)$ in the steady state, i.e.~the enzyme tends to concentrate in regions where its diffusion is slowest. This type of behaviour was recently reported experimentally in Ref.~\cite{jee17} for urease, in apparent conflict with older results in the literature, Ref.~\cite{seng13}, in which urease was observed to chemotax towards higher concentrations of urea. The existence of two distinct mechanisms for chemotaxis, namely phoresis and binding-induced changes in diffusion as just described, may explain the seemingly contradictory observations. It is interesting to note that chemotactic alignment can also contribute to the process of net chemotactic drift as well as enhanced diffusion on time scales longer than the rotational diffusion time \cite{AdelekeLarodo2019b}.

\section{Phoresis on the Slow Lane: Trail-following Bacteria}	\label{sec:trail}

When bacteria live in the bulk, they use swimming motility via flagella, and search strategies like run-and-tumble to provide for themselves, as single entities. The planktonic life of bacteria is lonely, because random walks in 3D are terse. When bacteria find a surface and interact with it mechanically, at some point they ``decide'' to settle on it. The decision making process is stochastic, and involves a cross-correlation between cAMP signal and type-IV pili (TFP) mechanical activity, which lasts through several generations of cell division \cite{Lee2018}. The settled bacteria lose their flagella and grow pili, and start exploring the 2D surface using surface attachment motility: pili elongate and retract through polymerization and depolymerization, and they stochastically attach to the surface and detach from it. On surfaces, bacteria live communal lives, because random walks in 2D are compact, which implies that bacteria meet very often even if they do not signal each other.

In practice, however, bacteria do signal each other and use it to self-organize much more efficiently than random walks can provide for them. They achieve this by leaving behind a metabolically expensive polysaccharide trail. While this might appear to be an inefficient use of individual resources, the communal lifestyle that follows from it benefits them immensely. This process belongs to the general class of motility with chemical sensing, in the extreme limit where the chemicals do not diffuse. Here we describe the detailed derivation of a set of stochastic equations of motion for bacteria that phoretically interact via non-diffusing trails they leave behind themselves. The theoretical formulation is based on a microscopic motility model that takes into account stochastic attachment and detachment of pili, as well as friction coefficients and pili contraction forces that depend on the amount of polysaccharide \cite{Kranz-trail-PRL:2016,Gelimson-trail-PRL:2016}.

\begin{figure}[b]
\includegraphics[width =  0.4 \columnwidth]{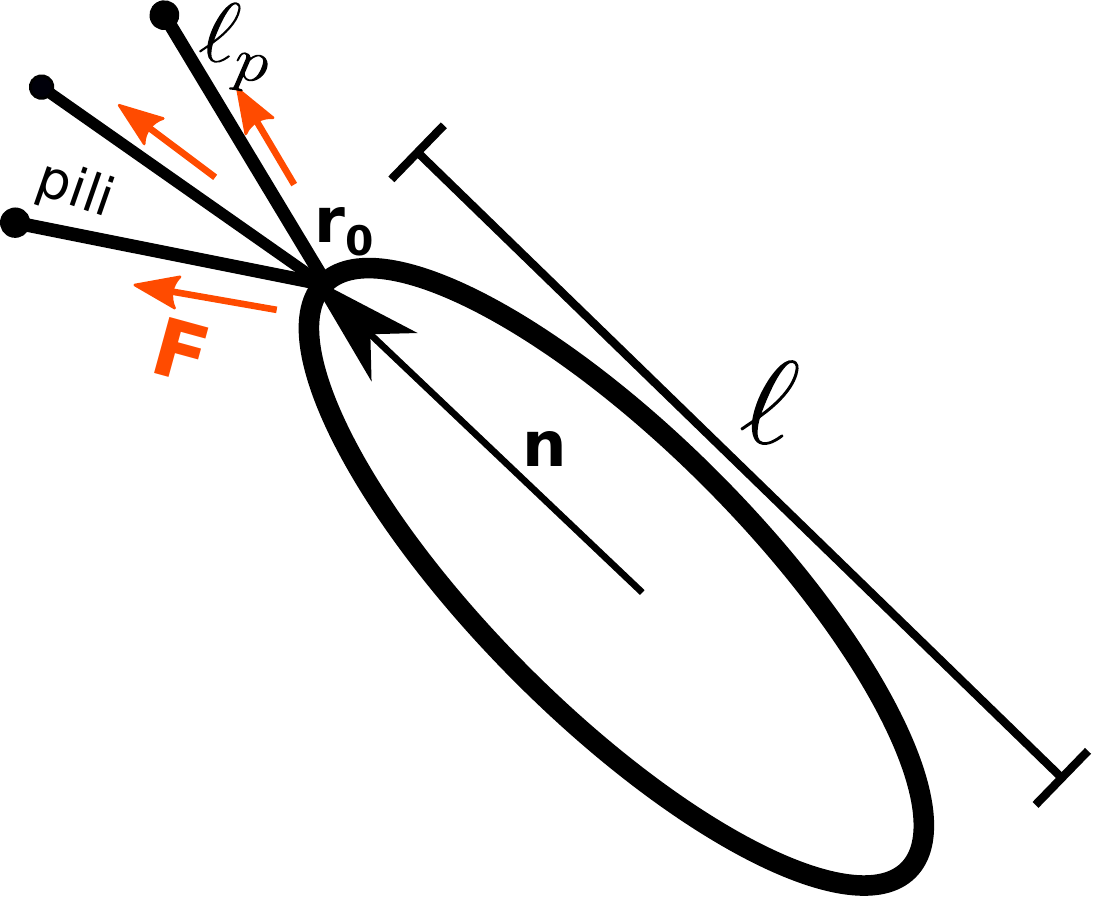}
\caption{Microscopic model: the bacterium has type-IV pili of length $\ell_p$, which can attach to a surface and pull the bacterium forward with a generically polysaccharide-dependent force. Originating at a bacterial pole $\bm{r}_0$, the pili point in different directions $\hat{\bm{e}}_i, \; i = 1,\dots,N$, such that the pili tips are located at the positions $\bm{r}_0 + \ell_p \hat{\bm{e}}_i$. A pilus pulling force $f(\psi_i) \hat{\bm{e}}_i$ on the bacterial tip $\bm{r}_0$ is only generated if the pilus is attached to the surface and in general the attachment and detachment rates $\lambda(\psi_i)$ and $\mu(\psi_i)$ will be dependent on the polysaccharide concentration $\psi_i = \psi( \bm{r}_0 + \ell_p \hat{\bm{e}}_i )$ at the end of pilus $i$ \label{model-schematics}}
\end{figure}

\subsection{Systematic Derivation of Stochastic Dynamical Equations}

We consider a bacterium of length $\ell$ (Fig. \ref{model-schematics}). We will assume that each of the $N$ pili located at one tip (or at both tips) of a bacterium will attach to, and detach from, the surface and pull in a polysaccharide-dependent way. The direction unit vectors of the pili $\bm{e}_i, \; i = 1,\dots, N$ are assumed to originate from the bacterial tip $\bm{r}_0$. One could consider variants of this model where the pili distribution is different. A single pilus tip, located at $\bm{r}_0 + \ell_p \hat{\bm{e}}_i$, will randomly attach to the surface with a polysaccharide-dependent rate $\lambda(\psi_i)$, detach with a rate $\mu(\psi_i)$ and pull in its direction with a force $f(\psi_i)$ if it is attached to the surface. Here, $\psi_i$ is the polysaccharide concentration at the tip of pilus $i$, $\psi_i = \psi( \bm{r}_0 + \ell_p \hat{\bm{e}}_i )$. For each pilus $i$ we define $\Theta_i = 1$ if the pilus is attached and $\Theta_i = 0$ if it is detached. $\Theta_i$ will have a mean
\begin{equation}
\bar{\Theta} = \langle \Theta_i \rangle = \frac{\lambda(\psi)}{\lambda(\psi) + \mu(\psi)},
\end{equation}
and a variance
\begin{equation}
\sigma^2 = \langle (\Theta_i - \bar{\Theta})^2 \rangle = \frac{\lambda(\psi)\mu(\psi)}{[\lambda(\psi) + \mu(\psi)]^2}.
\end{equation}
The total force exerted at the bacterial tip will then be
\begin{equation}
\bm{F} =  \sum_i \hat{\bm{e}}_i f(\psi_i) \Theta_i.
\end{equation}

Using the form $\Theta_i = \bar{\Theta} + \delta \Theta_i$, and the Taylor expansion $\psi_i = \psi( \bm{r}_0 + \ell_p \hat{\bm{e}}_i ) \approx \psi( \bm{r}_0) + \ell_p ( \bm{\nabla} \left. \psi \right|_{\bm{r}_0} \cdot \hat{\bm{e}}_i)$, we obtain
\begin{equation}
\bm{F} =  \sum_i \hat{\bm{e}}_i f(\psi) \bar{\Theta}(\psi) + \sum_i \hat{\bm{e}}_i \ell_p (\bm{\nabla} \psi \cdot \hat{\bm{e}}_i) \partial_\psi( f \bar{\Theta})+ \sum_i \hat{\bm{e}}_i f(\psi_i) \delta \Theta_i.
\end{equation}

We can express the pili orientation vectors in terms of the bacterial body orientation $\hat{\bm{n}}$ and the orientation $\hat{\bm{n}}_\perp = \hat{\bm{e}}_z \times \hat{\bm{n}}$  orthogonal to it as $\hat{\bm{e}}_i = \cos\vartheta_i \hat{\bm{n}} + \sin \vartheta_i \bm{n}_\perp$. It is reasonable to assume that the pseudomonas bacteria mainly contributing to surface-mediated chemotaxis will have a pili distribution that is approximately symmetrical with respect to the body orientation (if this was not the case the bacterium would generate a torque in one preferred direction, rotate around itself and effectively stay in one point). Therefore, we can neglect terms in the sum over the pili that are odd in $\sin \vartheta_i$, like $\langle \sin \vartheta_i \rangle = \frac{1}{N} \sum_i \sin \vartheta_i$ or $\langle \cos \vartheta_i \sin \vartheta_i \rangle = \frac{1}{N} \sum_i \cos \vartheta_i \sin \vartheta_i$. The force can then be written as
\begin{eqnarray}
\bm{F} &= &N \langle \cos \vartheta_i \rangle f(\psi) \bar{\Theta}(\psi) \hat{\bm{n}} + N \langle \cos^2 \vartheta_i \rangle \ell_p \partial_\psi \left[ f(\psi) \bar{\Theta}(\psi) \right]  \left(\bm{\nabla} \psi \cdot \hat{\bm{n}} \right) \hat{\bm{n}} + \hat{\bm{n}} \overbrace{\sum_i \cos \vartheta_i f(\psi_i) \delta \Theta_i}^{\delta F_{\parallel}}  \nonumber \\
&&+  N \langle \sin^2 \vartheta_i \rangle \ell_p \partial_\psi \left[ f(\psi) \bar{\Theta}(\psi) \right] \left(\bm{\nabla} \psi \cdot \hat{\bm{n}}_\perp \right) \hat{\bm{n}}_\perp+ \hat{\bm{n}}_\perp \underbrace{\sum_i \sin \vartheta_i f(\psi_i) \delta \Theta_i}_{\delta F_{\perp}}.
\label{f-final}
\end{eqnarray}
If the pili attachment events of two different pili are independent of each other we have $\langle \delta \Theta_i \delta \Theta_j \rangle = \sigma^2(\psi) \delta_{ij}$. Moreover, we can calculate the auto-correlation of the attachment as $\langle \delta \Theta_i(t) \delta \Theta_i(t') \rangle = \sigma^2 \, e^{-(\mu+\lambda) |t-t'|} \,\delta_{ij}$. If we focus on time scales that are longer than the average attachment/detachment time, we can represent this auto-correlation as a delta function
\begin{equation}
\langle \delta \Theta_i(t) \delta \Theta_i(t') \rangle = \frac{\sigma^2(\psi)}{[\lambda(\psi) + \mu(\psi)]} \,\delta_{ij} \delta(t-t')=\frac{\lambda(\psi)\mu(\psi)}{[\lambda(\psi) + \mu(\psi)]^3}\,\delta_{ij} \delta(t-t').
\end{equation}
With this, we get the mean square fluctuations of the parallel force as
\begin{eqnarray}
\big\langle \delta F_{\parallel}^2 \big\rangle  =  N \langle \cos^2 \vartheta_i \rangle f^2(\psi) \sigma^2(\psi) + O \left(\bm{\nabla} \psi \cdot \ell_p\right).
\end{eqnarray}
Analogously, for the mean square fluctuations of the perpendicular force we obtain
\begin{eqnarray}
\big\langle \delta F_{\perp}^2 \big\rangle  =  N \langle \sin^2 \vartheta_i \rangle f^2(\psi) \sigma^2(\psi) + O \left(\bm{\nabla} \psi \cdot \ell_p\right).
\end{eqnarray}
The attachment/detachment of pili on the surface can be considered a set of $N$ random events and in case of $N \gg 1$ we can expect that the overall fluctuations in the force can be well approximated by a Gaussian with a variance of $N \langle \cos^2 \vartheta_i \rangle f^2(\psi) \sigma^2(\psi)$ in the direction $\hat{\bm{n}}$ and a variance of $N \langle \sin^2 \vartheta_i \rangle f^2(\psi) \sigma^2(\psi)$ in the direction $\hat{\bm{n}}_\perp$.

The pili-generated force will propel the bacterium and also generate a torque. Therefore, we obtain the following translational equation of motion
\begin{equation}
\frac{d \bm{r}}{d t} = \frac{1}{\gamma_\parallel} \, \bm{F}_\parallel +  \frac{1}{\gamma_\perp} \, \bm{F}_\perp = v(\psi) \hat{\bm{n}} + A(\psi) (\bm{\nabla} \psi \cdot \hat{\bm{n}}_\perp) \hat{\bm{n}}_\perp + B(\psi) (\bm{\nabla} \psi \cdot \hat{\bm{n}}) \hat{\bm{n}} +  \sqrt{ 2 D_\parallel } \; \eta^\parallel \hat{\bm{n}} + \sqrt{2 D_\perp} \; \eta^{\perp} \hat{\bm{n}}_\perp,
\label{eq:position-nonoise}
\end{equation}
and the following rotational equation of motion
\begin{equation}
\frac{d \hat{\bm{n}}}{d t} = \bm{\omega} \times \hat{\bm{n}} = - \frac{1}{\gamma_{\text{rot}}} \, \hat{\bm{n}} \times  \bm{\tau} =  -  \chi(\psi) \hat{\bm{n}} \times[\hat{\bm{n}} \times \bm{\nabla} \psi] +  \sqrt{2 D_{\rm r}(\psi)} \; \eta^{\perp} \hat{\bm{n}}_\perp.
\label{n-final}
\end{equation}
The parameters are known functions of the rates, but there are only two independent parameters
\begin{eqnarray}
&&v(\psi) = N \, c_1\, \frac{f(\psi) }{\gamma_\parallel} \, \frac{\lambda(\psi)}{\lambda(\psi) + \mu(\psi)}, \label{v-psi}\\
&&D_{\rm r} =  \frac{1}{8} N \ell^2 (1-c_2)\, \frac{f^2(\psi) }{ \gamma_{\text{rot}}^2} \, \frac{\lambda(\psi)\mu(\psi) }{ [\lambda(\psi) + \mu(\psi)]^3 }. \\
\end{eqnarray}
The rest of the parameters can be expressed in terms of $v(\psi)$ and $D_{\rm r}(\psi)$ in the following way
\begin{eqnarray}
&&D_\parallel(\psi) = \left[\frac{4 \gamma_{\text{rot}}^2  c_2}{  \ell^2 \gamma_\parallel^2 (1-c_2)}\right] D_{\rm r}(\psi), \\
&&D_\perp(\psi) = \left( \frac{2 \gamma_{\text{rot}} }{ \gamma_\perp \ell } \right)^2 D_{\rm r}(\psi),\\
&&\chi(\psi) = \left[\frac{\gamma_{\parallel} \ell \ell_p (1-c_2)}{2 \gamma_{\text{rot}} c_1} \right] \partial_\psi v(\psi)  +  \theta(0)  \partial_\psi D_{\rm r}(\psi),\\
&&A(\psi) =  \left[\frac{\ell_p (1-c_2) \gamma_\parallel}{c_1 \gamma_\perp} \right] \partial_\psi v(\psi)  + \theta(0) \left( \frac{2 \gamma_{\text{rot}} }{ \gamma_\perp \ell } \right)^2 \, \partial_\psi D_{\rm r}(\psi),\\
&&B(\psi) =   \left[\frac{\ell_p c_2}{c_1}\right] \partial_\psi v(\psi) +\theta(0) \left[\frac{ 4 \gamma_{\text{rot}}^2 c_2}{\ell^2 \gamma_\parallel^2 (1-c_2)  }\right] \partial_\psi D_{\rm r}(\psi),\label{Dr}
\end{eqnarray}
where $c_1 = \langle \cos \vartheta_i \rangle$ and $c_2=\langle \cos^2 \vartheta_i \rangle$ are assumed to be polysaccharide independent. 

The noise terms $\eta_\parallel$, $\eta_{\perp}$ are Gaussian noise components parallel and orthogonal to $\hat{\bm{n}}$ with $\langle \eta^{\parallel / \perp}\rangle = 0$, $\langle {\eta}^{\parallel}(t) {\eta}^{\parallel}(t')\rangle = \delta (t-t')$, $\langle {\eta}^{\perp}(t) {\eta}^{\perp}(t')\rangle =  \delta (t-t')$, and $\langle {\eta}^{\parallel}(t) {\eta}^{\perp}(t')\rangle = 0$. Our formulation contains multiplicative noise and therefore needs an additional interpretation rule as discussed in \cite{van1981ito}: the It\^{o} interpretation corresponds to $\theta(0)= 0$ whereas $\theta(0)= 1/2$ corresponds to Stratonovich. 

\subsection{Single-particle Dynamics}

The microscopic derivation provides us with a powerful framework that can be used towards a deeper understanding of the behaviour of trail-following bacteria. The following two aspects are particularly interesting for consideration.

\subsubsection{Dependence of the Parameters on the Chemical Concentration}

We are interested to address generic questions about this system, such as the influence of the polysaccharide on the speed of bacteria. From the expression for $v(\psi)$ in Eq. (\ref{v-psi}), we can see that the answer to this question is not trivial. The expression for speed involves the force $f(\psi)$ that is likely to increase with $\psi$ due to stronger pulling (although it does not necessarily need to change), the friction coefficient $\gamma(\psi)$ that is likely to increase with $\psi$ due to the stickiness of the polysaccharide, the attachment rate $\lambda(\psi)$ that is likely to increase due to preferential attachment, and the detachment rate $\mu(\psi)$ that is likely to decrease to preferential attachment. Given the combination with which these parameters appear in the expression for speed in Eq. (\ref{v-psi}), it is not straightforward to predict the overall trend, as there are competing contributions. 

By fitting single-particle trajectories from the experiment to Eqs. (\ref{v-psi})--(\ref{Dr}), it was possible to deduce that $v(\psi) \propto \frac{f}{\gamma_\parallel}\frac{\lambda}{\lambda+\mu}$ increases significantly with $\psi$, whereas $D_{\rm r}(\psi) \propto  \frac{f^2}{ \gamma_{{\rm rot}}^2} \frac{\lambda \mu}{ [\lambda + \mu]^3 }$ is relatively insensitive to $\psi$ at the lowest order \cite{Kranz-trail-PRL:2016,Gelimson-trail-PRL:2016}. While this information is not sufficient to help us extract all the functions, it suggests that the observations are consistent with the following pattern: $\lambda \approx \kappa_0+ \kappa_1 \psi$, $\mu \approx \kappa_0- \kappa_1 \psi$, and $\frac{f}{\gamma} \approx {\rm constant}$ \cite{Kranz-trail-PRL:2016,Gelimson-trail-PRL:2016}. This provides us with novel insight into the behaviour of trail-following bacteria.

\subsubsection{Perpendicular Alignment Trail-following Strategy}

\begin{figure}[t]
\includegraphics[width =  0.6 \columnwidth]{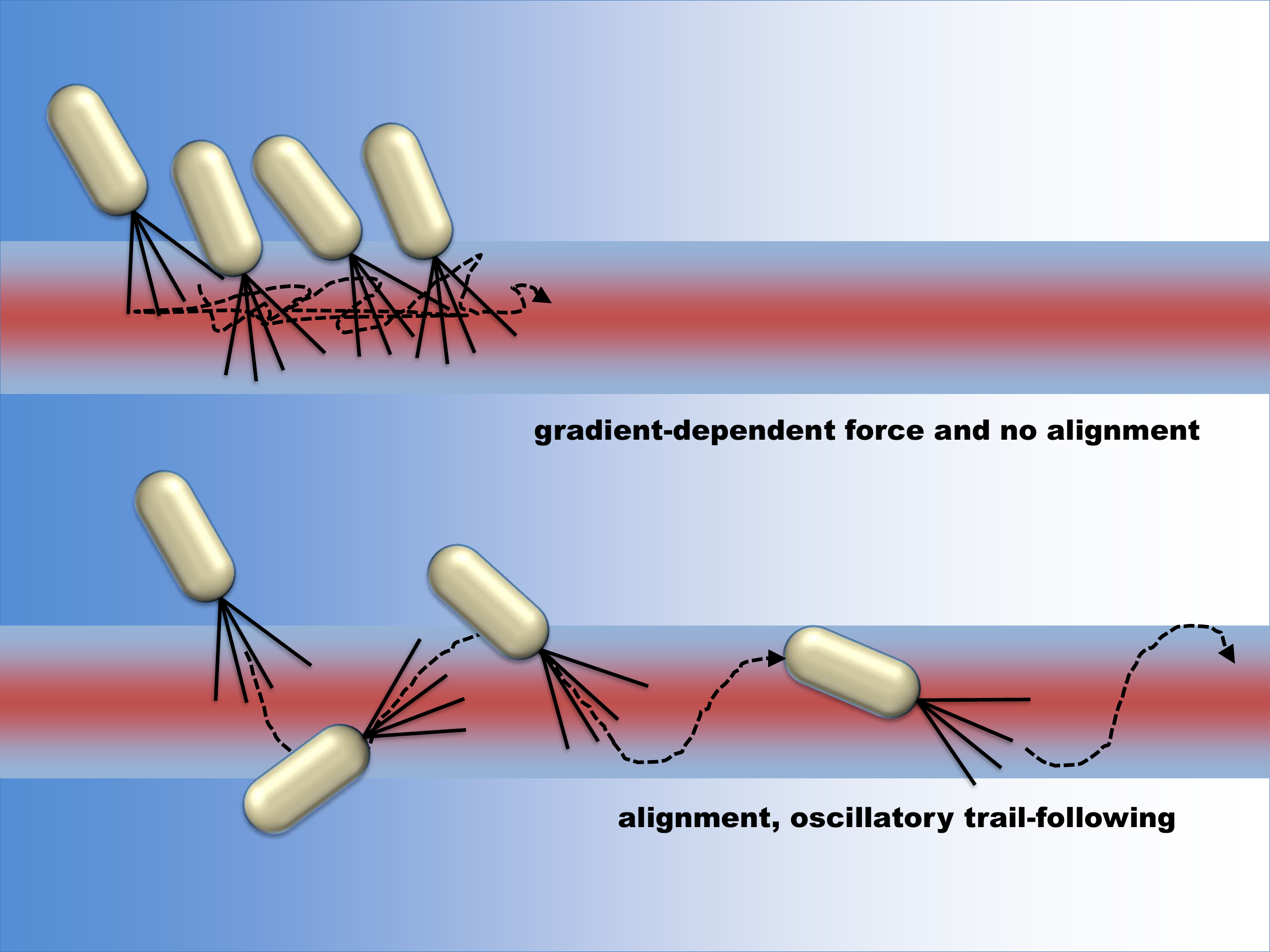}
\caption{Translational gradient sensing versus orientational gradient sensing: The perpendicular alignment strategy of chemotaxis allows for an efficient trail-following mechanism via oscillatory trajectories. \label{trail}}
\end{figure}

For trail-following bacteria, it is not clear how a good strategy for following an existing trail can be devised using gradient sensing, in line with common search strategies of microorganisms. The reason is that the main gradient in a trail is setup perpendicularly to the main axis (see Fig. \ref{trail}). Therefore, a conventional or ``translational'' gradient sensing will only ensure that the bacterium is attracted to the centre of the trail cross section; to move along trail, it will need to rely on random fortuitous alignment of the body that will give it the appropriate velocity along the trail via self-propulsion. Our model reveals an alternative and more efficient strategy, via alignment along the gradient. This ``orientational'' gradient sensing strategy ensures that the bacterium is always aligned perpendicularly to the trail heading towards its centre. When crossing through the maximum at the centre, the alignment strategy will make the bacterium turn, resulting in an oscillatory or zigzag trajectory along the trail. This provides the microorganism with an efficient trail-following strategy without any need for a complex feedback mechanism between motility and sensing. This remarkable scenario is a natural consequence of the microscopic interactions that are incorporated in the model, even though it might appear to be counter-intuitive at first.

\subsection{Many-particle Dynamics}

The above equations need to be solved self-consistently together with the equation that describes trail deposition at rate $k$ by the moving bacteria
\beq
\partial_t \psi - \xcancel{{\cal D}_{\rm p} \nabla^2 \psi}=k \sum_a \delta(\bfr-\bfr_a(t)),
\eeq
where the second term can be ignored as for the polysaccharide trails ${\cal D}_{\rm p} \approx 0$. Solving the above equations for the many-body bacterial system provides a quantitative account of the process of early biofilm nucleus formation that agrees well with experiments \cite{Gelimson-trail-PRL:2016}. In particular, it unravels a hierarchical self-organization of the multi-cellular system at very low densities, which exhibits a power law visit distribution of space, whose prediction from the theory is quantitatively matched with the experimental observations without any fitting parameters for the many-body dynamics.

\subsection{Chemotactic Localization Transition}

An intriguing outcome of the above formulation for trail-following bacteria is the emergence of a localization transition when a microorganism can interact with its own trail  \cite{Kranz-trail-PRL:2016}. Let us consider the trail excreted from a microorganism as characterized by the equation
\beq
\partial_t \psi(\bfr, t)=k \delta_R^2\left(\bfr - \bfr(t)\right), 
\eeq
where $\delta_R^2\left(\bfr - \bfr(t)\right)\equiv\theta(R^2 -r^2)/\pi R^2$ is a ``regularized delta function'' that accounts for the finite size $R$. Integrating this equation, we find for the trail profile at time $t$ and position $\bfx$ as
\begin{equation}
  \label{eq:2}
  \psi(\bfx, t) = k\int_0^t \dd t'\,\delta_R^2\left(\bfx - \bfr(t')\right)=\frac{k}{\pi R^2} \int_0^t \dd t'\,\theta(R^2 - |\bfx - \bfr(t')|^2).
\end{equation}
The trail width $2R$ defines a microscopic time scale $\tau = R/v_0$, which gives the trail crossing time. 

The orientation dynamics written in terms of the angle $\varphi(t)$ defined via $\bfn= (\cos\varphi,\sin\varphi)$ is governed by the following Langevin equation
\begin{equation}
  \label{eq:3}
  \partial_t\varphi(t) = \chi\partial_{\perp} \psi(\bfr(t), t) + \xi(t),
\end{equation}
where $\partial_{\perp} \psi = \bfn_{\perp}(t)\cdot\nabla\psi(\bfr(t), t)$, with the lateral unit vector given as $\bfn_{\perp}= (-\sin\varphi,\cos\varphi)$. Here $\xi(t)$ is a Gaussian random variable obeying $\aver{\xi(t)\xi(t')} = 2D_{\rm r}^0 \delta(t - t')$ and $D_{\rm r}^0$ is the microscopic rotational diffusion coefficient controlling the persistence time $1/D_{\rm r}^0$. 

The lateral gradient can be calculated by projecting the gradient of the trail profile onto the lateral unit vector. From the definition of the trail field we have
\begin{equation}\label{eq:30}
\nabla_{\bfx}\psi(\bfx, t) = -\frac{2k}{\pi R^2}\int_0^t \dd t' [\bfx - \bfr(t')] \delta(R^2 - |\bfx - \bfr(t')|^2),
\end{equation}
and with a change of variable $t'\to t - t'$, we find
\begin{equation}  \label{eq:40}
\partial_{\perp}\psi(\bfr(t), t)= -\frac{2k}{\pi R^2}\int_0^t \dd t'[\bfr(t) - \bfr(t-t')] \cdot \bfn_{\perp}(t)\delta(R^2 - |\bfr(t) - \bfr(t-t')|^2).
\end{equation}

The translational dynamical equation described by
\begin{equation}\label{eq:1}
\dt{\bfr(t)}= v_0\bfn(t),
\end{equation}
can be recursively integrated as
\begin{equation}\label{eq:9}
\bfr(t-\tau) - \bfr(t) = v_0 \int_t^{t-\tau} \dd u \, \bfn(u) = v_0 \int_t^{t-\tau} \dd u \left[\bfn(t)+\int_t^{u} \dd w \, \dot{\bfn}(w)\right],
\end{equation}
one finds $[\bfr(t-\tau) - \bfr(t)]^2 = v_0^2\tau^2 + {\mathcal O}(\tau^3)$ and
\begin{equation}  \label{eq:10}
 [\bfr(t-\tau) - \bfr(t)]\cdot\bfn_{\perp}(t)= - v_0\int_0^{\tau} \dd u \int_0^{u} \dd w \left\{\chi \hat{\bm e}_z \cdot \left[\bfn(t-w)\times \nabla\psi(t-w)\right] + \xi(t-w)\right\} \bfn_{\perp}(t-w)\cdot\bfn_{\perp}(t).
\end{equation}
An identical iteration shows $\bfn_{\perp}(t-w)\cdot\bfn_{\perp}(t) = 1 + \mathcal O(w)$ and thus
\begin{equation}  \label{eq:11}
[\bfr(t-\tau) - \bfr(t)]\cdot\bfn_{\perp}(t) = - v_0\int_0^{\tau} \dd u \int_0^{u}\dd w \left[\chi\partial_{\perp}\psi(t-w) +\xi(t-w)\right]+ \mathcal O(\tau^3).
\end{equation}
Using the above two approximations in Eq.~(\ref{eq:40}) and performing one of the integrals yields a stochastic integral equation
\begin{equation} \label{eq:4}
\partial_{\perp} \psi(t) = \frac{\Omega}{\tau}\int_0^{\tau} \dd u \, (\tau-u) \, \left[\partial_{\perp} \psi(t-u) + \xi(t- u)/\chi \right],
\end{equation}
where $\Omega = k\chi\tau/(\pi R^3)$ is an effective turning rate, which increases for more intense trails (larger $k$) and for more sensitive organisms (larger $\chi$). The delay $\tau$ reflects the memory imparted by the trail. The closed set of equations (\ref{eq:3}), (\ref{eq:1}), and (\ref{eq:4}) constitute our effective dynamical description of the system.

\begin{figure}[t]
\includegraphics[width =  0.6 \columnwidth]{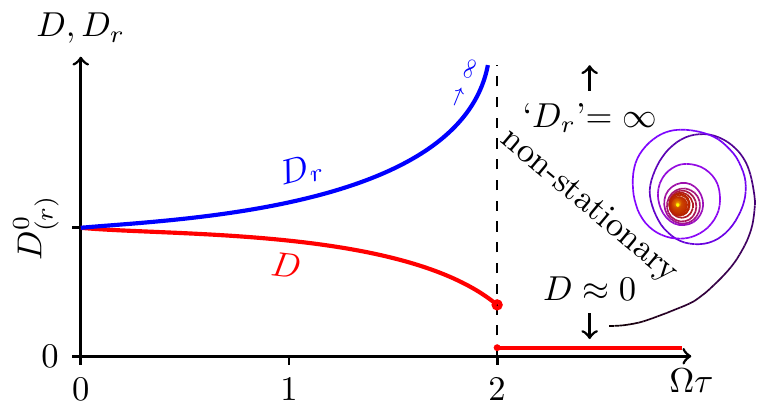}
\caption{Phase diagram of the dynamics of the microorganism with trail-mediated self-interaction, the effective translational diffusion coefficient $D$, and the effective rotational diffusion coefficient $D_{\rm r}$, as functions of the dimensionless turning frequency $\Omega \tau$ that is a measure of the rate of trail deposition. A location transition is observed at $\Omega \tau=2$. \label{localization}}
\end{figure}

For the average gradient one finds $\aver{\partial_{\perp}\psi}\sim e^{\alpha t}$ where the rate $\alpha$ is given implicitly as the solution of $\lambda(\alpha) = 0$ where
\begin{equation} \label{eq:5a}
\lambda(\alpha) = 1 - \frac{\Omega\tau}{\alpha\tau}\left[1 + \frac{1}{\alpha\tau}\left(e^{-\alpha\tau} - 1\right)\right].
\end{equation}
The relevant dimensionless parameter controlling the behaviour of our system is $\Omega\tau={k\chi}/{(\pi R v_0^2)}$. For $\Omega\tau < 2$, $\alpha < 0$ and Eq.~(\ref{eq:4}) defines a random process with zero mean that leads to a stationary dynamics which is time-translation invariant.  For $\Omega\tau > 2$ one finds $\alpha>0$, i.e., the gradient (angular velocity) diverges exponentially in time. This implies that the trajectory develops a spiral form and converges to a localized point. We thus find that there is a maximum value of the product of trail deposition rate and sensitivity, $k\chi$, that allows steady-state motion. The phase diagram of the system is shown in Fig. \ref{localization}.

\section{Chemotaxis and Cell Division}		\label{sec:growth}

Chemotaxis plays a crucial role in living systems even at time scales relevant to cell division. Examples of such cases include cancer metastasis, the early stages of bacterial colony formation, wound healing, and development of embryos. However, the underlying mechanisms of these important processes are not fully understood due to the high complexity of these living many-body systems. 

It is possible to shed some light on the dynamics of a system of chemically interacting cells, by taking into account cellular growth and death (Fig. \ref{growth}). Using a stochastic field theoretical framework, one can construct a dynamical equation for the density fluctuations of such a system as follows
\begin{equation}
\partial_t \rho = D_\mathrm{c} \nabla^2 \rho - \theta \rho - \nu_1 \nabla \cdot \left[\rho \nabla \left(\frac{1}{\nabla^2}\right) \rho \right] -\frac{\nu_2}{2}  \rho^2+ \eta.
\label{eq:rho-eq}
\end{equation}
In this equation, $\nu_1$ is the coupling constant for the term that represents chemotaxis and $\nu_2$ is the coupling constant for the term that represents growth and death. One observes that the two seemingly unrelated nonlinearities have the same strength in terms of power counting, but possess different symmetries: the chemotactic term is conserved and the growth term is not \cite{gelimson+golestanian15oth}. The noise correlator in Fourier space 
\beq
\langle \eta(\mathbf{k}, \omega) \eta(\mathbf{k}', \omega') \rangle= 2 \left[D_0 + D_2 k^2\right] (2 \pi)^{d+1} \delta(\mathbf{k} + \mathbf{k}')\delta(\omega + \omega'),
\eeq
highlights conserved and non-conserved contributions to the noise strength as well. The large scale behaviour of a population of cells that grow and interact through the concentration field of the chemicals they secrete can be studied using the method of dynamical renormalization group (RG). The combination of the effective long-range chemotactic interaction and lack of number conservation leads to a rich variety of phase behaviour in the system, including anomalous diffusion and number fluctuations at the critical point.

\begin{figure}[t]
\includegraphics[width =  0.3 \columnwidth]{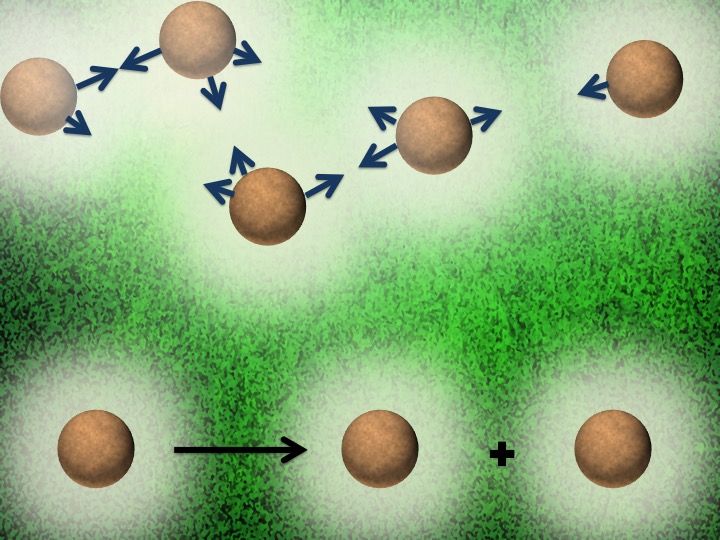}
\caption{The stochastic process of birth and death can compete with chemical interactions.\label{growth}}
\end{figure}

\section{Concluding Remarks}		\label{sec:concl}

A collection of interacting active colloids could serve as a promising model system to study collective non-equilibrium dynamics, as both the single-particle activity and the interactions could be controlled by construction. This approach will be directly relevant to studies of chemically active processes in biological systems, from enzyme cluster formation to multi-cellular organization. Moreover, it helps solve one of the challenges in making synthetic autonomous microscopic systems with non-trivial collective dynamics, which is the injection of energy at the level of individual objects and the ability to engineer the emergent properties of a collection of such objects.

\section*{Acknowledgements}
I would like to acknowledge my many collaborators over the last few years with whom I have had the pleasure of working on Phoretic Active Matter: Tunrayo Adeleke-Larodo, Jaime Agudo-Canalejo, Armand Ajdari, Andrew Campbell, Jack Cohen, Stephen Ebbens, Anatolij Gelimson, Jon Howse, Yahaya Ibrahim, Pierre Illien, Richard Jones, Till Kranz, Tannie Liverpool, K.R. Prathyusha, Sriram Ramaswamy, Suropriya Saha, Rodrigo Soto, and Gerard Wong.



\end{document}